\documentclass[showpacs,aps]{revtex4-2}
\usepackage{amsmath}
\usepackage{graphicx}

\begin{document}

\title{\bf Poisson type conformastat  spherically symmetric anisotropic fluid spacetimes}








\author{Gonzalo Garc\'{\i}a-Reyes}

\email[e-mail: ]{ggarcia@utp.edu.co}
\affiliation{Departamento de F\'{\i}sica, Universidad Tecnol\'ogica de Pereira,
 A. A. 97, Pereira, Colombia}

\begin{abstract}

We construct   conformastat  spherically symmetric  spacetimes   representing  anisotropic fluid ma\-tter distributions  
from given  solutions of the Poi\-sson's equation of Newtonian gravity   and its  corres\-ponding circular speed profile. 
As simple examples, we present  three  families of  spherically symme\-tric spacetimes which we apply in constructing new models of  relativistic anisotropic  thick spherical shells, and  of relativistic galaxy models   composite by  a central spherical  bulge, the thick disk and the dark matter halo, writing in this case the metric  in cylindrical coordinates.  
Moreover,  the geodesic motion of test particles in stable circular orbits around such structures is studied.  We build anisotropic  fluid sources for these spacetimes  which
satisfy   all the energy conditions and  the principal  stresses are positive quantities  (pressure).
  

\end{abstract}

\maketitle

\section{ Introduction}

Spherically symmetric  distributions of matter  are important   in relativistic astrophysics as models of 
neutron stars, highly dense stars,
gravastars, dark energy stars, quark stars and  galactic nuclei.  Spherical shells  are also   useful in astrophysics as models of supernovas,   in general relativity 
to analyze basic issues of gravitational collapse and  as
sources of vacuum gravitational fields, and in cosmology 
\cite{Schmidt}. 
Anisotropic thick spherical shell models 
as sources of  static spherically symmetric spacetimes 
built  from  Newtonian  potential-density pairs in isotropic coordinates  were studied  in Refs. \cite{Let-shell1,Let-shell2} for a Schwarzschild type  metric and  in  Ref. \cite{GGR-shell} for Majumdar-Papapetrou type fields.   This same approach has also been  used in building perfect fluid sources for these spacetimes \cite{GGR-perfect} and 
 regarding static axisymmetric fields  in the investigation  of anisotropic galaxy relativistic  models   for a  Schwarzschild type  metric \cite{Let-galaxy}
and in Ref.  \cite{GGR-K} in constructing
three-dimensional  sources for   Majumdar-Papapetrou type fields.

In this work, we build   conformastat spherically symmetric spacetimes     which we apply in constructing new relativistic  models of  anisotropic  thick spherical shells  and  galaxies  from Newtonian potential-density pairs   and its  corresponding  circular speed profile. 
As in the  references  cited above,
one of the metric functions,  the spatial components of the  metric, is obtained of the fact 
that one of the field equations is a nonlinear Poisson
type equation. This equation  is solved for a particular  energy density profile which is chosen
by demanding that    in Newtonian limit  it reduces to its Newtonian value. The other metric function, the temporal  component of the metric,  is obtained analytically in this work  by kinematic considerations  by also  requiring  that in the Newtonian limit   the  relativistic circular speed profile   reduces to its  Newtonian value.   

The paper  is structured as follows. 
In Section II, in order to calculate the spatial components of the metric,  we summarize the method to build different compact sources  for  conformastat spherically symmetric   spacetimes  
 from   given solutions of  Poisson's  equation. On the other hand, to compute the temporal component of the metric, 
we also  analysis the geodesic circular motion of  test particles
moving on a static spherically  symmetric spacetime  in isotropic coordinates 
and the stability of the orbits against radial perturbations. 
In Section III,    for a given specific    circular speed profile,  three  families of  spherically symmetric  anisotropic fluid spacetimes are obtained, and,   as a simple  application,  new models of  relativistic thick spherical shells   are constructed  for some  members of the  families.  

In Section IV,   such metrics are  also used  in building  two three-component relativistic models
of   galaxies (bulge, disk and dark matter halo)  using as seed Newtonian potential-density  pairs  the  Miyamoto-Nagai  potentials for the central bulge and the disk  \cite{Miyamoto,Nagai} and for the dark matter halo the well-known 
Navarro-Frenk-White (NFW) model \cite{NFW}. 
Since the disk-like component  are structures with  axial symmetry,  in this case the metric is written in cylindrical coordinates. Such spacetimes can be interesting  
not only from
the merely theoretical point of view but also, for example,  
in describing   galactic nuclei  where relativistic effects are expected to be important.
Finally, in Section  V  we summarize  the results obtained.


\section{ Poisson type spacetimes  and motion of particles }

The metric for  a  conformastat   spherically symmetric  spacetime   is given by \cite{Synge,Kramer}
\begin{equation}
ds^2 =  -e^{2\nu(r)}  dt^2 + e^{2\lambda(r)}  (dr^2 + r^2d\Omega^2),  \label{eq: met}
\end{equation}
where   $d\Omega^2 = d\theta^2 +  \sin ^2 \theta  d\varphi ^2$.  The term ``conformastat'' means that spatial metric is conformally flat. The units are chosen so that we have for the speed  of light in vacuum $c = 1$. 
Einstein's gravitational  field equations $G_{ab}= 8 \pi G  T_{ab} $ yield the following non-zero components
of the energy-momentum tensor
\begin{subequations}\begin{eqnarray}
T^t_{\ t} & = &  \frac{1}{4 \pi G} e^{ -2\lambda} \left [\nabla^2 \lambda +  \frac1 2 \nabla \lambda \cdot \nabla \lambda \right ], \label{eq: Ttt-sph} \\
&  &  \nonumber \\
 T^r_{\ r} & = & \frac{1}{8 \pi G} e^{ -2\lambda} \left [ (\lambda')^2 + 2\lambda' \nu' + \frac {2} {r} (\lambda' + \nu' )  \right ]  ,  \label{eq: p1}  \\
& &  \nonumber  \\
T^\theta_{ \ \theta} & = & T^\varphi_{ \ \varphi}  = \frac{1}{8 \pi G} e^{ -2\lambda} \left [
\lambda '' + \nu '' + (\nu')^2 + \frac {1} {r} (\lambda' + \nu' ) \right] ,    \label{eq: p2} 
\end{eqnarray} \end{subequations} 
where primes indicate differentiation with respect to $r$.   The expression (\ref{eq: Ttt-sph}) for the $ T ^ t \ _t $ component of the energy-momentum tensor holds even if spatial symmetry is not assumed for the gravitational field, in other words, in the case that  metric functions  be a function of all spatial coordinates.
 
With respect to the orthonormal  tetrad or  locally Minkowskian observer
\begin{subequations}\begin{eqnarray}
{\bf V} & = & \frac{1}{\sqrt{-g_{tt}}} ( 1, 0, 0, 0 ),	 \label{eq: tetrad0} \\
	&	&  \nonumber	\\
{\bf X} & = & \frac{1}{\sqrt{g_{rr}}}  ( 0, 1, 0, 0 ),  \\
	&	&  \nonumber	\\
{\bf Y} & = &  \frac{1}{\sqrt{g_{\theta \theta}}}(0  , 0 , 1,  0  ),  \\
	&	&  \nonumber	\\
{\bf Z} & = &  \frac{1}{\sqrt{g_{\varphi \varphi}}}   (0  , 0 , 0,  1  ), \label{eq: tetrad3}
\end{eqnarray} \end{subequations} 
whose  components are denoted as  ${{\rm e}_{ (a)}}^b = \{V^b , X^b,Y^b,  Z^b \}$, where Latin indices $a$ and $b$  run from 0 to 3,
the relativistic energy density is given by 
$\rho  =  - T^t_{\ t}$,
 and the principal stresses (pressures or tensions)  by $p_i= T^i_{\ i}$. 
 
 By setting   
\begin{equation}
e^{2 \lambda} = \left( 1- \frac{\phi(r)}{2}    \right)  ^{4}, \label{eq: lamb}
\end{equation}
 we get, even without spatial symmetry,  for the energy density  the following    nonlinear  Poisson type equation
\begin{equation}
\nabla ^2 \phi = 4 \pi G  \rho  \left( 1- \frac{\phi}{2} \right)^{5} , \label{eq: nonlinear}
\end{equation}
In fact, in Newtonian limit when  $\phi  \ll 1$   it reduces to Poisson's equation,   $\nabla ^2 \Phi = 4 \pi G \rho_N$. 
For  a  given physical energy density  profile $\rho$,
the metric function $\phi$ can be obtained 
by resolving this equation.  A physically reasonable way  to choose $\rho$ is  
by requiring  that in the Newtonian limit it   reduces to its  Newtonian value $\rho_N$. 
A  simple particular form of   $\rho$ which satisfies such condition  is
\begin{equation}
\rho  =   \frac{\rho_0 }{   \left( 1- \frac{\phi}{2 } \right)^{5} }.  \label{eq: rho1}
\end{equation}

Replacing   this expression in (\ref{eq: nonlinear}) one finds in this case  that the pair  $(\phi,\rho_0 )$ is a  solution of the Poisson's equation. 
In conclusion, for each Newtonian solution there is a relativistic
one such that  $(\Phi,\rho_N) \rightarrow (\Phi,\rho_N/(1-\Phi/2)^5)$, where the first
pair solves the Poisson equation while the latter one solves (\ref{eq: nonlinear}).

Accordingly, such geometries  can be called Poisson type spacetimes. For example, the pair 
$(\phi,\rho_0 )= (-\frac{MG}{r}, 0)$  corresponds  to the external Schwarzshild solution in isotropic coordinates. To obtain the  other metric function $\nu$  an  additional assumption  must be imposed and will be obtained  in next   section  by kinematic considerations. 

\subsection{ Stable circular orbits}

With respect to the comoving frame of reference  (\ref{eq: tetrad0}) - (\ref{eq: tetrad3}),  the 4-velocity  ${\bf u}$ of a test particle moving on the spacetime (\ref{eq: met})  has components
\begin{equation}
u^{(a)} = e^{(a)}_{\ \ \ b} u^b,
\end{equation}
 and   the  3-velocity ${\bf v}$
\begin{equation}
v^{(i)} =  \frac { u^{(i)} } { u^{(t)} }  =    \frac { e^{(i)}_{\ \  \ a} u^a }{  e^{(t)}_
{ \ \ \ b}  u^b  }. 
\end{equation}
 
 Because of   its mathematical simplicity and  astrophysical relevance, we shall consider circular orbits. 
For instance, the stars of the   disk in spiral galaxies  travel in nearly circular orbits around the galactic center \cite{Binney}. In addition, due to the spherical symmetry and also for  simplicity, we shall analyze equatorial orbits on the plane $\theta =\pi /2$. Thus, for equatorial, circular   orbits $dr/dt = d\theta/dt=0$, then  ${\bf u} =u^t(1,0,0, \omega)$, where $\omega = u^\varphi/u^t =d\varphi/dt$ is the angular speed of the test particles. In this  case  $v^{(\varphi)}$ is the only nonvanishing velocity component and is  given by
\begin{equation}
  [v^{(\varphi)}]^2 = v_c^2=  - \frac{ g_{\varphi \varphi} }{ g_{tt} } \omega^2.
\label{eq:vc2}
\end{equation}
$v_c$ represents   the circular speed   (rotation profile)   of the particles measured by an inertial
observer far from the source, and it is a physical parameter
of  interest,  especially in the description of galaxies, related to the circular motion of test particles along geodesics on the galactic plane.

The  angular speed   $\omega^2$ can be calculated  considering  the geodesic  motion of the particles.
For equatorial orbits, the Lagrangian  for a massive test particle is  
\begin{equation}
2{\cal L} = g_{ab} \dot x^a \dot x^b =  -e^{2 \nu}  \dot t^2 +  e^{2 \lambda}  (\dot r^2 + r^2 \dot  \varphi ^2),
\end{equation}
where  the overdot denotes derivative with respect to the proper time  $\tau$.   The Lagrange's  equations 
\begin{equation}
\frac{d}{d\tau} \left ( \frac{ \partial {\cal L}} {\partial \dot x^a} \right )
- \frac{ \partial {\cal L}} {\partial x^a} = 0 
\end{equation} 
yield two  constants of motion  
\begin{subequations} \begin{eqnarray}
E & = &  - p_t/m =   e^{2 \nu}  \dot t,   \label{eq: cons1} \\
L & = &  p_{\varphi }/m =   e^{2 \lambda}  r^2  \dot \varphi ,    \label{eq: cons2}
\end{eqnarray}\end{subequations}
where $E$ represents  the relativistic specific energy and $L$ the   specific angular momentum. Therefore, the Lagrangian can be written as  
\begin{equation}
2{\cal L} =   e^{2 \lambda} \dot r ^2  +  e^{-2 \lambda} \frac{L^2}{r^2}  - e^{-2 \nu} E^2.
\end{equation}
 Normalizing  $u^a$, that is requiring  $g_{ab}u^au^b=-1$, 
we obtain
\begin{equation}
e^{2 \nu +2 \lambda} \dot r^2 + e^{2 \nu} \left (  1 + e^{-2 \lambda} \frac{L^2}{r^2} \right) = E^2.  \label{eq: condnorm}
\end{equation}

This expression allows to define an effective potential $V_{eff}$  as
\begin{equation}
V_{eff} =  e^{2 \nu} \left (  1 + e^{-2 \lambda} \frac{L^2}{r^2} \right) . 
\end{equation}

For a circular motion have  that
\begin{equation}
E^2 =  e^{2 \nu} \left (  1 + e^{-2 \lambda} \frac{L^2}{r^2}
\right)    \label{eq: E1}.
\end{equation}
Assuming the Lagrangian as $\tilde {\cal L} = 2 {\cal L} + 1 $
and using   the condition of normalizing  (\ref{eq: condnorm}),  the  radial motion equation for circular orbits   (extreme motion) reads 
\begin{equation}
\frac{d V_{eff}}{d r} =0.
\end{equation}
This condition implies
\begin{equation}
\frac{L^2}{r^2} =\frac{ r \nu_{,r}  e^{2 \lambda - 2 \nu} E^2 }
{1 + r \lambda _{,r}}.  \label{eq: L1}
\end{equation}

From Eqs.  (\ref{eq: E1}) and (\ref{eq: L1}) we find
\begin{subequations} 
\begin{eqnarray}
E^2 & = &  \frac { e^{2\nu}  }{1 - \frac{r \nu _{,r}}{1+ r\lambda_{,r}} }, \\
L^2 & = & \frac {r^3 \nu _{,r}  e^{2\lambda}} {1 + r(\lambda_{,r} - \nu _{,r})}. 
\end{eqnarray} 
\end{subequations} 

By using the conserved quantities  (\ref{eq: cons1}) and
  (\ref{eq: cons2}),  and the expression (\ref{eq: L1}) we obtain
\begin{equation}
\omega^2
= \frac { \dot  \varphi ^2 }{ \dot t ^2 } = e^{ 4 (\nu - \lambda)} \frac{L^2}{ r^4 E^2} = - \frac{g_{tt,r}}{ g_{\varphi \varphi,r }}.
\end{equation}

Thus, for motion of particles in  circular orbit   the tangential velocity  is given by 
\begin{equation}
v_c ^2  = \frac{ r \nu _{,r} }{ 1+ r \lambda _{,r}}, \label{eq: vc}
\end{equation}
and the angular momentum can be cast  as
\begin{equation}
L^2 = \frac{r^2 e^{2 \lambda } v_c^2}{1- v_c^2}.
 \label{eq: L2}
\end{equation}

In addition, in order to have  stable circular orbits against radial perturbations the following condition must be satisfied
\begin{equation}
\left. \frac{d^2V_{eff}}{d r^2}  \right |_{extr} > 0 ,
\end{equation}
or explicitly
\begin{equation}
\nu_{,rr} + \nu_{,r}\left (   2 \lambda_{,r}   -  2 \nu_{,r} + \frac 3 r -  \frac{  \lambda_{,r}  + r  \lambda_{,rr} }{1 + r \lambda_{,r} } \right ) > 0. \label{eq: stab1}
\end{equation}
In terms of the angular momentum the stability condition (\ref{eq: stab1})   reads 
\begin{equation}
\frac{d L^2}{dr} > 0,
\end{equation}
which is  an extension of the Rayleigh criteria of stability of a fluid in rest in a gravitational field \cite{RAYL,LAND6,Let-estab}.

For  non-circular orbits on the equatorial plane   the velocity ${\bf v}$ of the particles has two components, the radial velocity $[v^{(r)}]^2 = v_r^2 =   - \left (g_{rr} / g_{tt} \right )   \left ( dr/dt \right )^2$ and the tangential velocity $v^{(\varphi)}=v_c$, then for the geodesic  motion of the particles the  speed $v$ is given by
\begin{equation}
 v^2 =   v_r^2 + v_c^2 
 = \frac{ r \nu _{,r} }{ 1+ r \lambda _{,r}}
 + \frac{\left ( g_{\varphi \varphi} g_{rr,r } -  g_{\varphi \varphi,r} g_{rr }  \right)}{g_{tt }  g_{\varphi \varphi ,r }}  \left (  \frac{dr}{dt} \right)^2
 - \frac{2  g_{\varphi \varphi} g_{rr } }{ g_{tt } g_{\varphi \varphi,r}} \left (  \frac{d^2r}{dt^2} \right ).  \label{eq: vnc}
\end{equation} 




\section{Anisotropic fluid spacetimes}

For a given physical  circular speed profile
 $v_c$,
the metric function   $\nu$   can be obtained,  for circular geodesics,  by integrating  (\ref{eq: vc}) which,
for the metric potential $\lambda$   (\ref{eq: lamb}) and with $\phi=\Phi$,    can be cast as  
\begin{equation}
\nu_{,r }= \frac {v_c^2} r \frac { (1 - \frac \Phi 2- v_N^2) } {(1 - \frac \Phi 2)}, \label{eq: nur}
\end{equation}
where $v_N^2 = r \Phi_{,r}$ is the Newtonian circular speed. Like the energy density,   $v_c$ can be chosen     
by also requiring  that in the Newtonian limit 
it   reduces to its  Newtonian value $v_N$.
Any tangential velocity profile of the form
\begin{equation}
v_c^2 = \frac {v_N^2   (1 - \frac \Phi 2) }{( 1 - \frac \Phi 2 - v_N^2) }  F(\Phi) ,
\end{equation}
with  $F(0)=1$ satisfies such condition.
It follows that 
$\nu = \int{ F(\Phi) d \Phi + C}$, where $C$ is
a constant of integration which is chosen so that the metric  is    asymptotically flat.
The form of the tangential speed and also the parameters
must    be chosen so that the matter fields,
in order to be physically meaningful,   satisfy    the energy conditions:   $\rho\geq 0$ (weak  energy
condition),  $|\rho|  \geq |p_i|$ (dominant energy condition) and 
$ \rho+ p_r + p_\theta + p_\varphi \geq 0$ (strong energy condition). Moreover, we must obtain positive
values (pressure) for the  principal stresses and stable circular orbits. In general, the spacetimes built using this approach  represent   anisotropic fluid matter configurations. To obtain  a perfect fluid source, equality  between equations (\ref{eq: p1}) and (\ref{eq: p2})  (condition of isotropy pressure)  must also be required. Accordingly, only in some special cases   spacetimes are obtained  that have as source an isotropic fluid.  

A particular expression for  $F$ 
 for which   $\nu$  can also be obtained analytically is
$F=P_{\alpha}(\cos \Phi)$ ($\alpha$-type spacetimes), where $P_{\alpha}$ are the Legendre polynomials for which  $P_{\alpha}(1)=1$.  Powers of these functions are also  possible spacetimes.  Integrating  (\ref{eq: nur}),  we obtain, for instance,  for the first four 
Legendre polynomials
\begin{subequations}\begin{eqnarray}
\nu_0 & = & \Phi,  \\
\nu_1 & = & \sin \Phi,  \\
\nu_2 & = & \frac 1 4 \left ( \Phi +\frac 3 2 \sin 2 \Phi  \right),  \\
\nu_3 & = &  \frac  1 6 (5 \cos^2 \Phi + 1) \sin \Phi.
\end{eqnarray}\end{subequations}
  For the first   member of the family     $\alpha=0$ and  weak gravitational fields $\Phi \ll 1$, $e^{2 \nu}  \approx 1 + 2 \Phi$. Therefore,  this spacetime has a well-defined Newtonian limit. 

Another simple expression for $F$ is 
$F(\Phi)=(1-\Phi/2)^{\beta-2}$ ($\beta$-type spacetimes),  where $\beta$ is a constant.
In this case 
\begin{equation}
\nu = \begin{cases}
 - 2 \ln(1 - \frac \Phi 2),  &  \beta=1,  \label{eq: nuani}  \\
\frac 2 {\beta-1} \left [1 -   (1 - \frac \Phi 2)^{\beta-1} \right ] ,  &  \text{otherwise}. 
  \end{cases}
\end{equation}
The case $\beta=1$ was studied in  Ref. \cite{GGR-shell} and describes
a Majumdar-Papapetrou type spacetime.
The second  member of the family $\beta=2$   
is the same as the above spacetime   $\alpha=0$. 

A third family of spacetimes is when   $F(\Phi)=\frac{(1-\Phi/2)^{\gamma-2}}{(1+\Phi/2)^{\gamma}}$  ($\gamma$-type spacetimes).
 We obtain
\begin{equation}
\nu = \begin{cases}
  \ln \left [ (1+\frac\Phi 2) (1-\frac\Phi 2)^{-1} \right ],  &  \gamma=1,  \label{eq: nuani}  \\
\frac 1 {\gamma-1} \left [1 -  \left (  \frac {1 - \Phi/ 2}{1 + \Phi/ 2} \right )^{\gamma-1}  \right ] ,  &  \text{otherwise}. 
  \end{cases}
\end{equation}

The case $\gamma=1$ 
with  $\Phi = - MG /\sqrt{a^2 + r^2}$ (the Plummer model \cite{Plummer}), being   $a$  a constant with the dimensions of length, corresponds to  the Buchdahl's perfect fluid solution \cite{Buchdahl}, a relativistic  analog of a classical polytrope of index 5.   Anisotropic spacetimes  for this member of the family were  studied in Refs. \cite{Let-shell1,Let-shell2}.

 

Thus, given a seed Newtonian potential-density pair ($\Phi$, $\rho_N$)  and its  corresponding circular speed profile $v_N$  we can  construct,  for circular geodesics,  the relativistic version of different  spherically symmetric physical structures.  
For noncircular geodesics the expression for the speed $v$ of the particles,
Eq. (\ref{eq: vnc}), 
differs
substantially from  (\ref{eq: vc}) and consequently the compute of $\nu$ from an imposed speed profile $v$ which in Newtonian limit approach its Newtonian counterpart  is more involved.

\subsection{Relativistic thick  spherical  shell models}

We consider the potential-density pair 
\begin{subequations}\begin{eqnarray}
\Phi &=&  -  \frac{GM} { b + ( a^n + r^n )^{1/n}},  \\
 \rho_N & = & \frac{M}{4 \pi} \frac{r^{n-2} \left [ 2br^n +  (n+1) (b + d ) a^n \right ] }{ d^{2n-1} (b+d)^3},
\end{eqnarray} \end{subequations}
where
\begin{equation}
d =  ( a^n + r^n )^{1/n} ,
\end{equation}
$n \geq 1$ and $a$ and $b$ are a non-zero constants with the dimensions of length. The models with   $n > 2$ describe  a shell-like
matter distribution. The case  $a=b$  corresponds
to the  generalized isochrone models presented in reference  \cite{Evans2006} and $b=0$ are  generalized Plummer models \cite{Veltmann}. 
The latter  include
the Hernquist  model ($n = 1$) \cite{Hernquist} and the Plummer model ($n = 2$)  \cite{Plummer} as particular cases.  Such
potential-density pairs  have been used as successful   analytic models for elliptical galaxies and  bulges of disk galaxies.   The Newtonian circular speed is given by
\begin{equation}
 v^2_N =  \frac {MG r^n}{d^{n-1} (b+d)^2}.
\end{equation}

In the relativistic  case, 
we show below  the main physical quantities associated to the systems to the first member $\alpha = 0$ of the 
first family of spacetimes.  Since the expression are huge, other cases   will be analyzed graphically.
Thus, when $\alpha = 0$ we have 
\begin{subequations}\begin{eqnarray}
\tilde \rho & = &  \frac{  8 \tilde r^{n-2} (\tilde b+ \tilde d)^2 \left [ 2\tilde b \tilde r^n +  (n+1) (\tilde b + \tilde d ) \tilde a^n \right ] }{\pi  \tilde d^{2n-1} \left [  2 (\tilde b+\tilde d) +1  \right ]^5}, \\
\tilde p_r & = & \frac{8  \tilde r^{n-2} (\tilde b  + \tilde d) 
\left [ (\tilde b - \frac 1 2)\tilde r^n + (2\tilde b + \tilde d + \frac 12)\tilde a^n  + \tilde b (\tilde b + \frac 1 2 ) \tilde d^{n-1}       \right ]  }
{ \pi    \tilde d^{2(n-1)} \left [  2 (\tilde b+\tilde d) +1  \right ]^6  }, \\
\tilde p_\varphi & =  & \frac {4 \tilde r^{n-2}} 
{ \pi   \tilde d^{2n-1} \left [  2 (\tilde b+\tilde d) +1  \right ]^6} 
\left [  n \tilde a^n \tilde d^3
+ \left (  (\tilde b + \frac 3 2)\tilde r^n + 3 n (\tilde b+ \frac 1 6)\tilde a^n  \right )\tilde d^2  \right .  
\nonumber \\   
&& \left . +  \left (2 (\tilde b + \frac 1 2)^2 \tilde  r^n + 3 n \tilde b  (\tilde b + \frac 13) \tilde a^n  \right ) \tilde d  
+ \tilde  b^2 (\tilde b+ \frac 12 )\  \left (  \tilde r^n + n \tilde a^n  \right ) 
\right ] ,  \\
v_c^2 & = & \frac{\tilde r^n [2(\tilde b+\tilde d) +1]}{   (\tilde b+\tilde d) \left [ \left (  2(\tilde b+\tilde d) +1  \right ) \tilde  d^{n-1} (\tilde b+\tilde d)  - 2 \tilde r^n  \right ] }, \\
\tilde L^2 & = & \frac { \tilde r^{n+2} [ 2(\tilde b+\tilde d) + 1]^5 }
{16 (\tilde b+\tilde d)^4 
\left [  \left (  2(\tilde b+\tilde d) +1  \right ) \tilde  d^{n-1} (\tilde b+\tilde d)^2   -  \left ( 4(\tilde b+\tilde d) + 1 \right ) \tilde  r^n \right ]},
\end{eqnarray} \end{subequations}
where $\tilde \rho = G^3 M^2 \rho$,  $\tilde p_r = G^3 M^2 p_r$, $\tilde p_\varphi  = G^3 M^2 p_\varphi$,  $\tilde L = L / (GM)$,  $\tilde r = r / (GM)$, $\tilde a = a / (GM)$, $\tilde b = b / (GM)$ and  $\tilde d = d / (GM)$.
We see that the  energy density  always  is a positive quantity in according to the  weak  energy condition and  in order to have pressures everywhere we must take  $\tilde b \geq 1/2$.



\begin{figure} 
$$
\begin{array}{cc}
\includegraphics[width=0.38\textwidth]{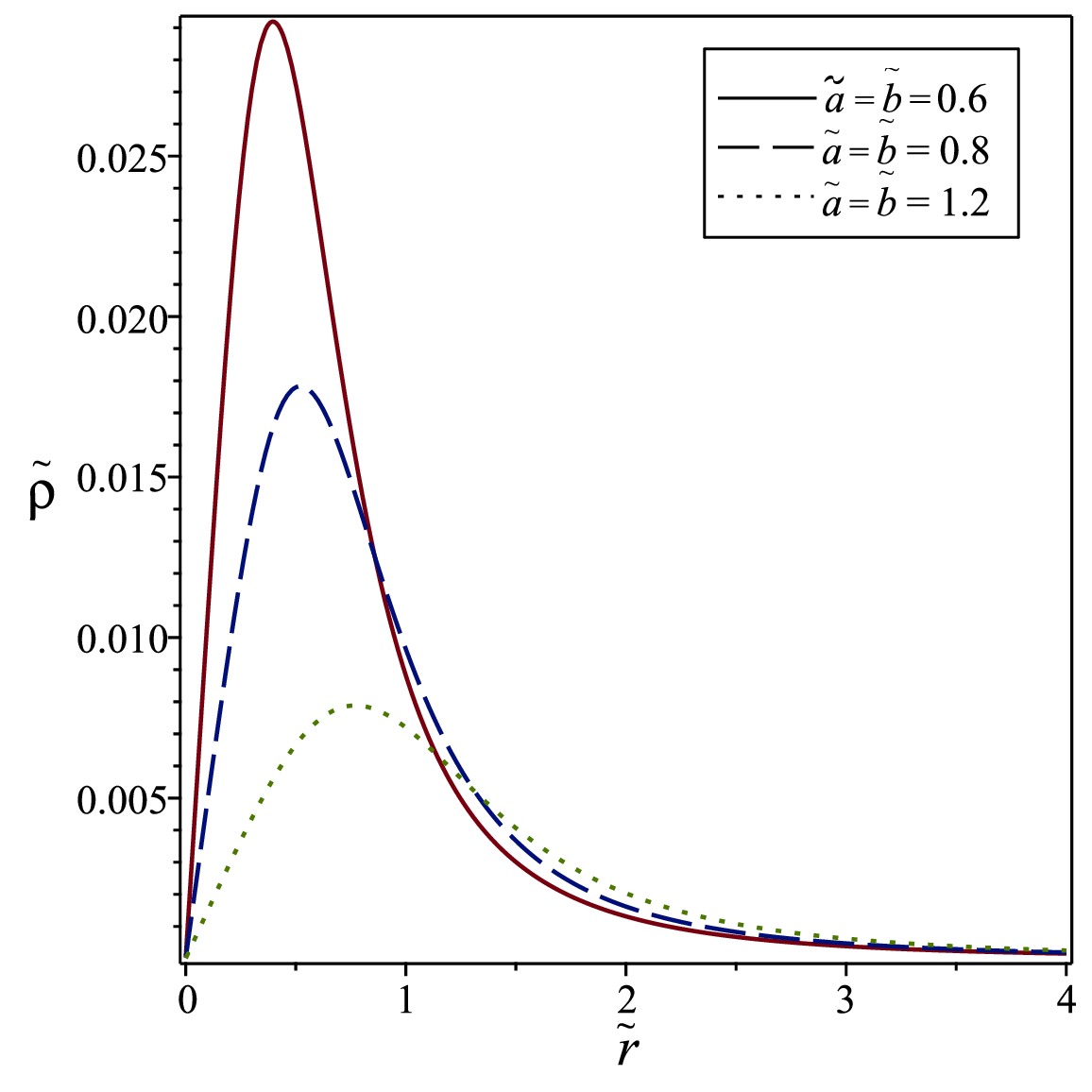} &
\includegraphics[width=0.38\textwidth]{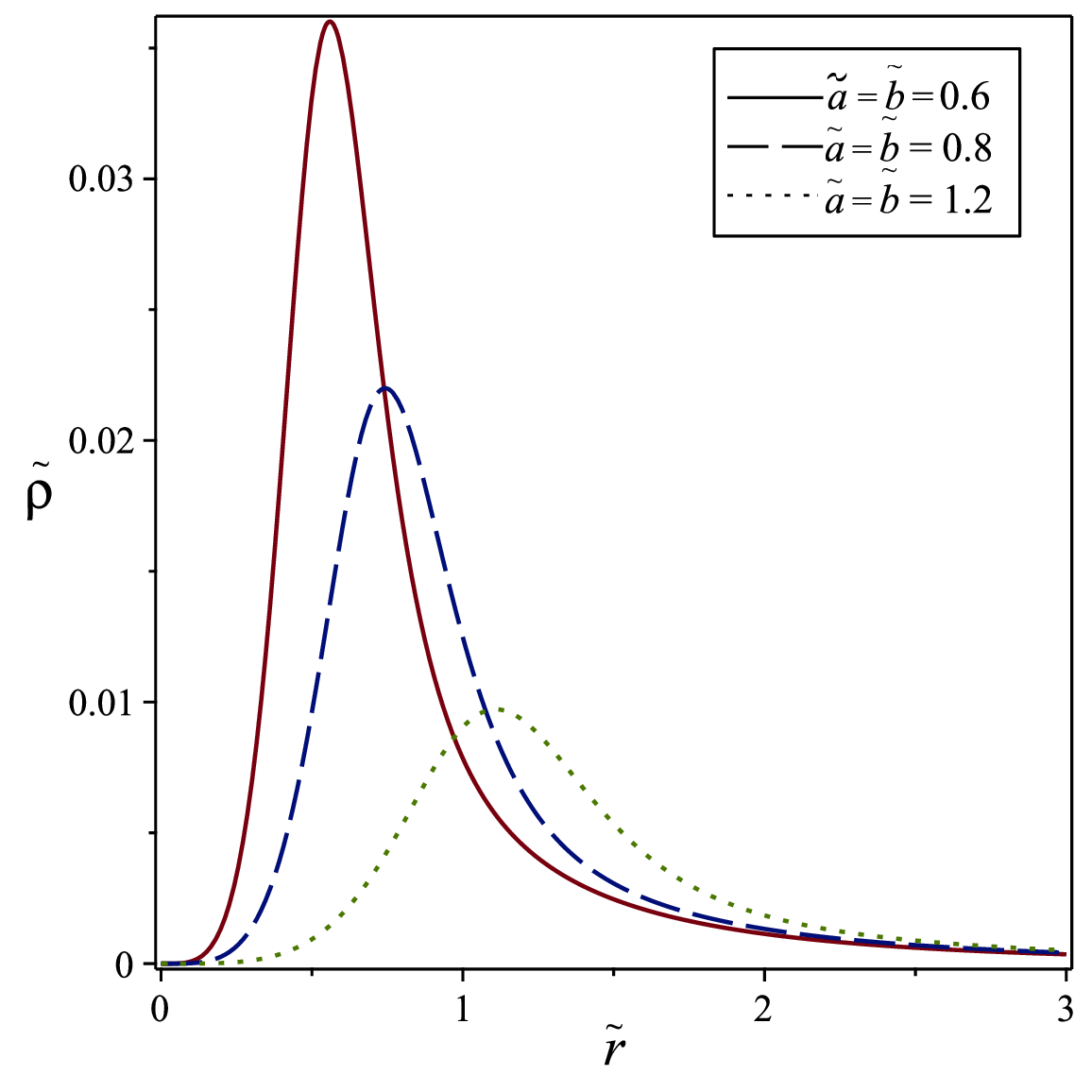}  \\
 (a) &  (b)  \\
&  \\
\includegraphics[width=0.38\textwidth]{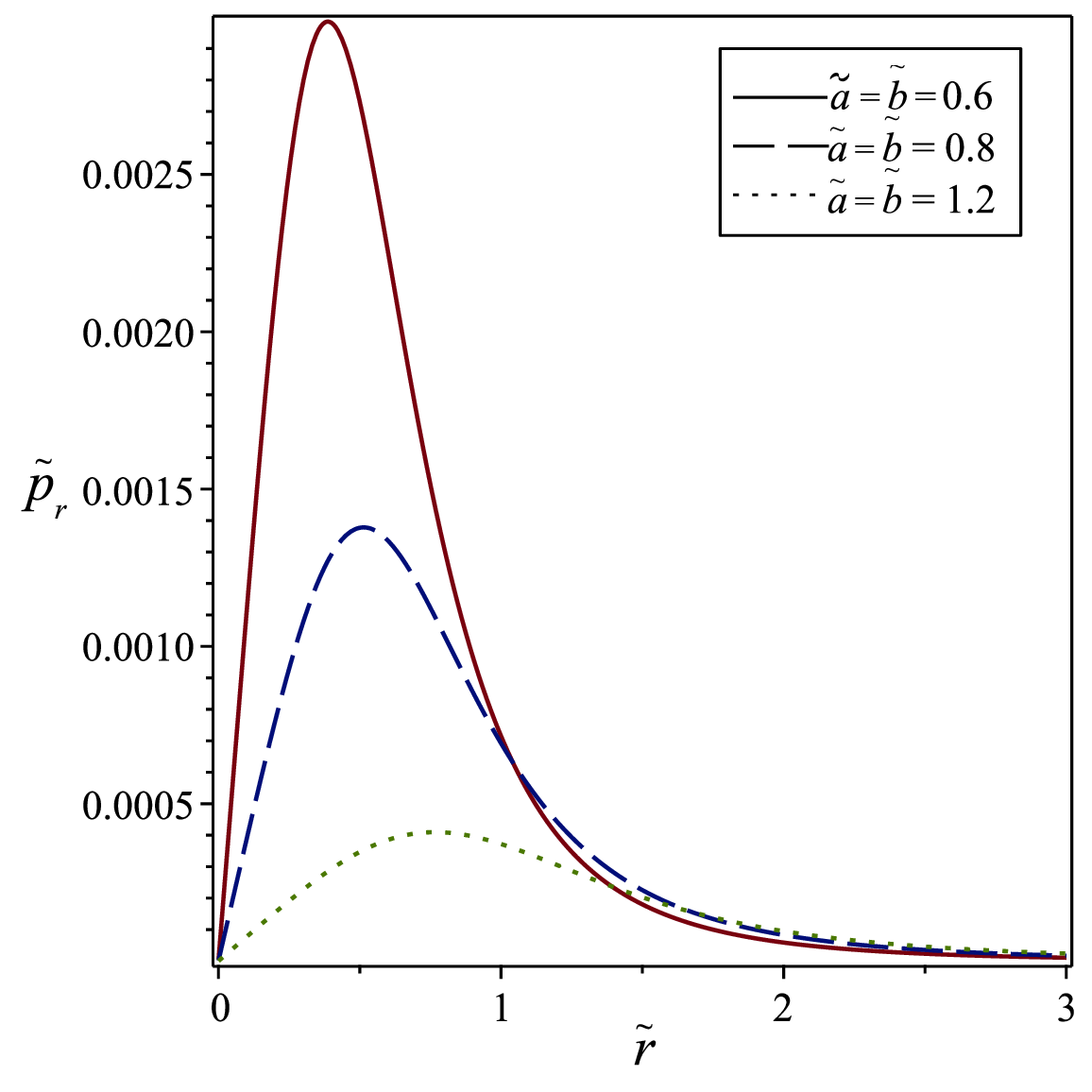} &
\includegraphics[width=0.38\textwidth]{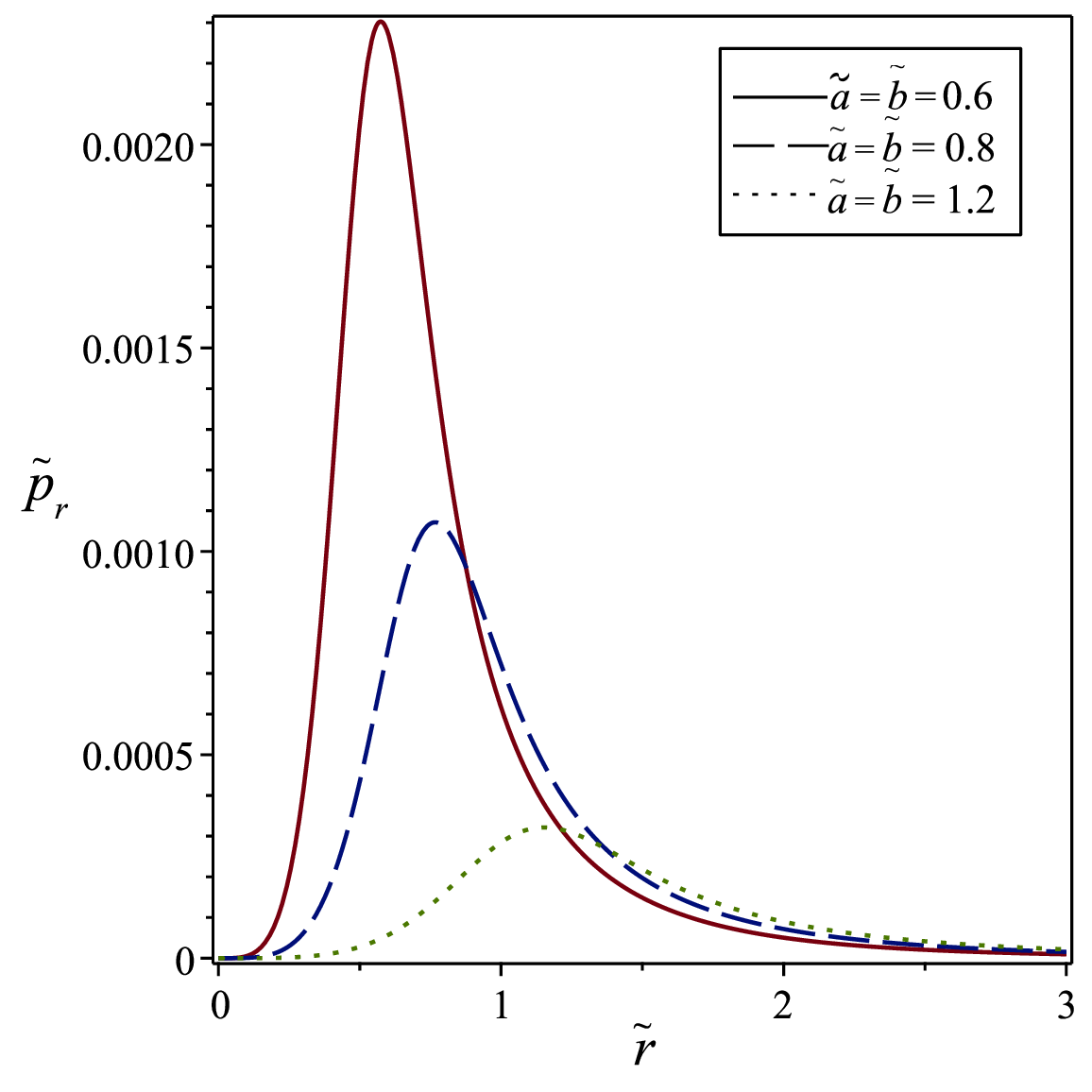}  \\
(c) &  (d)  \\
&  \\
\includegraphics[width=0.38\textwidth]{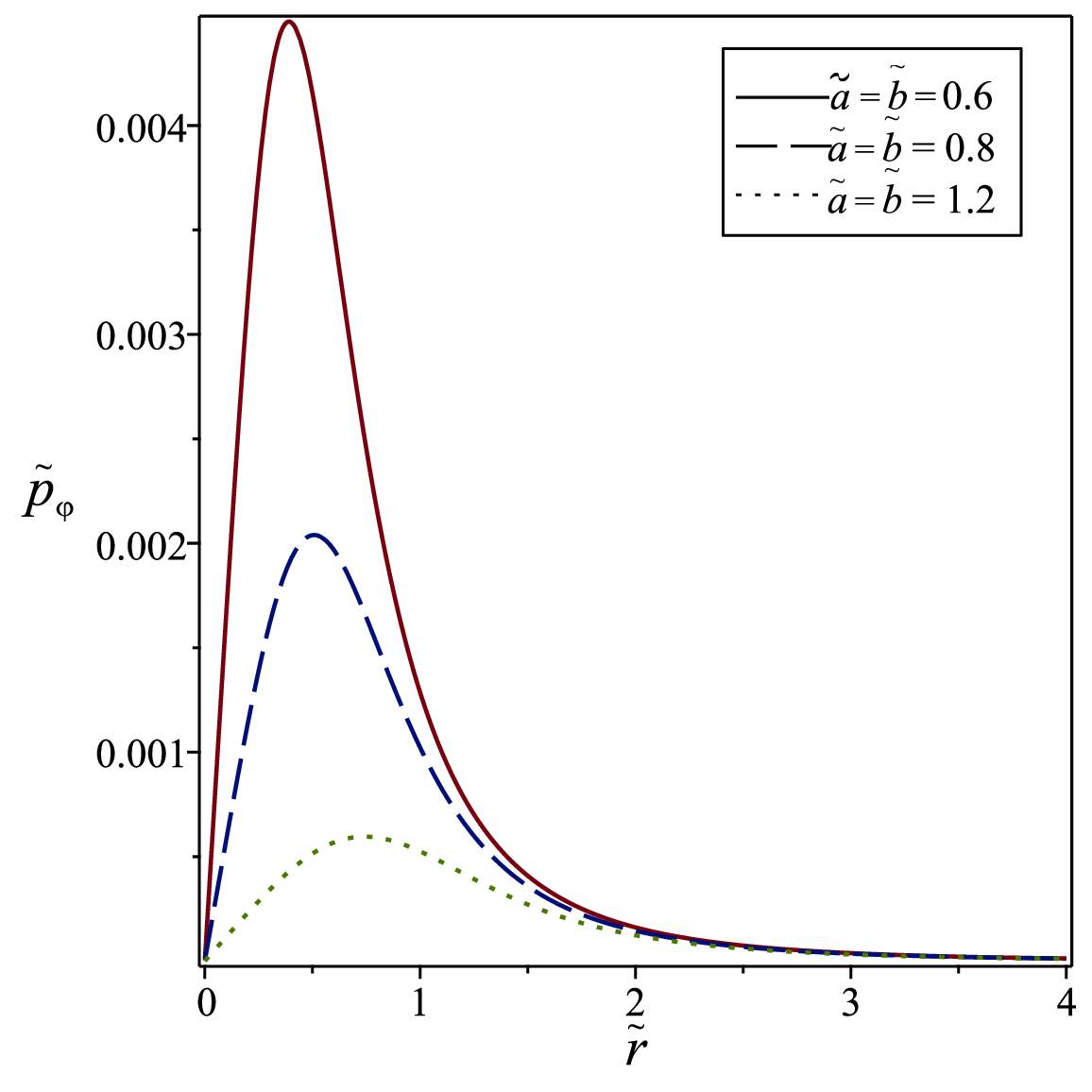} &
\includegraphics[width=0.38\textwidth]{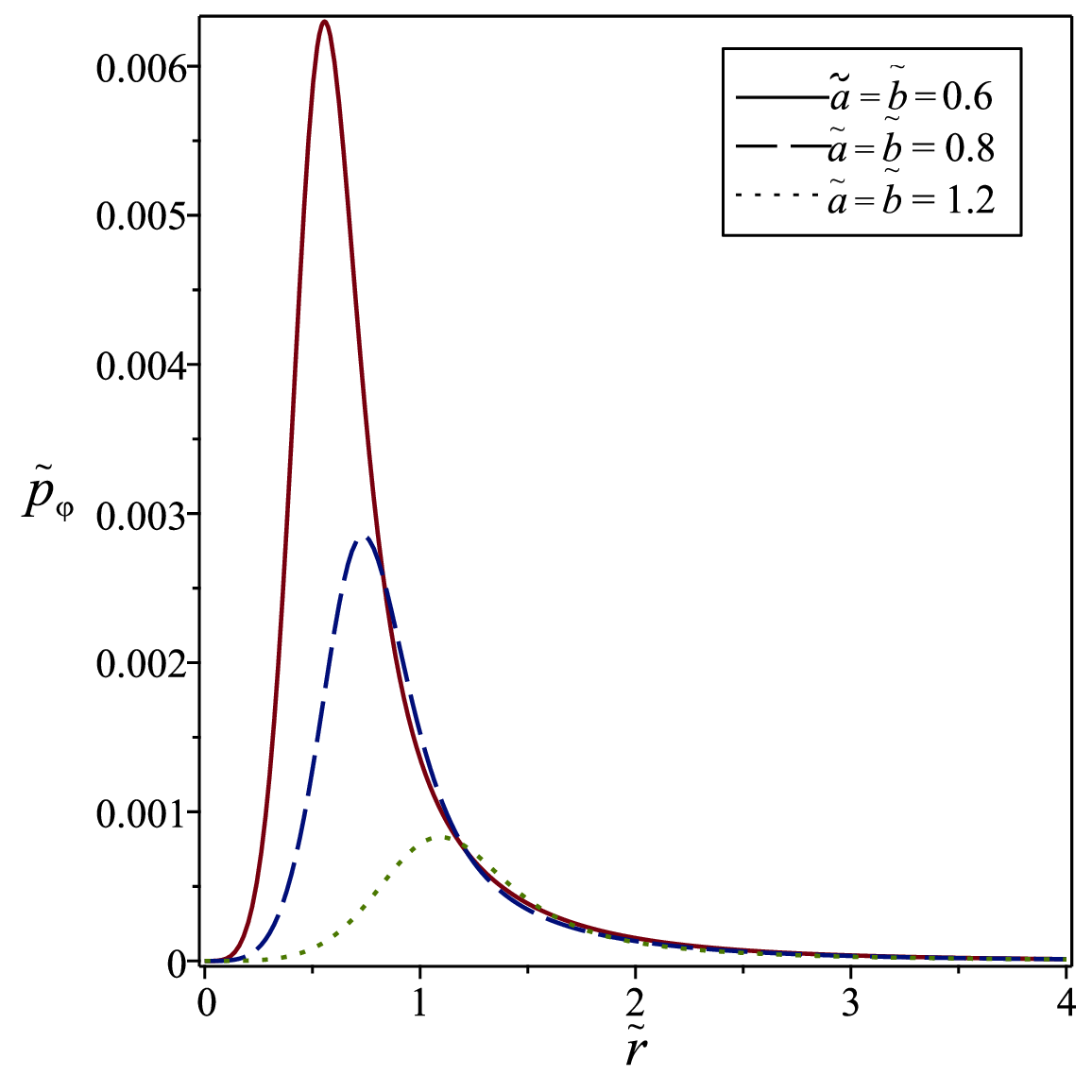} \\
 (e)   &    (f) \\
 &  \\
 \end{array}
$$	
\caption{$(a)$, $(b)$ The relativistic energy density  $\tilde \rho$,    $(c)$, $(d)$ the radial pressure $\tilde  p_r$, and $(e)$, $(f)$ the tangential  pressure $ \tilde  p_\varphi$  for the $\alpha$-type relativistic thick shells  with  $n=3$  (left curves) and  $n=6$  (right curves)  and parameters  $\alpha=0$, $\tilde a = \tilde b = 0.6$ (solid curves), $0.8$ (dashed curves),  and $1.2$ (dot curves), as functions of $\tilde  r$. }
\label{fig:fig1}
\end{figure}

In figures \ref{fig:fig1}, \ref{fig:fig3} and \ref{fig:fig5} we show, as functions of $\tilde  r$,  the curves of the relativistic energy density  $\tilde \rho$,  the radial pressure $\tilde  p_r$ and  the tangential  pressure $ \tilde  p_\varphi$ for the relativistic thick shells  with 
$\alpha = 0$, $1$, and  $\beta = 3$
and  parameters $ \tilde a = \tilde b = 0.6$, $0.8$, $1.2$, $ \tilde a = \tilde b = 2$, $3$, $4$, and  $ \tilde a = \tilde b = 0.7$, $0.9$, $1.2$, respectively,
with $n=3$ and $n=6$.    Such quantities vanish at the origin   
which suggests a shell-like matter distribution, then increase rapidly,  reach a maximum and later decrease rapidly with the radius   which  permits to define a cut off radius $r_c$ and, in principle, to consider these structures as  compact objects.  As the value of the parameters $\tilde a$ and $\tilde b$ are increased, the shells   are smoothed  out and  become more concentrated when the parameter $n$  is increased. We also observer that for the first family of spacetimes it is always possible  to find in all the cases parameters for which the principal stresses are positive quantities everywhere, but for the other two families,
 except for   $\beta = 2$ and $\gamma = 1$, 
it was  found that far from  the central region  the tangential stresses  take  very small negative values.
In this case,  the cut off radius of the structures could be defined just before the azimuthal stresses become negative.

\begin{figure}
$$
\begin{array}{cc}
 \includegraphics[width=0.38\textwidth]{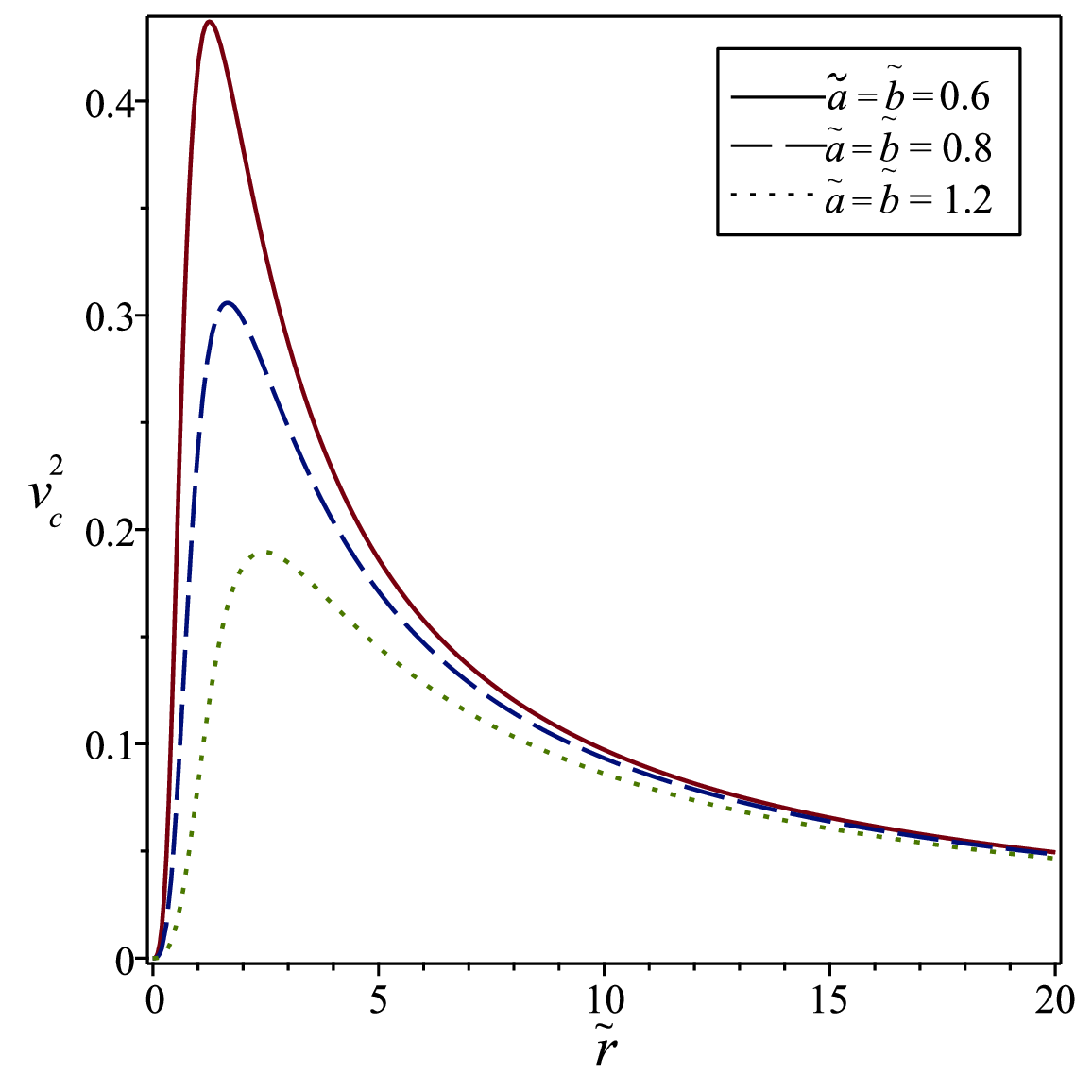} &
\includegraphics[width=0.38\textwidth]{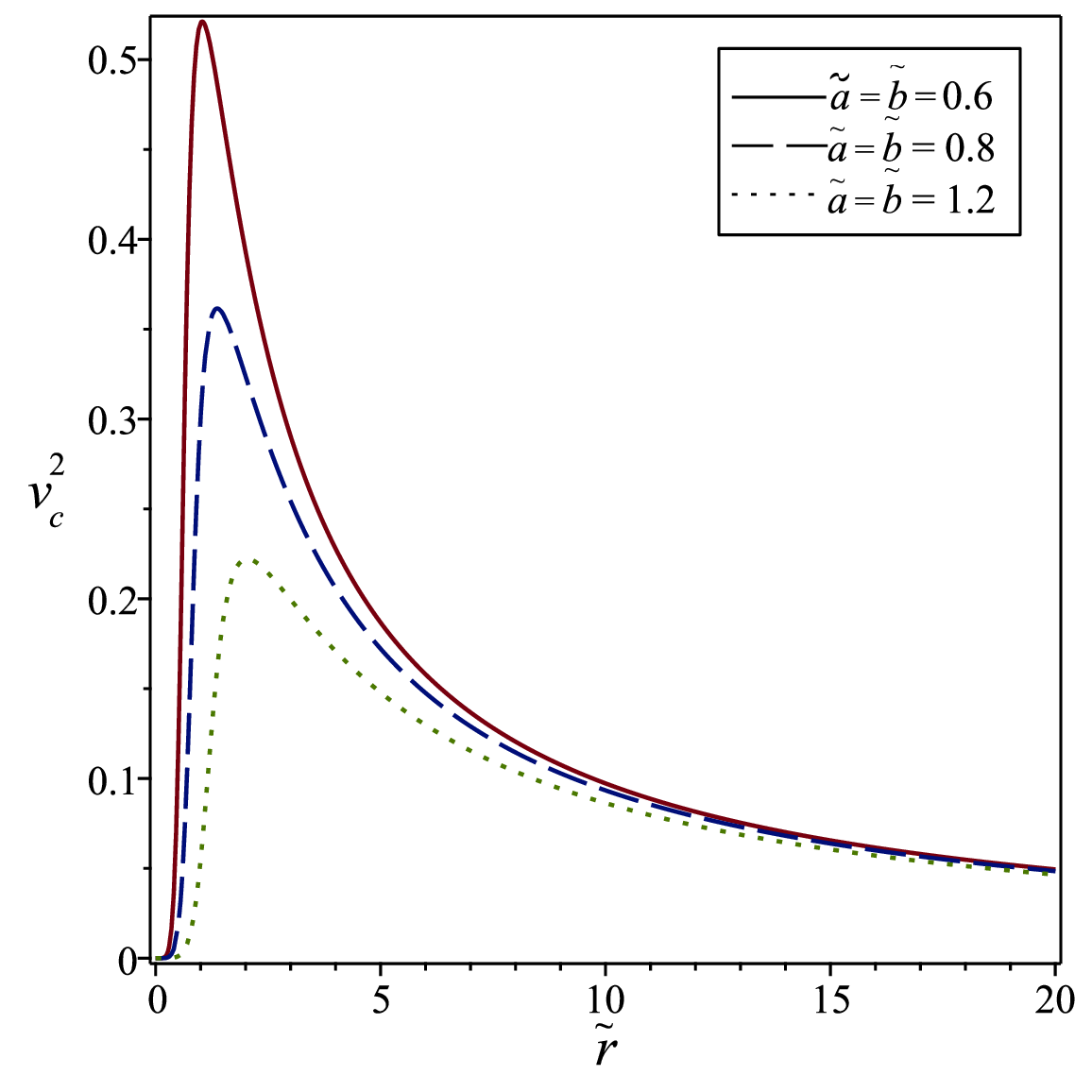} \\
 (a)   &    (b) \\
 &  \\
 \includegraphics[width=0.38\textwidth]{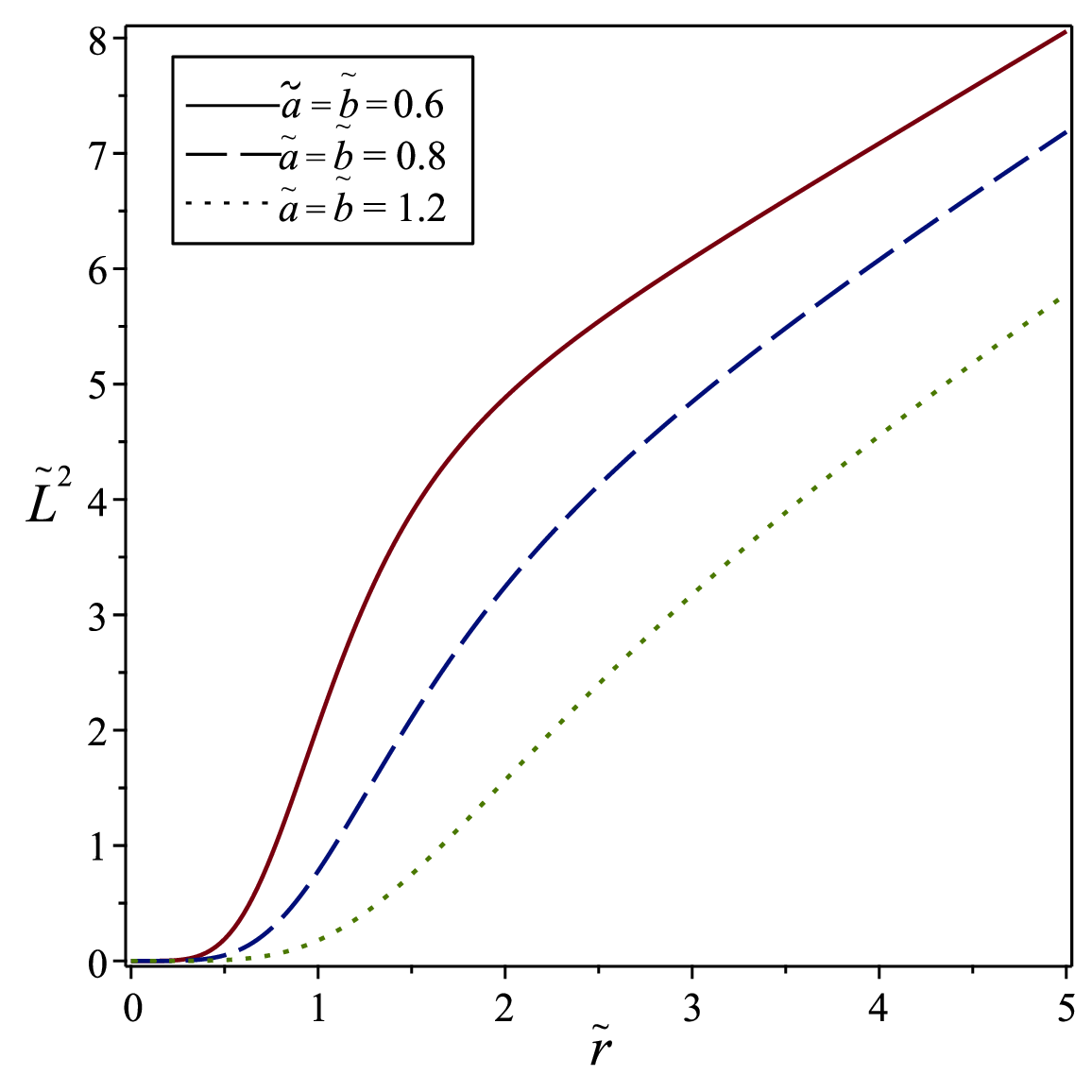} &
\includegraphics[width=0.38\textwidth]{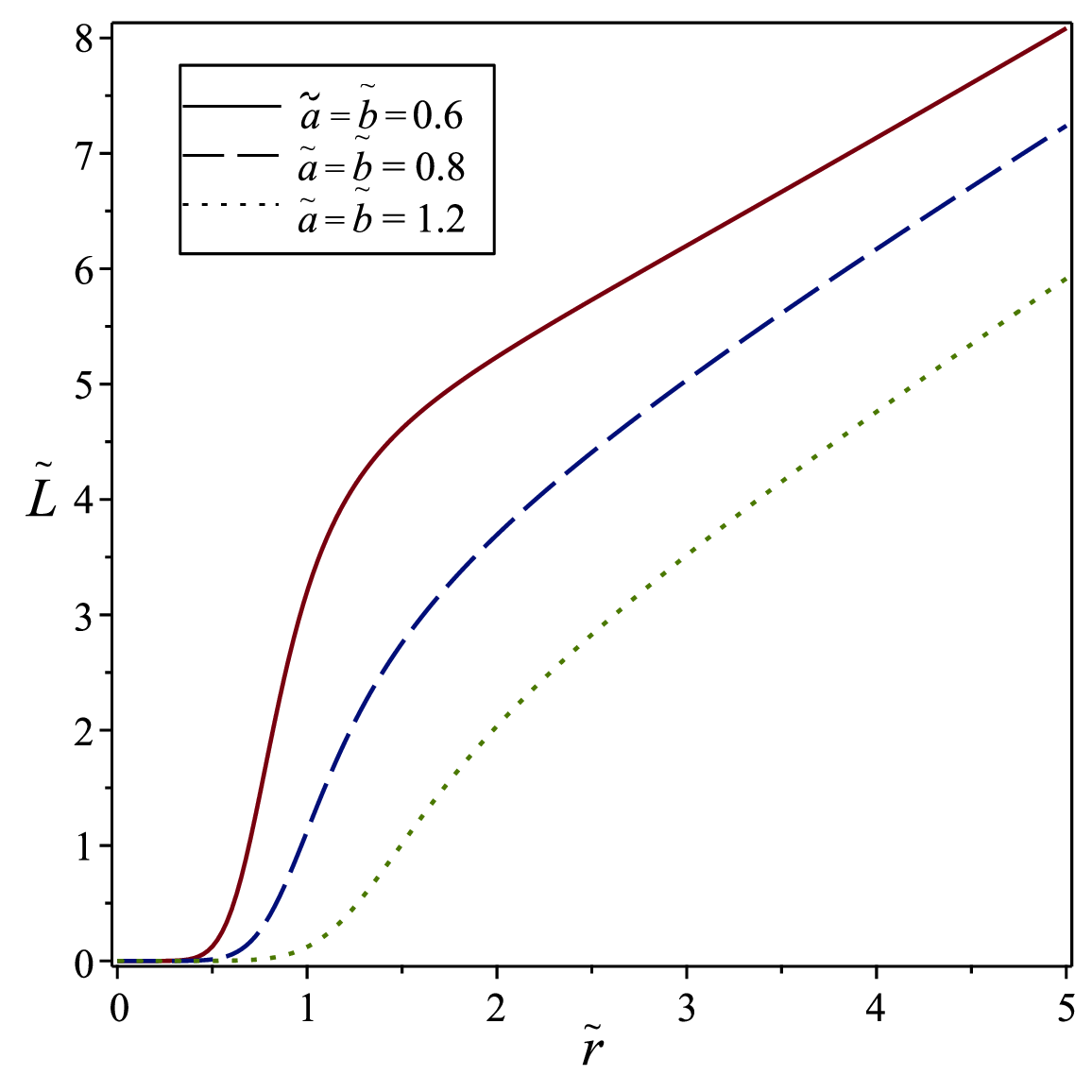} \\
 (c)   &    (d) \\
\end{array}
$$	
\caption{$(a)$, $(b)$ The rotation curves
$v_c^2$ and $(c)$, $(d)$  the specific angular momentum $ \tilde L^2$ for the $\alpha$-type relativistic thick shells  with  $n=3$  (left curves) and  $n=6$  (right curves)  and parameters  $\alpha=0$, $\tilde a = \tilde b = 0.6$ (solid curves), $0.8$ (dashed curves),  and $1.2$ (dot curves), as functions of $\tilde  r$. }
\label{fig:fig2}
\end{figure}

\begin{figure}
$$
\begin{array}{cc}
\includegraphics[width=0.38\textwidth]{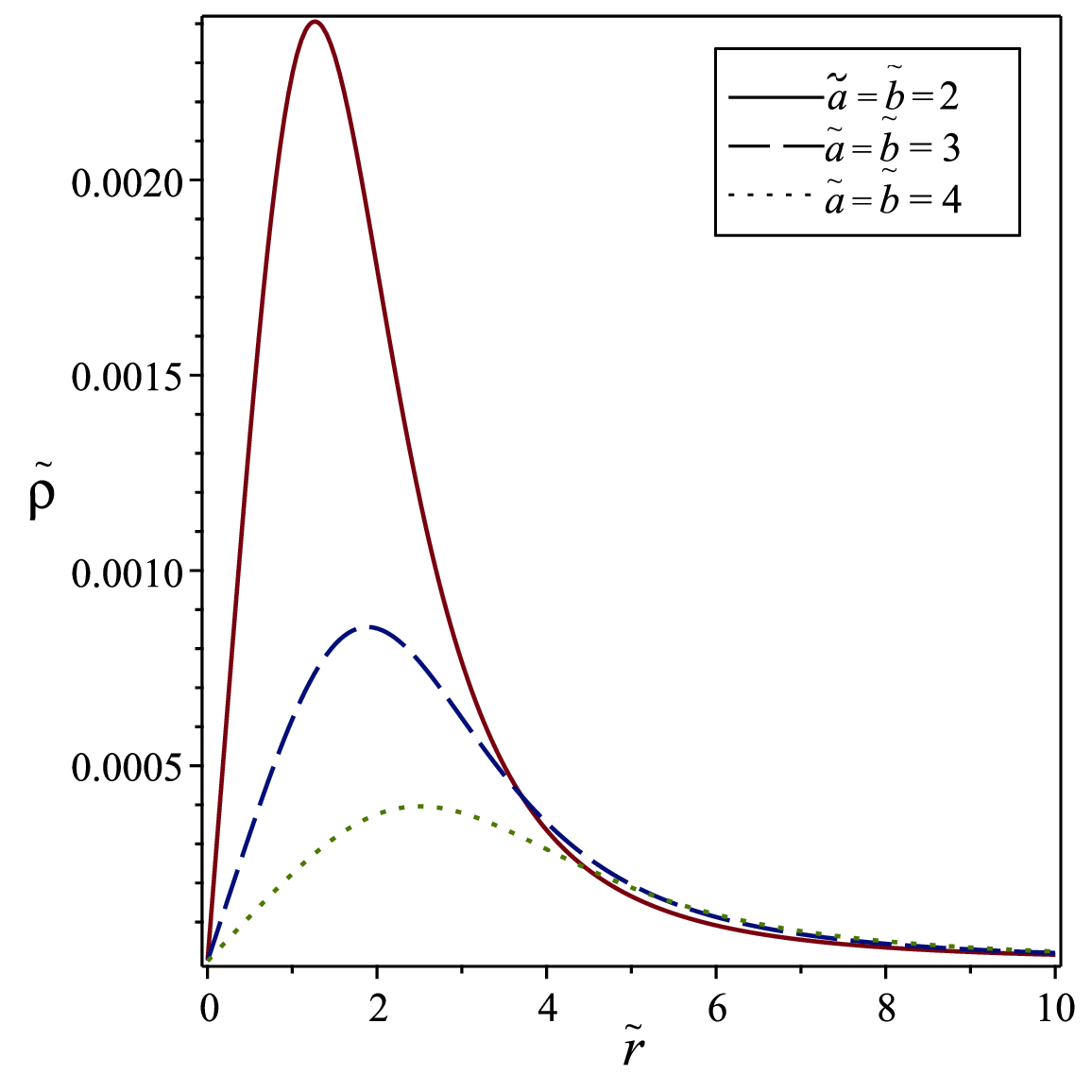} &
\includegraphics[width=0.38\textwidth]{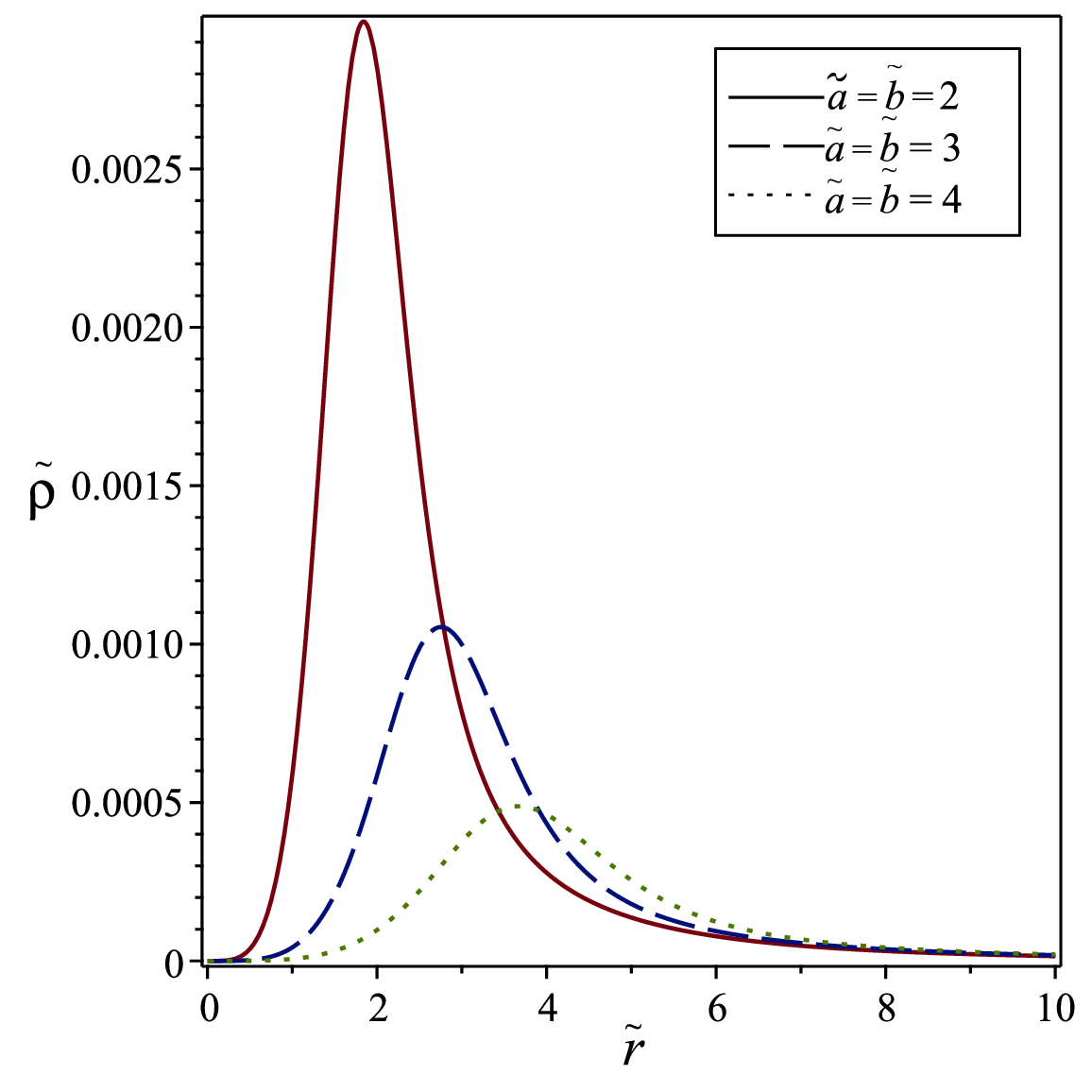}  \\
 (a) &  (b)  \\
&  \\
\includegraphics[width=0.38\textwidth]{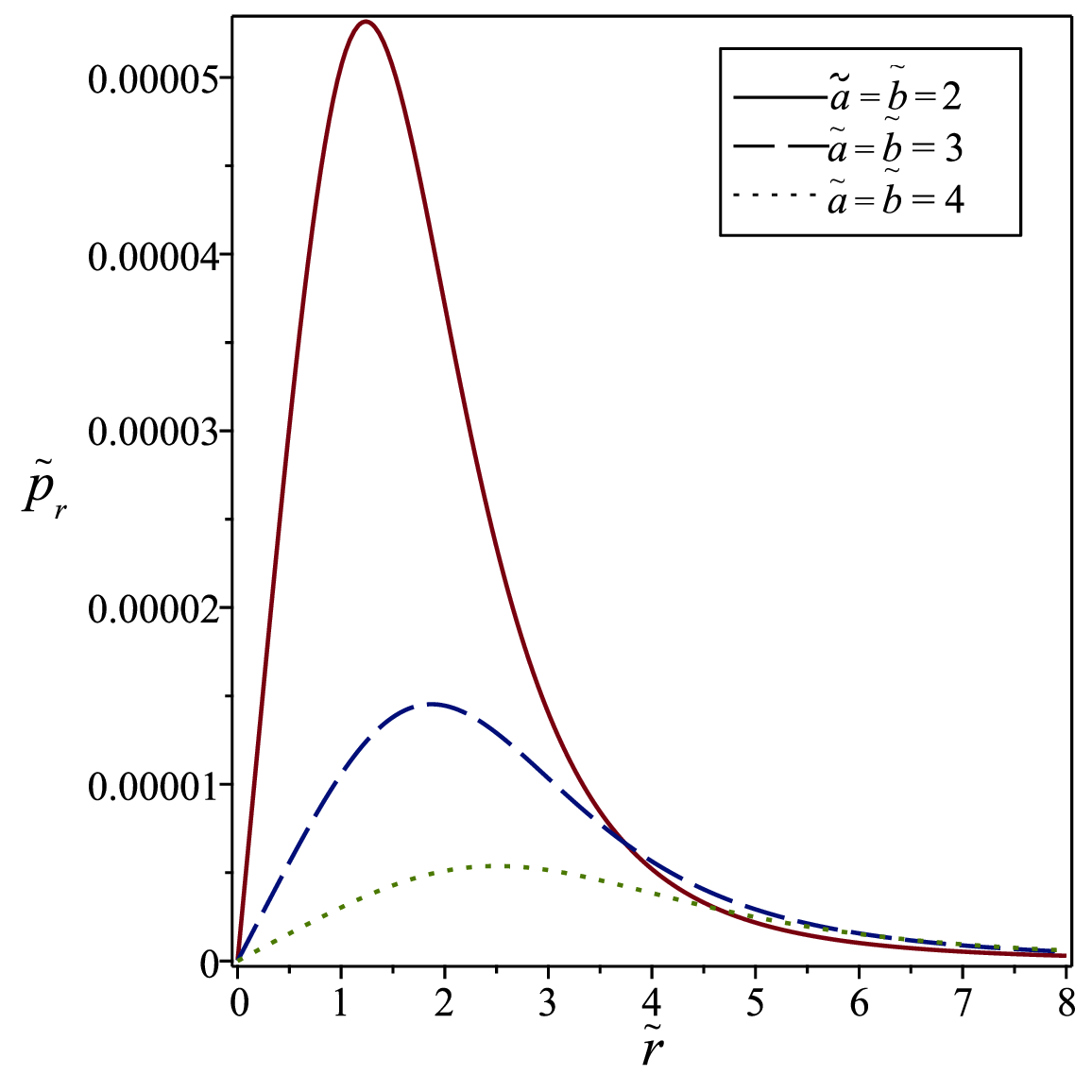} &
\includegraphics[width=0.38\textwidth]{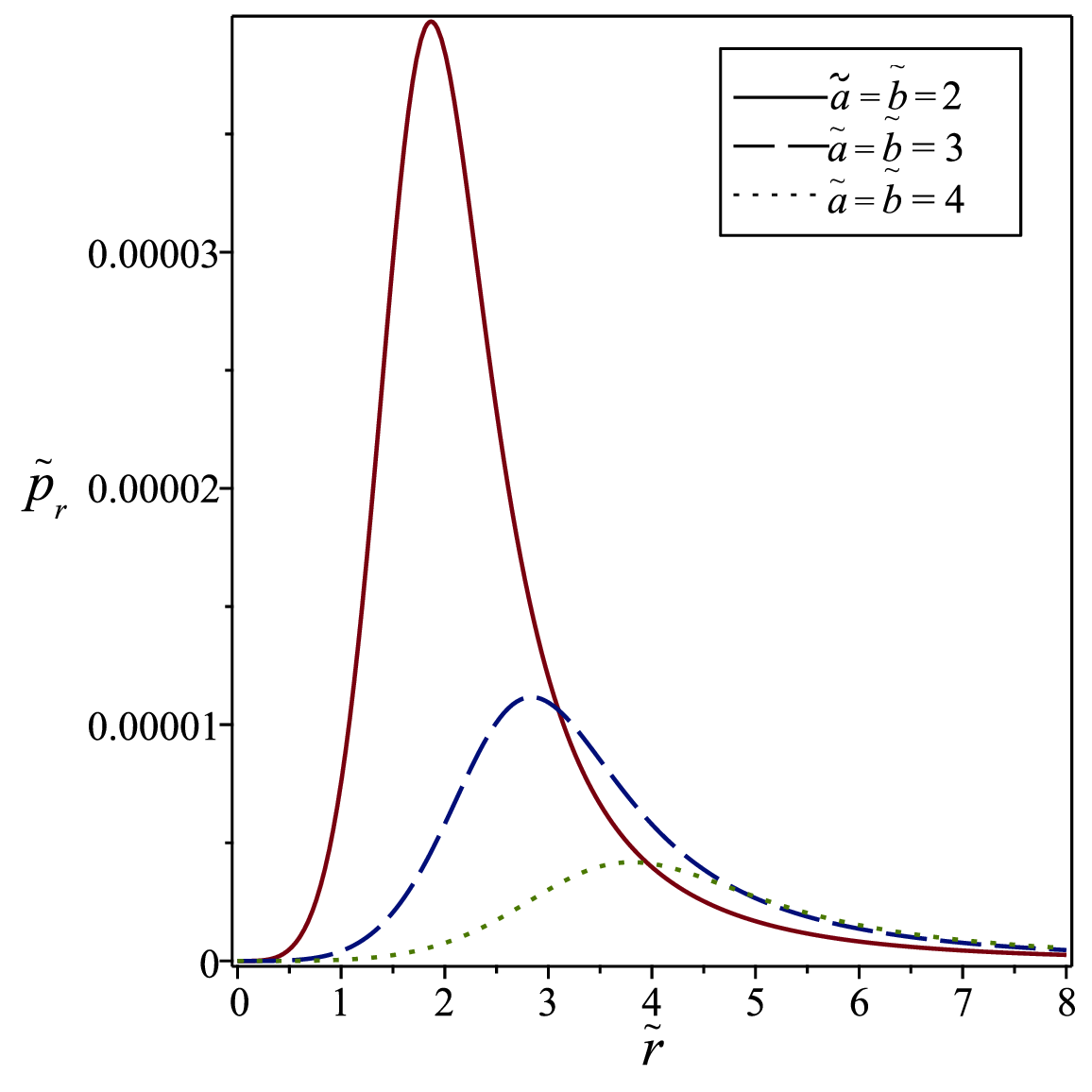}  \\
(c) &  (d)  \\
&  \\
\includegraphics[width=0.38\textwidth]{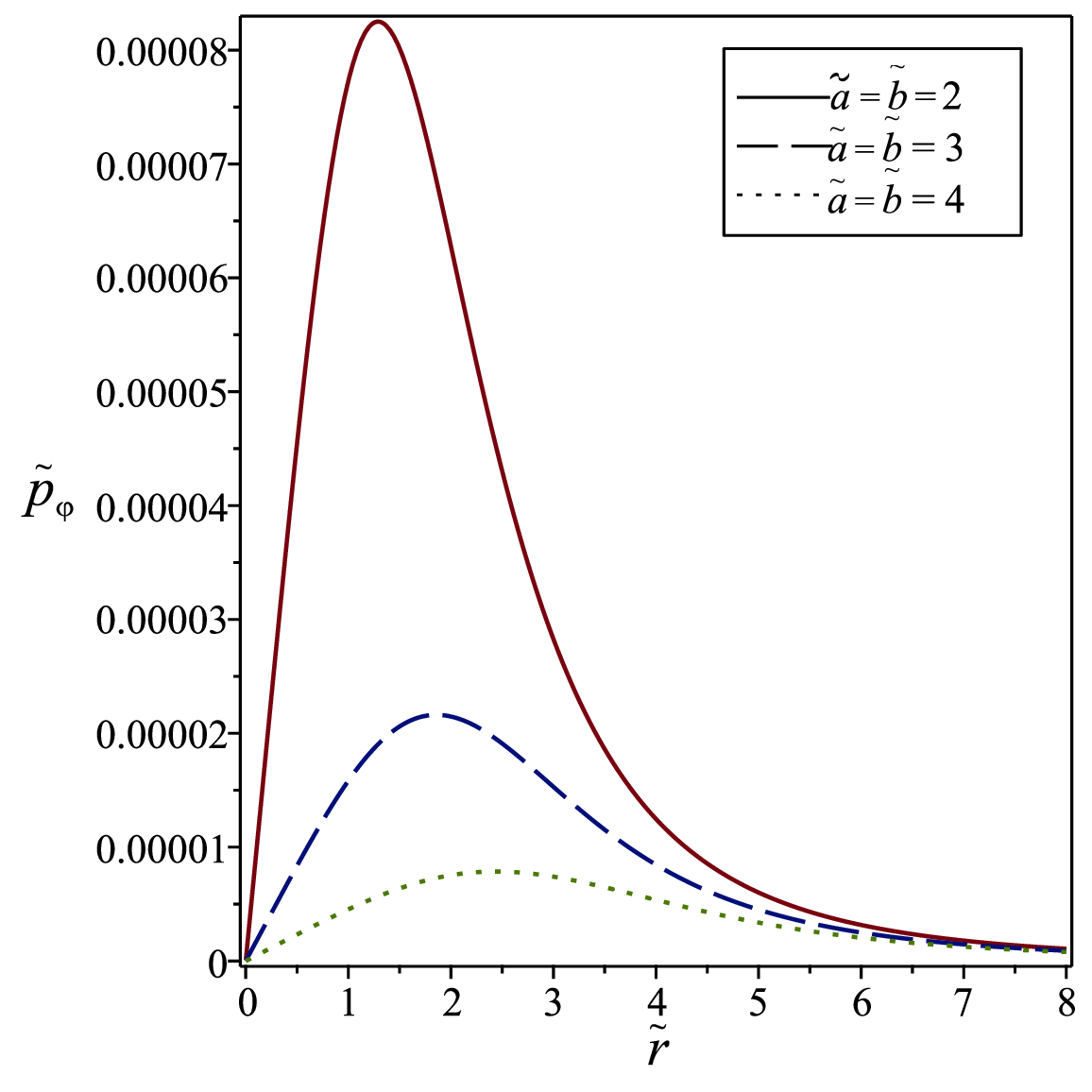} &
\includegraphics[width=0.38\textwidth]{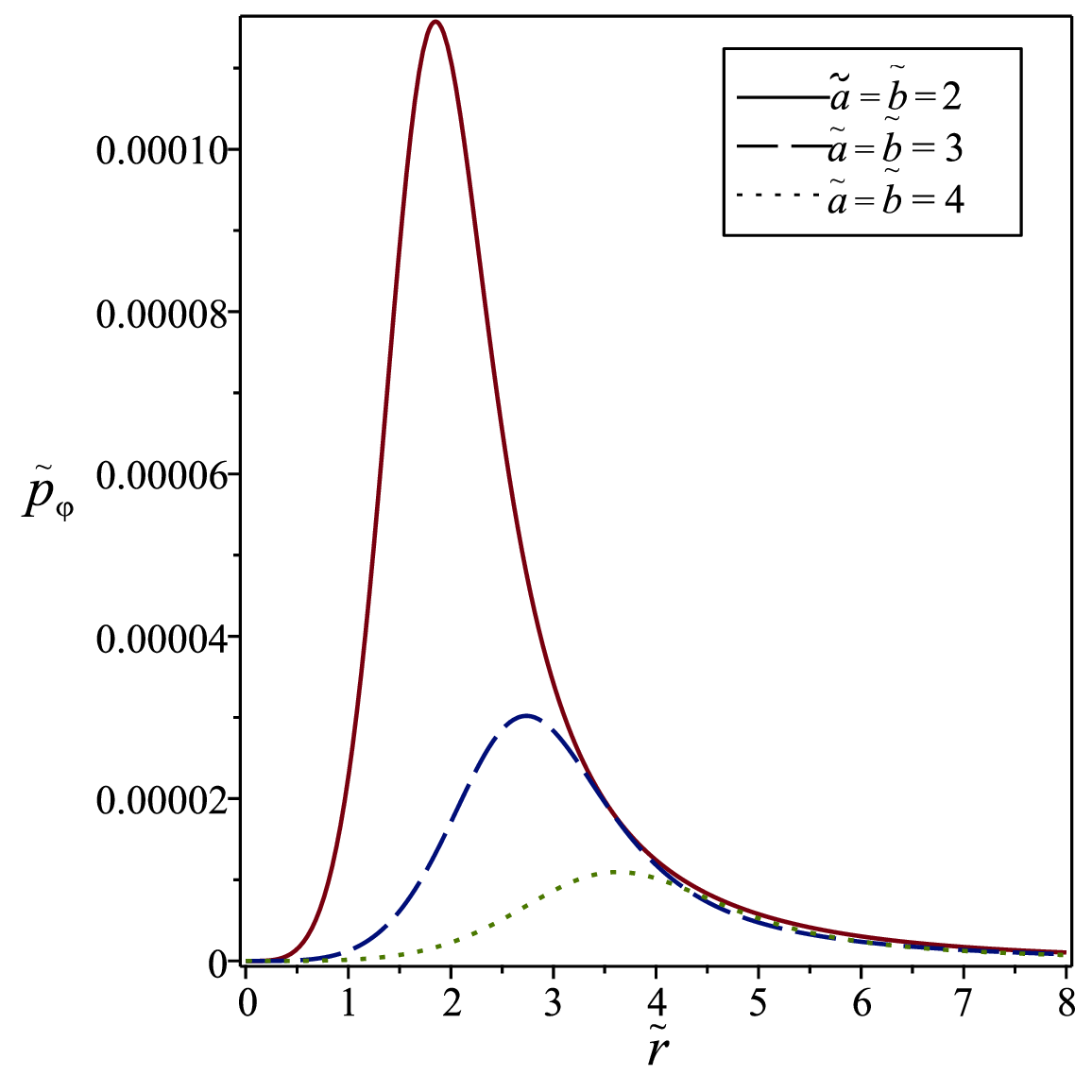} \\
 (e)   &    (f) \\
 &  \\
 \end{array}
$$	
\caption{ $(a)$, $(b)$ The relativistic energy density  $\tilde \rho$,    $(c)$, $(d)$ the radial pressure $\tilde  p_r$, and  $(e)$, $(f)$ the tangential  pressure $ \tilde  p_\varphi$ for the $\alpha$-type relativistic thick shells with  $n=3$  (left curves) and  $n=6$  (right curves) 
and parameters  $\alpha=1$, $\tilde a = \tilde b = 2$ (solid curves), $3$ (dashed curves), and $4$ (dot curves), as functions of $\tilde  r$. }
\label{fig:fig3}
\end{figure}

\begin{figure}
$$  
\begin{array}{cc}
\includegraphics[width=0.38\textwidth]{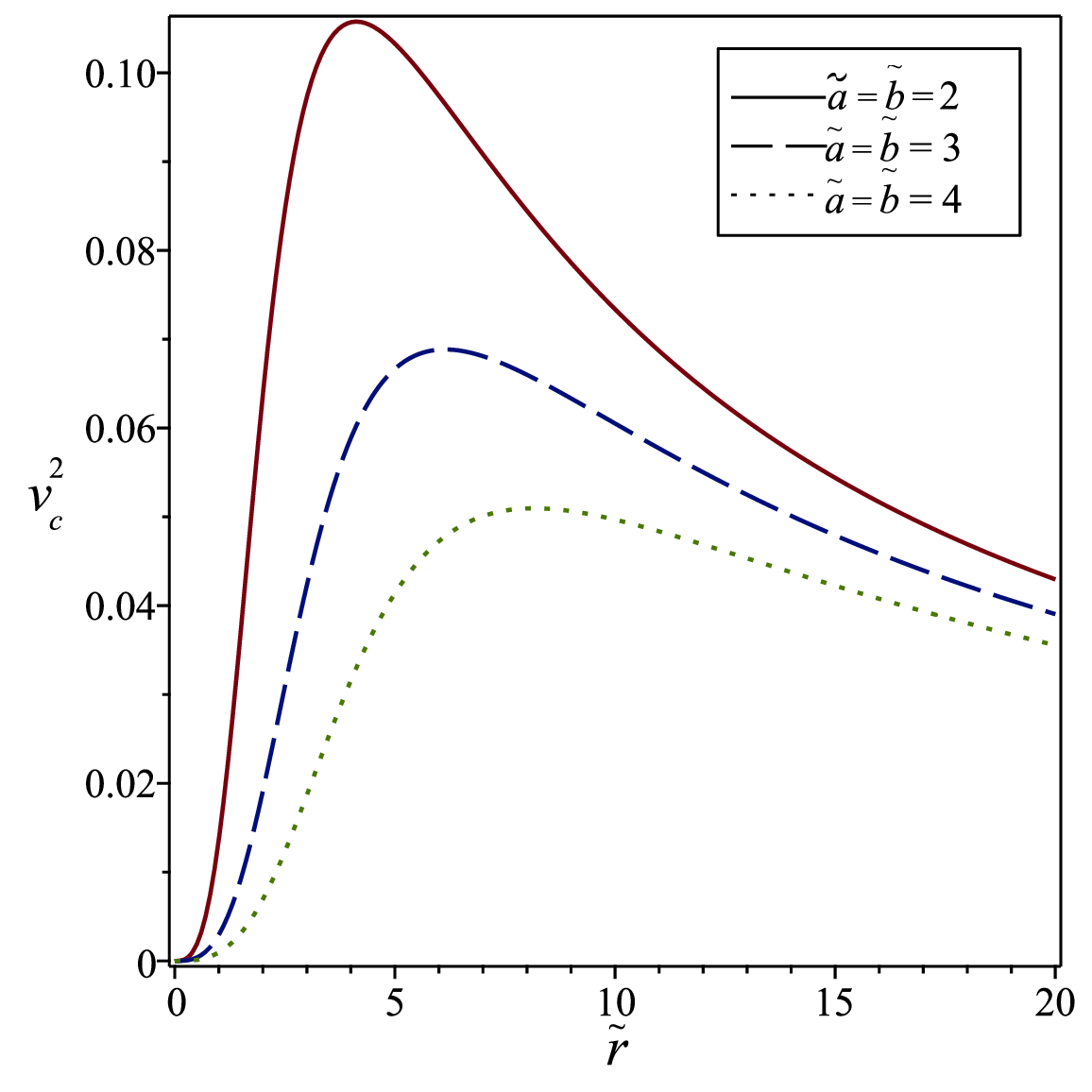} &
\includegraphics[width=0.38\textwidth]{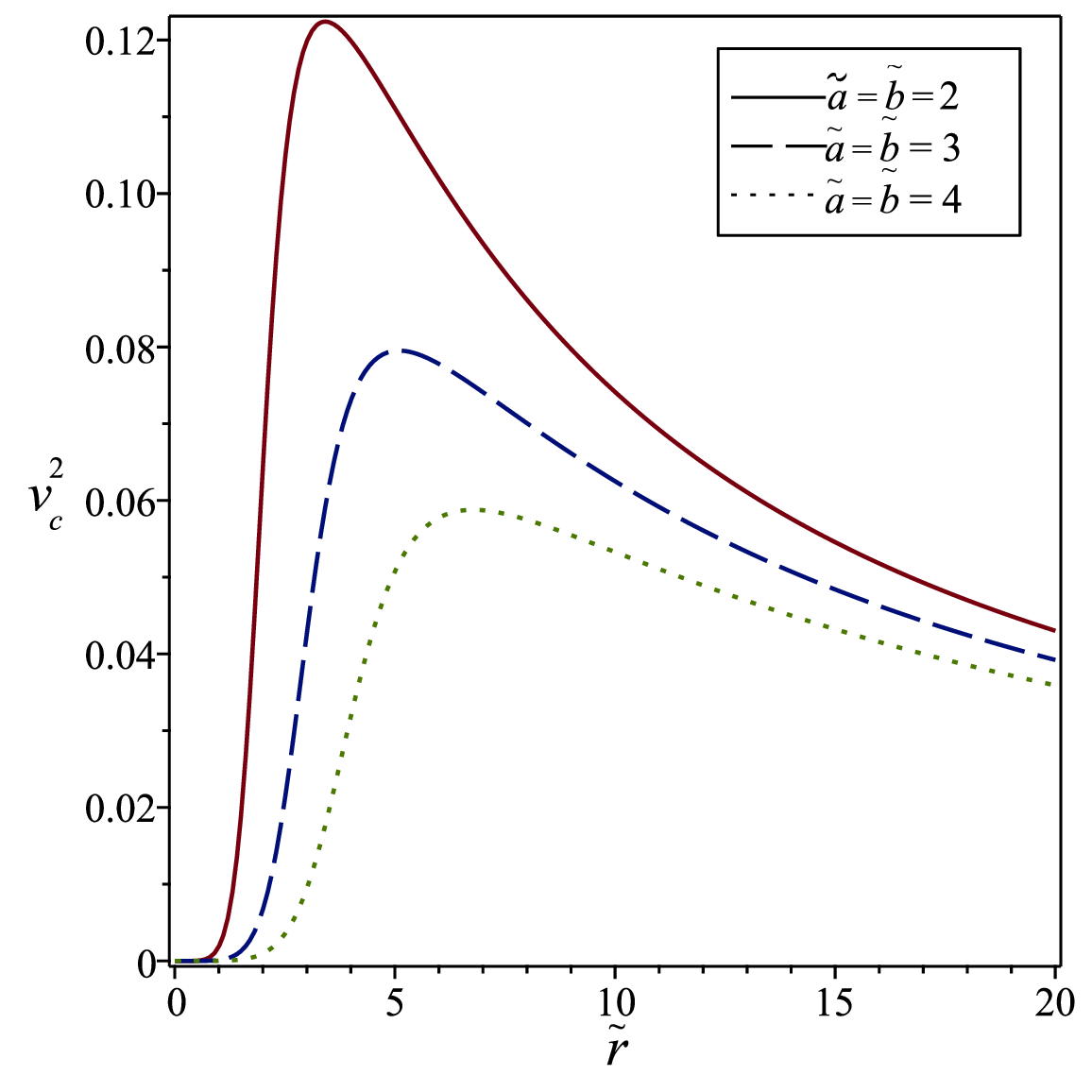} \\
 (a)   &    (b) \\
 &  \\
 \includegraphics[width=0.38\textwidth]{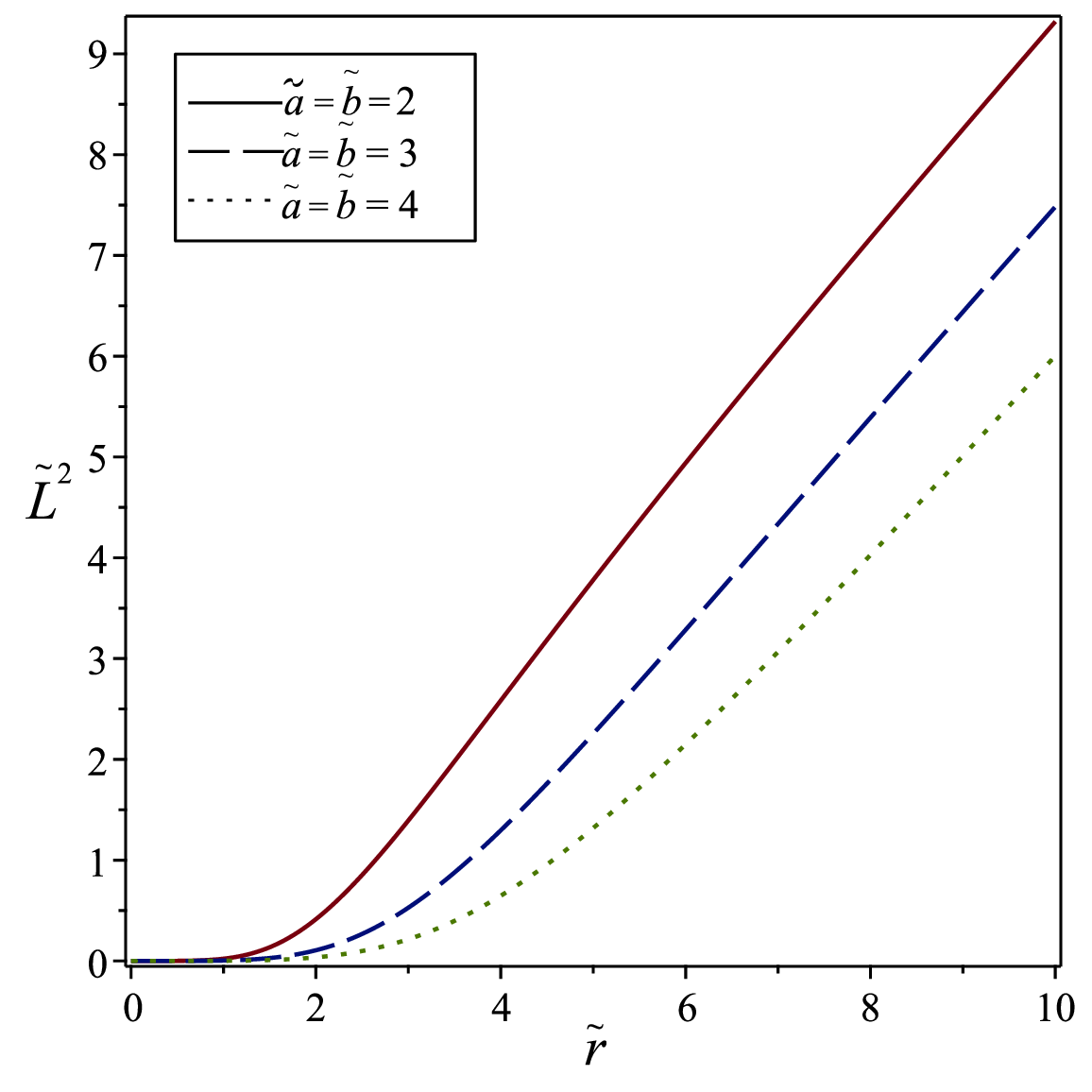} &
\includegraphics[width=0.38\textwidth]{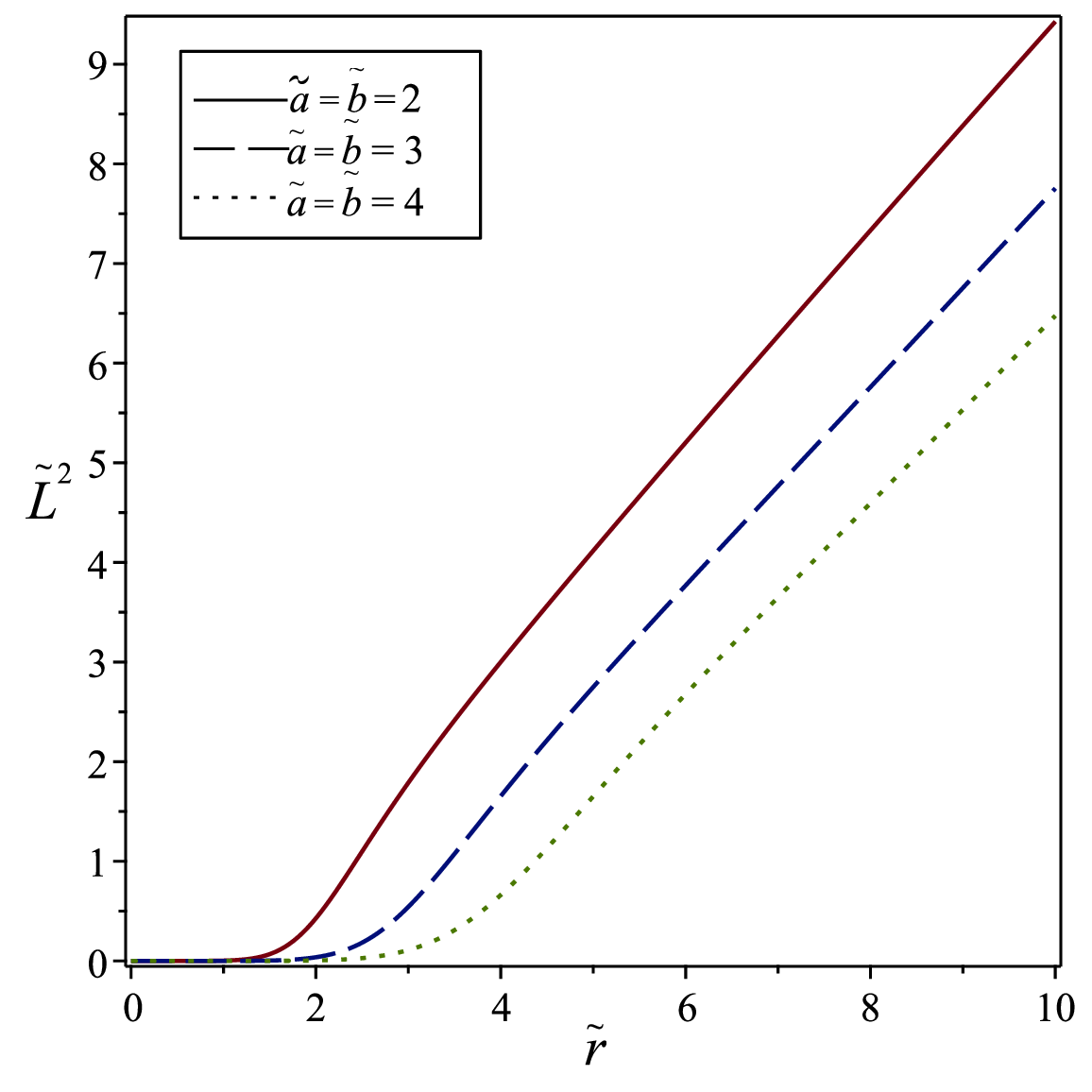} \\
 (c)   &    (d) \\
\end{array}
$$	
\caption{ $(a)$, $(b)$ The rotation curves
$v_c^2$ and $(c)$, $(d)$  the specific angular momentum $ \tilde L^2$ for the $\alpha$-type relativistic thick shells with  $n=3$  (left curves) and  $n=6$  (right curves) 
and parameters  $\alpha=1$, $\tilde a = \tilde b = 2$ (solid curves), $3$ (dashed curves), and $4$ (dot curves), as functions of $\tilde  r$. }
\label{fig:fig4}
\end{figure}

\begin{figure}
$$
\begin{array}{cc}
\includegraphics[width=0.38\textwidth]{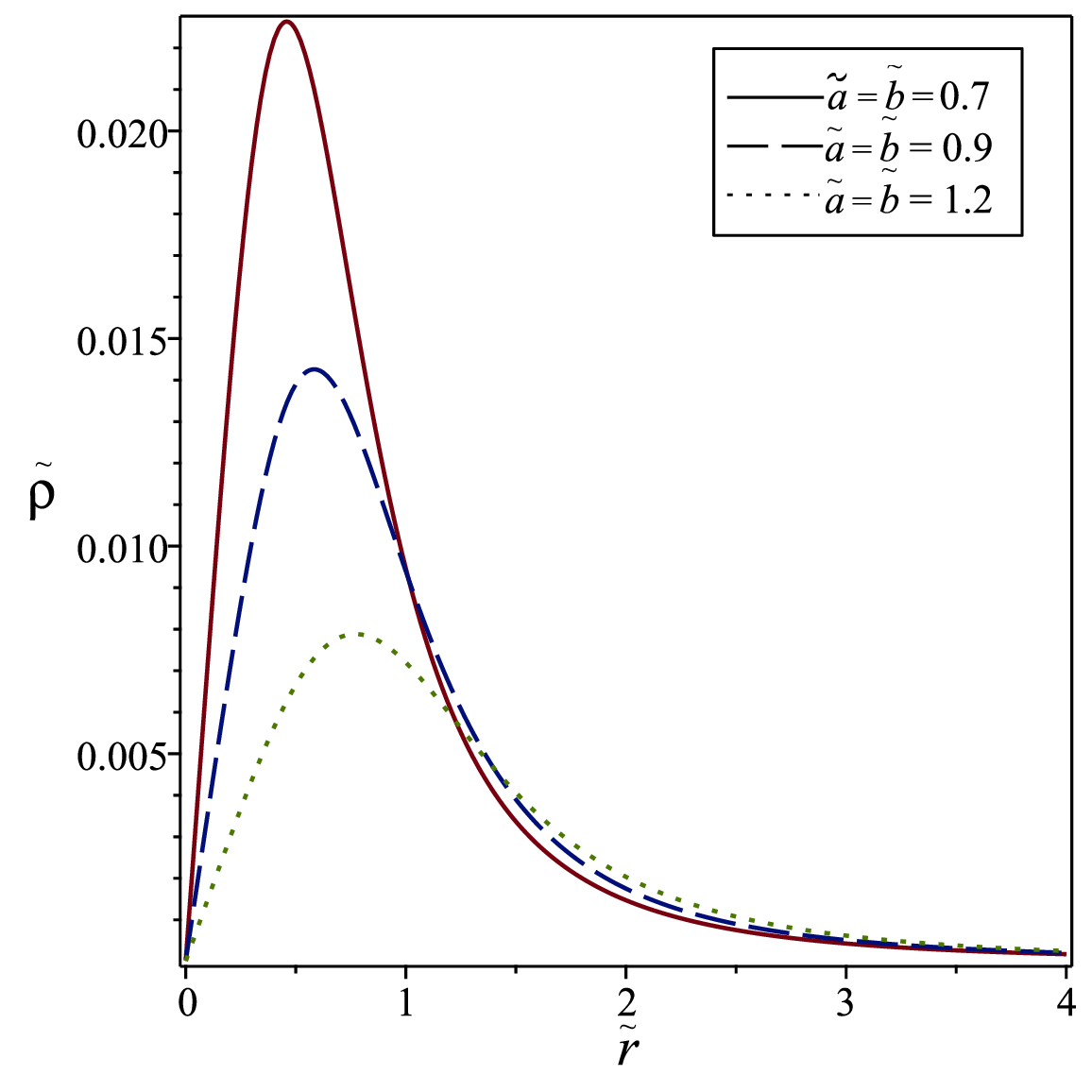} &
\includegraphics[width=0.38\textwidth]{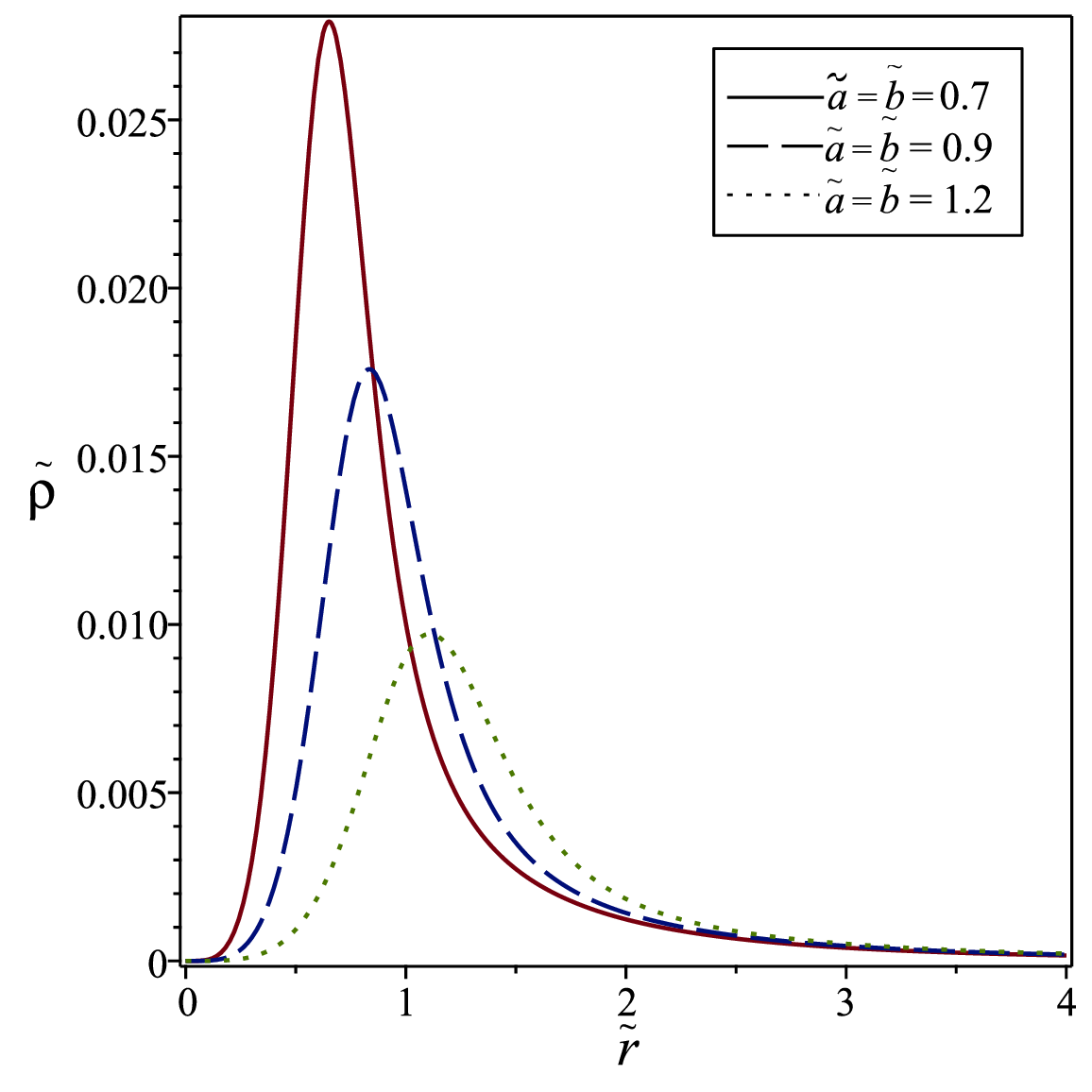}  \\
 (a) &  (b)  \\
&  \\
\includegraphics[width=0.38\textwidth]{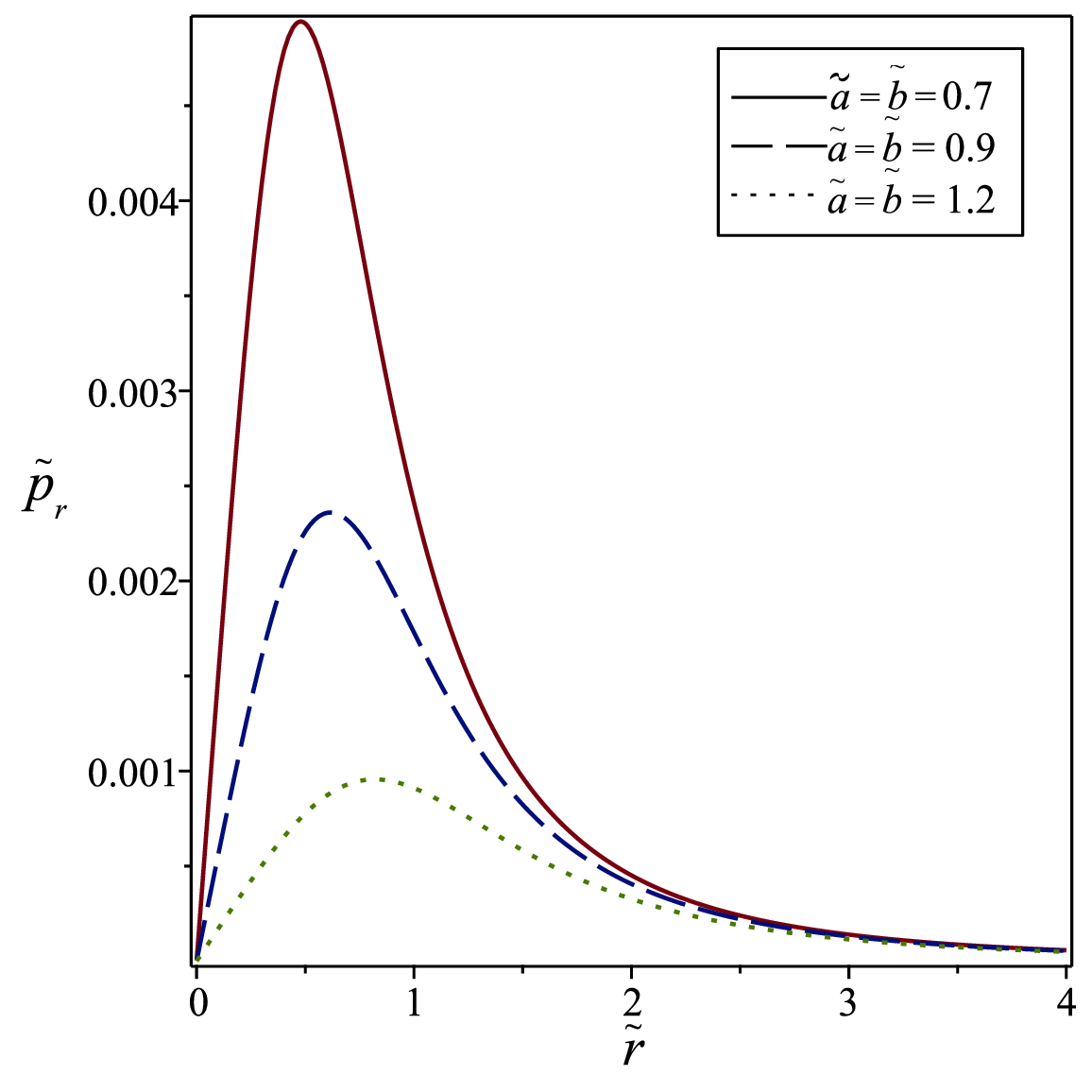} &
\includegraphics[width=0.38\textwidth]{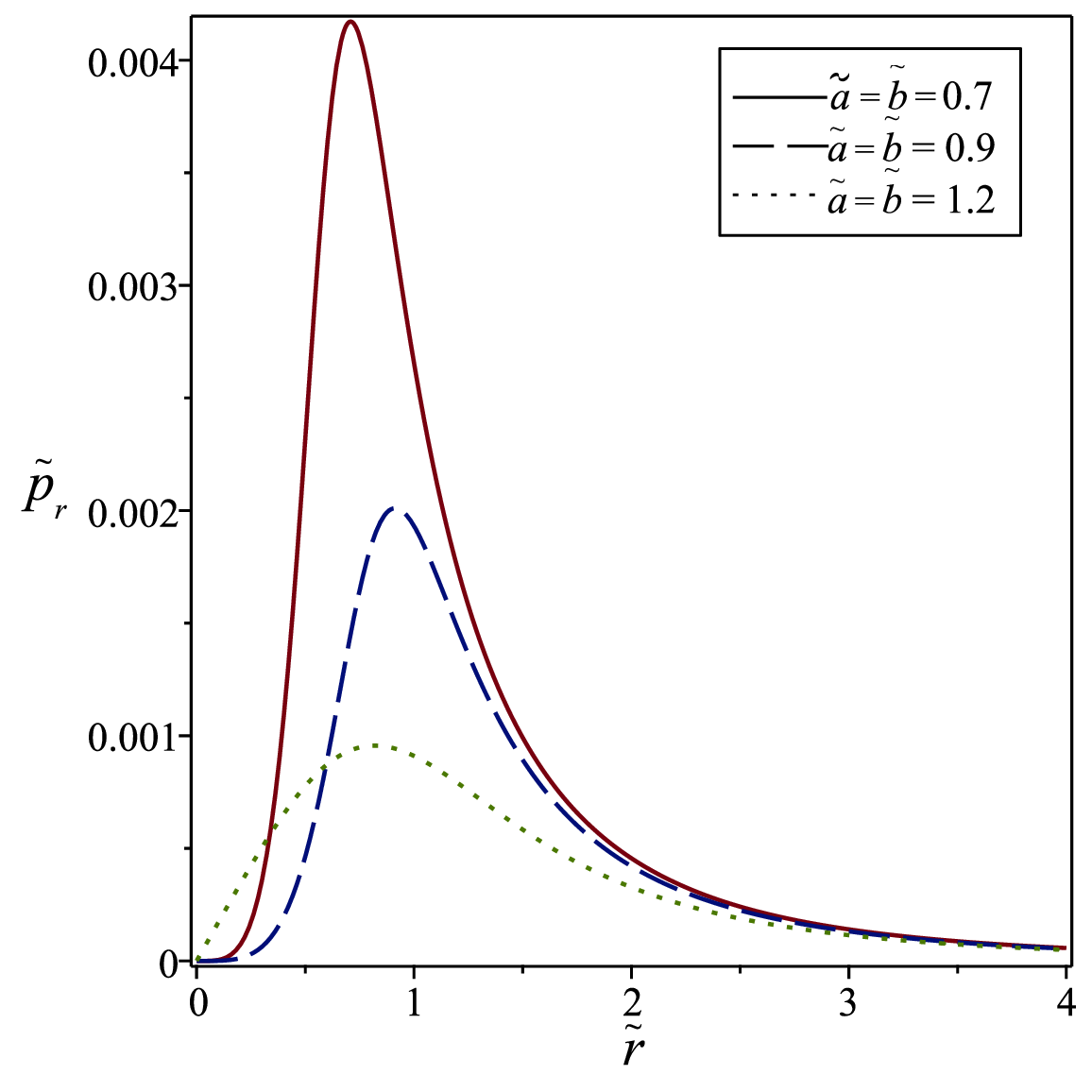}  \\
(c) &  (d)  \\
&  \\
\includegraphics[width=0.38\textwidth]{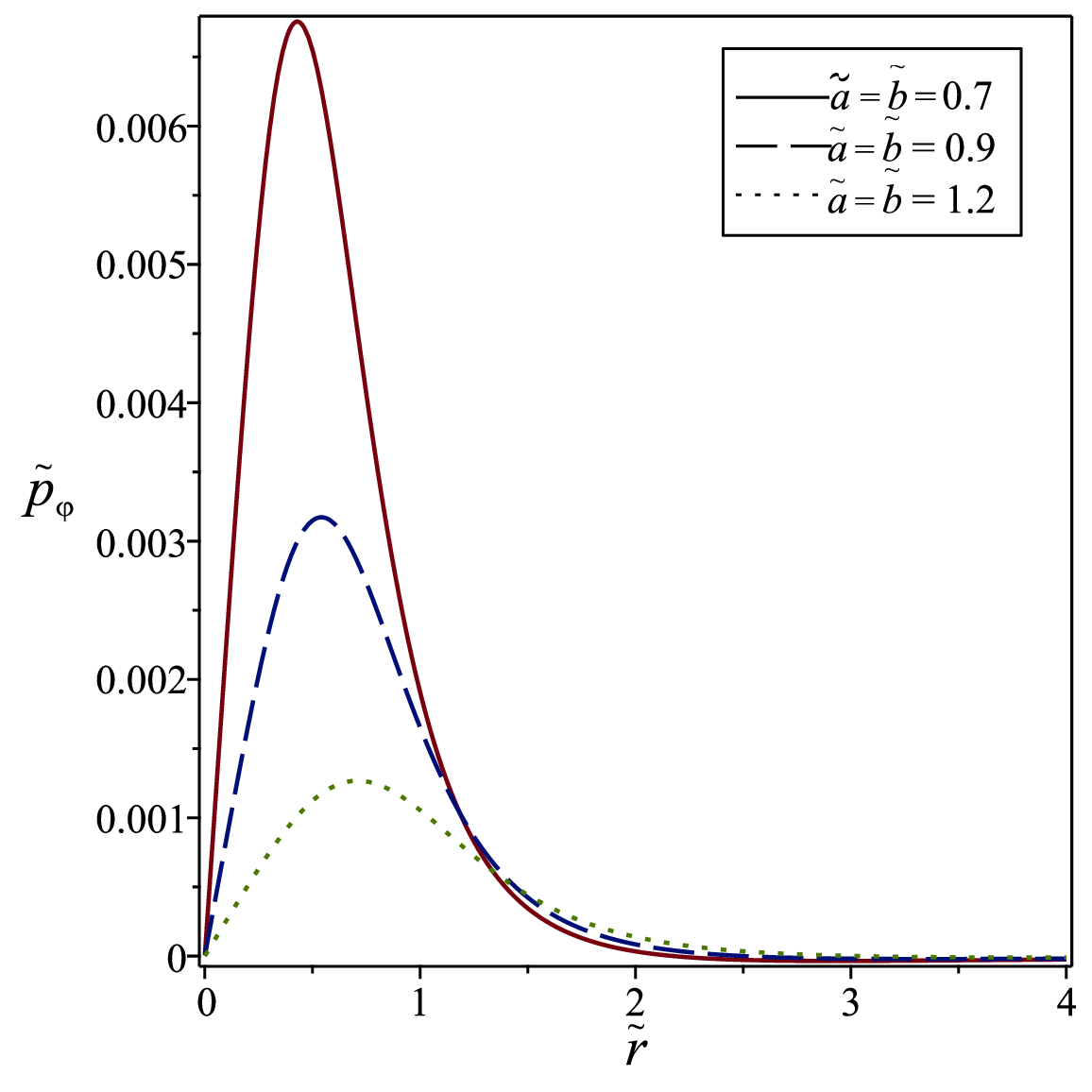} &
\includegraphics[width=0.38\textwidth]{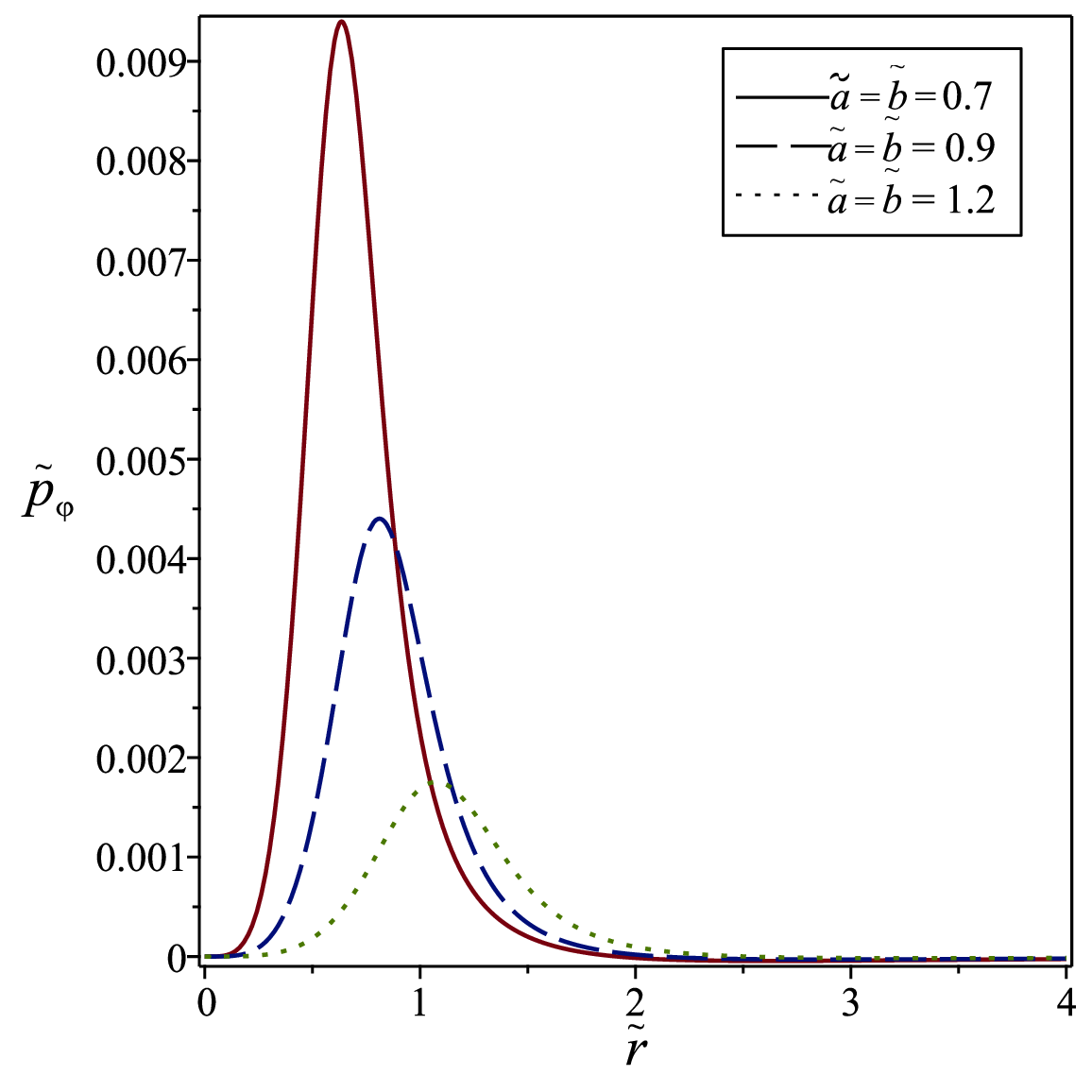} \\
 (e)   &    (f) \\
 &  \\
 \end{array}
$$	
\caption{ $(a)$, $(b)$ The relativistic energy density  $\tilde \rho$,    $(c)$, $(d)$ the radial pressure $\tilde  p_r$, and  $(e)$, $(f)$ the tangential  pressure $ \tilde  p_\varphi$ for the $\beta$-type relativistic thick shells with  $n=3$  (left curves) and  $n=6$  (right curves) 
and parameters  $\beta=3$, $\tilde a = \tilde b = 0.7$ (solid curves), $0.9$ (dashed curves), and $1.2$ (dot curves), as functions of $\tilde  r$. }
\label{fig:fig5}
\end{figure}

\begin{figure}
$$
\begin{array}{cc}
\includegraphics[width=0.38\textwidth]{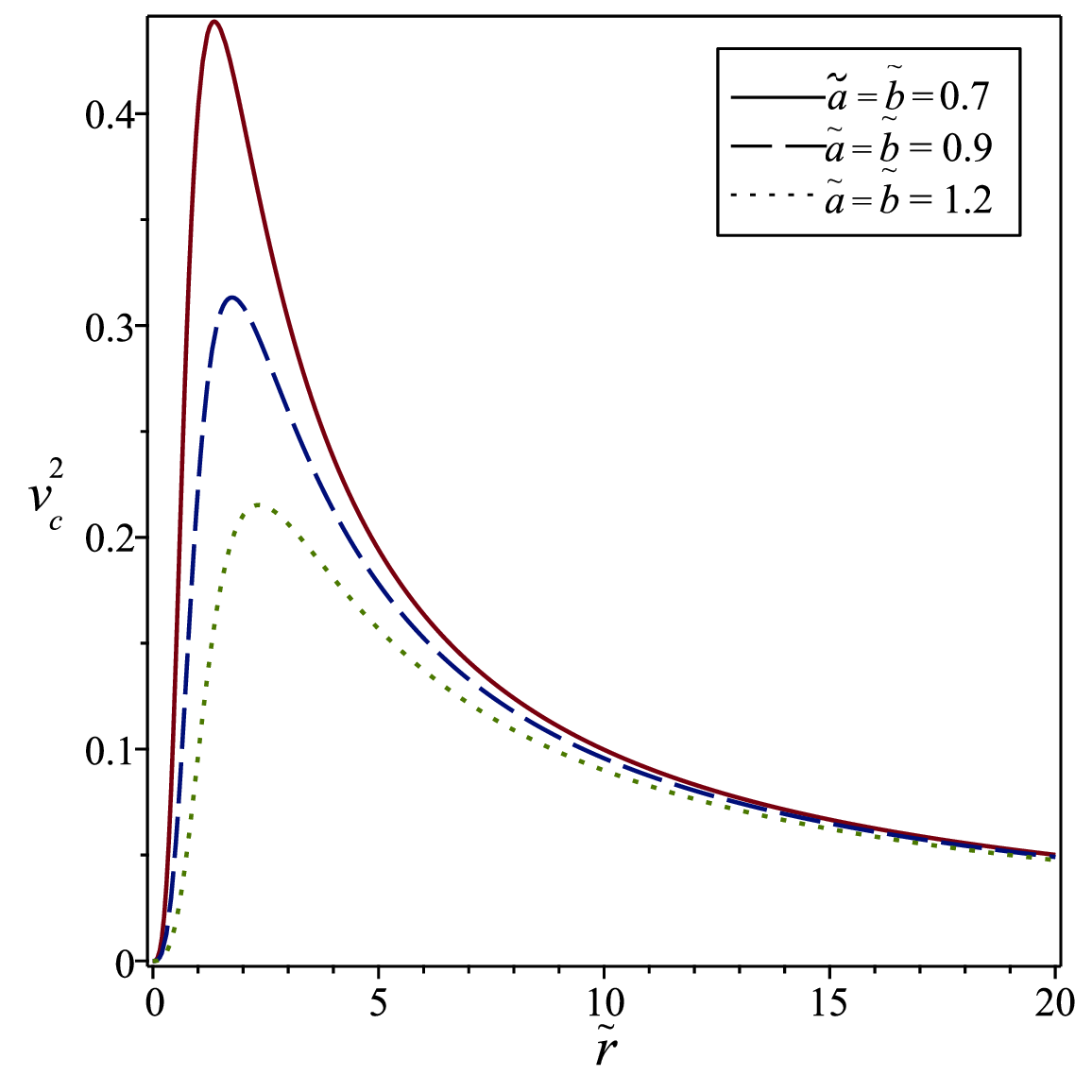} &
\includegraphics[width=0.38\textwidth]{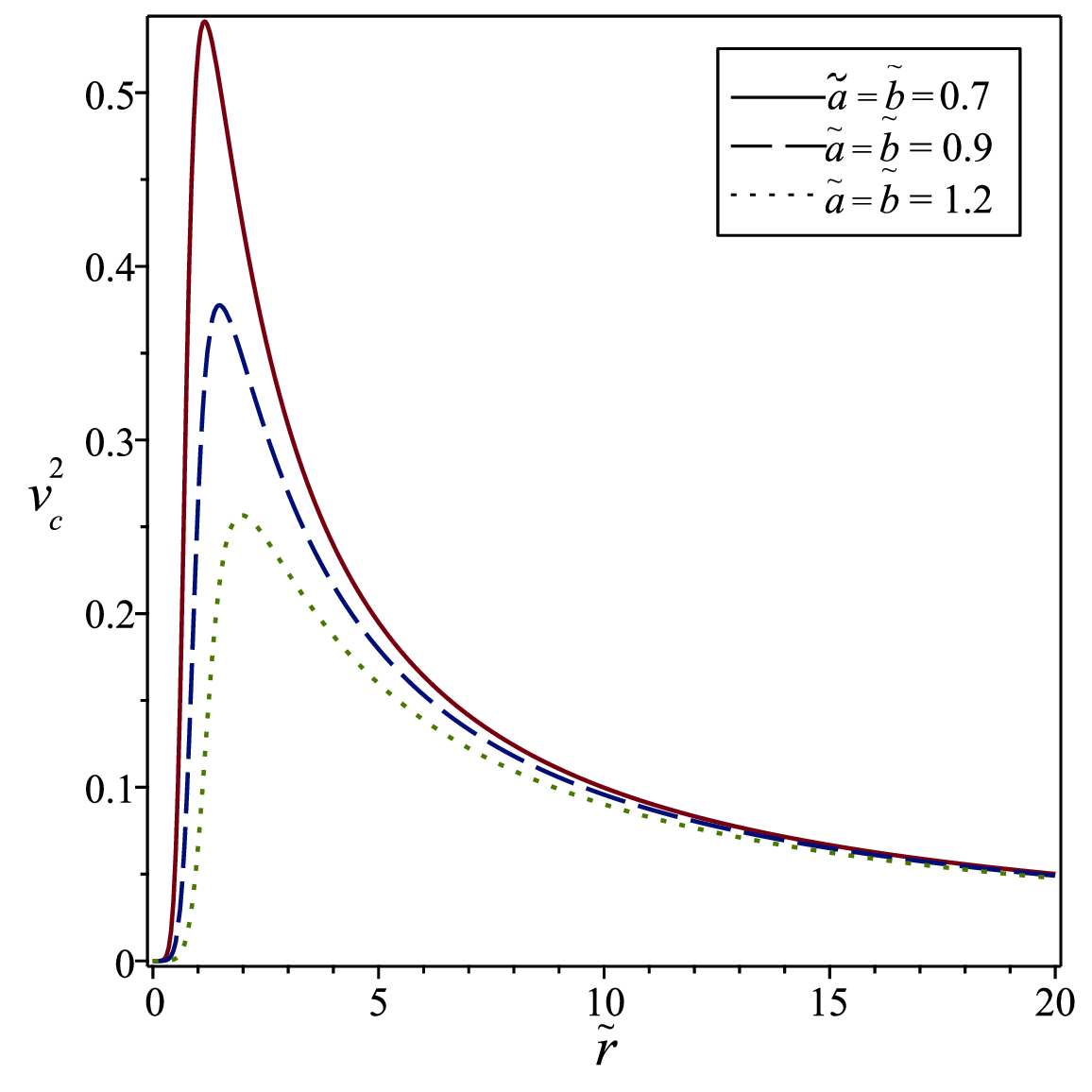} \\
 (a)   &    (b) \\
 &  \\
 \includegraphics[width=0.38\textwidth]{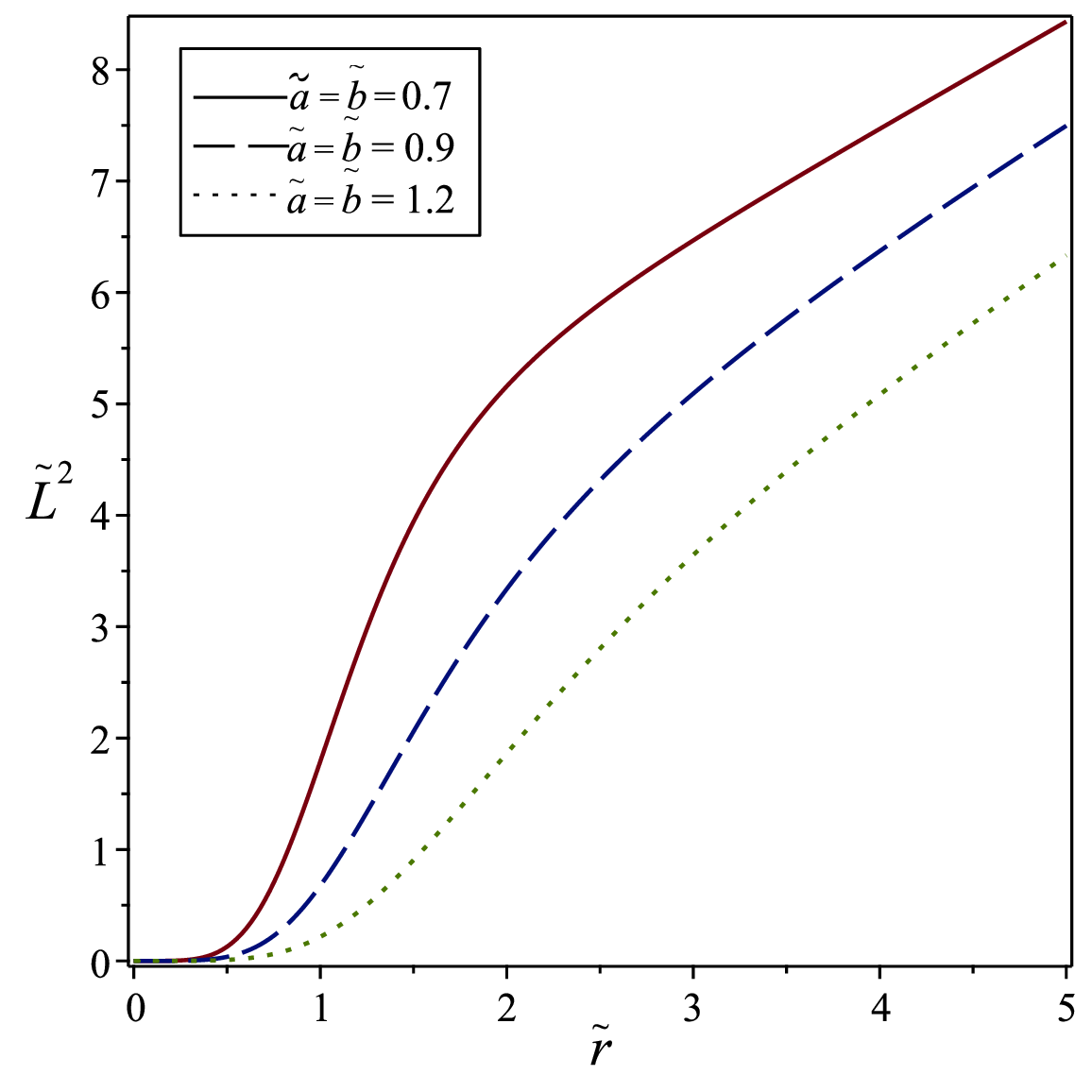} &
\includegraphics[width=0.38\textwidth]{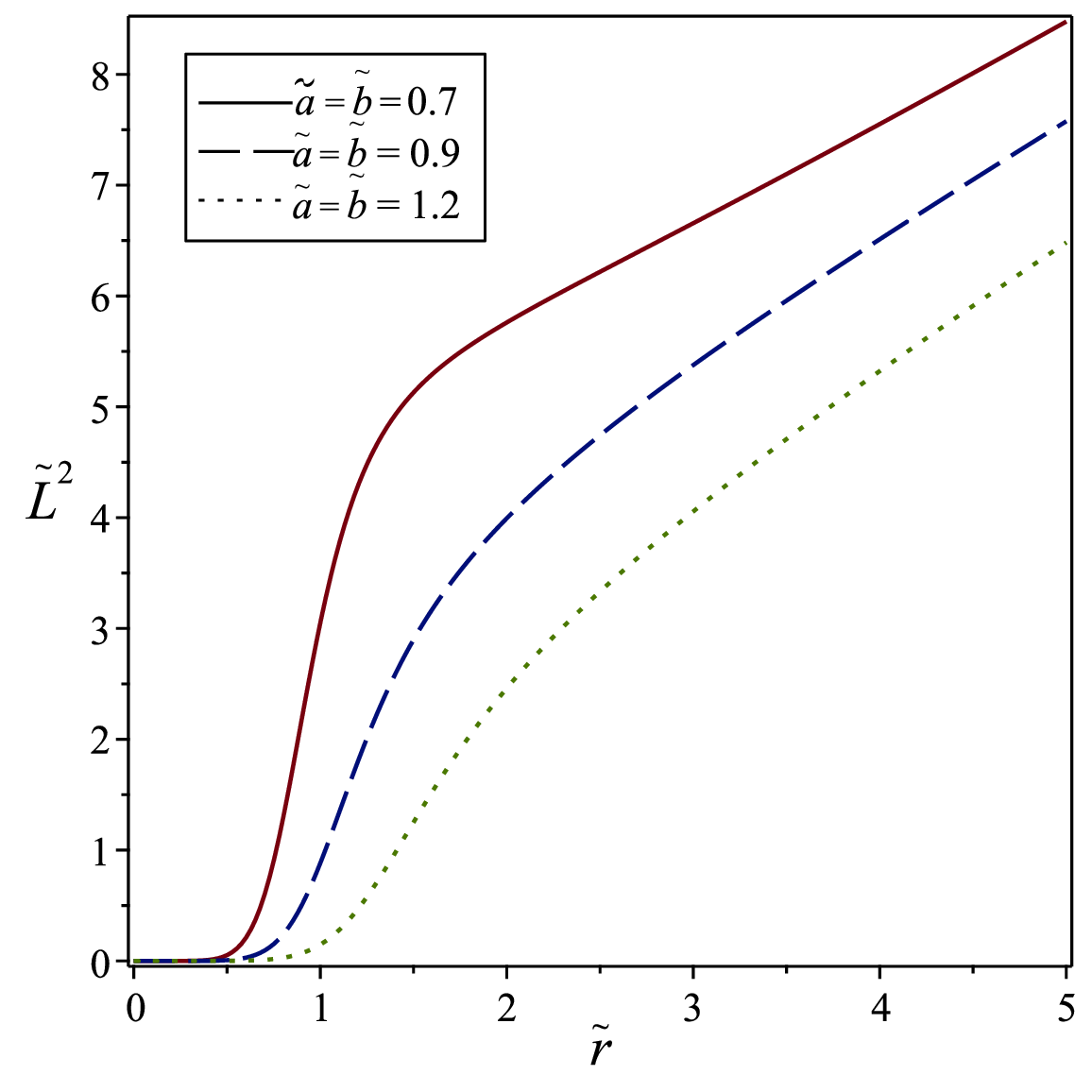} \\
 (c)   &    (d) \\
\end{array}
$$	
\caption{ $(a)$, $(b)$ The rotation curves
$v_c^2$ and $(c)$, $(d)$  the specific angular momentum $\tilde L^2$ for the $\beta$-type relativistic thick shells with  $n=3$  (left curves) and  $n=6$  (right curves) 
and parameters  $\beta=3$, $\tilde a = \tilde b = 0.7$ (solid curves), $0.9$ (dashed curves), and $1.2$ (dot curves), as functions of $\tilde  r$. }
\label{fig:fig6}
\end{figure}

In figures  \ref{fig:fig2}, \ref{fig:fig4} and \ref{fig:fig6} we plot  the rotation curves
$v_c^2$ and  the specific angular momentum $ \tilde L^2$ for the same value of parameters.   As the value of the parameters $\tilde a$ and $\tilde b$ are increased, the circular speed decreases  and increases  
with the increase of the  parameter $n$. For this value of parameters  also  the circular speed  of particles is 
a quantity less than the speed of light  in according to the dominant energy condition and the orbits of the particles are stable. However, the orbits can become unstable 
as the value of the  parameters  $\tilde a$ and $\tilde b$ decrease. 

\section{ A relativistic model of galaxy } 

In cylindrical coordinates  ($t$, $\varphi$, $R$, $z$) the metric $(\ref{eq: met})$ takes the form
\begin{equation}
ds^2 =  - e^{2\nu} dt^2 + e^{2\lambda} (R^2 d \varphi^2  + dR^2 + dz^2),  \label{eq:met}
\end{equation}
where $\phi$ is function of $R$ and $z$ only. The Einstein's gravitational  field equations  yield the following non-zero components
of the energy-momentum tensor
\begin{subequations}\begin{eqnarray}
T^t_{\ t} & = &   \frac{1}{4 \pi G} e^{ -2\lambda} \left [\nabla^2 \lambda +  \frac1 2 \nabla \lambda \cdot \nabla \lambda \right ],   \label{eq: Ttt} \\
&  &  \nonumber \\
 T^\varphi_{\ \varphi} & = &   \frac{1}{8 \pi G} e^{ -2\lambda} \left [   \nabla \nu \cdot \nabla \nu + \nabla^2 \nu +\nabla^2 \lambda
 -\frac 1 R ( \lambda_{,R} +  \nu_{,R} )
 \right ] ,   \\
& &  \nonumber  \\
T^R_{ \ R} & = &  \frac{1}{8 \pi G} e^{ -2\lambda} \left [
\lambda_{,zz} + \nu_{,zz} 
+ \lambda_{,R}^2  + \nu_{,z}^2
+2 \lambda_{,R}\nu_{,R}
+ \frac 1 R (\lambda_{,R} + \nu_{,R})
\right ]  ,  \nonumber   \\
& & \\
T^z_{ \ z} & = &   \frac{1}{8 \pi G} e^{ -2\lambda} \left [
\lambda_{,RR} + \nu_{,RR} 
+ \lambda_{,z}^2  + \nu_{,R}^2
+2 \lambda_{,z}\nu_{,z}
+ \frac 1 R (\lambda_{,R} + \nu_{,R})
\right ] ,  \nonumber \\
& & \\
T^R_{ \ z} & = & -  \frac{1}{8 \pi G} e^{ -2\lambda} \left [
\lambda_{,Rz} + \nu_{,Rz}
-(\lambda_{,z} + \nu_{,z})\lambda_{,R}
+ \nu_{,R}\nu_{,z} -  \lambda_{,z} \nu_{,R}
\right ].   \label{eq: TRz}
\end{eqnarray} \end{subequations} 

Now, in order to analyze the matter distributions is   necessary to
compute the eigenvalues and eigenvectors of  the energy-momentum tensor. The
eigenvalue problem for the   energy-momentum tensor   (\ref{eq: Ttt}) - (\ref{eq: TRz})
has the solutions
\begin{subequations}\begin{eqnarray}
\lambda_t \  &=& \  T^t \ _t  ,  \\
\lambda_\varphi \  &=& \  T^\varphi \ _\varphi,  \\
\lambda_{R,z} \  &=& \frac{T \pm \sqrt{D} }{2},
\end{eqnarray} \end{subequations} 
where 
\begin{subequations}\begin{eqnarray}
T &= & T^R \ _ R + T^z \ _ z, \\
D & = & (T^R \ _ R  -  T^z \ _ z)^2 + 
4 (T^R \ _ z)^2.
\end{eqnarray} \end{subequations} 

The corresponding eigenvectors are
\begin{subequations}\begin{eqnarray}
{\bf V}_{ax} & = & \frac{1}{\sqrt{-g_{tt}}} ( 1, 0, 0, 0 ),	 \label{eq: tetrada0} \\
	&	&  \nonumber	\\
{\bf X}_{ax} & = & \frac{1}{\sqrt{g_{\varphi  \varphi}}}  ( 0, 1, 0, 0 ),  \\
	&	&  \nonumber	\\
{\bf Y}_{ax} & = & \frac{1}{\sqrt{g_{RR} ( 1+ n_+^2 )}} (0  , 0 , 1,  n_+  ),  \\
	&	&  \nonumber	\\
{\bf Z}_{ax} & = &  \frac{1}{\sqrt{g_{RR} ( 1+ n_-^2 )}}   (0  , 0 , 1,  n_-  ),  \label{eq: tetrada3}
\end{eqnarray} \end{subequations} 
where
\begin{equation}
n _\pm = \frac { T^z \ _ z  -  T^R \ _ R \pm \sqrt{D}  }{2 T^R \ _z} .
\end{equation}

In terms of the  above proper observer,  the energy density is given by  $\rho= - \lambda_t$
and the principal stresses  by  $p_i = \lambda _i $. For the  particular form  of $\lambda$
(\ref{eq: lamb})  the energy density is given by the nonlinear Poisson type  equation (\ref{eq: nonlinear})
and  again taking the   energy density profile as (\ref{eq: rho1}),  then it follows $(\phi,\rho_0 )= (\Phi,\rho_N )$.
Similarly,  for  equatorial,  circular orbit  
the  circular speed and the specific angular momentum  are given by 
\begin{subequations}\begin{eqnarray}
v_c ^2  &=& \frac{ R \nu _{,R} }{ 1+ R \lambda _{,R}}, 
\\
L^2 & = & \frac{R^2 e^{2 \lambda } v_c^2}{1- v_c^2},
\end{eqnarray} \end{subequations} 
and  the stability condition   reads 
\begin{equation}
\frac{d L^2}{dR} > 0.
\end{equation}
Following the approach presented  in the above section,
the metric potential  
$\nu$ can also be take   in the particular form    (\ref{eq: nuani})  in cylindrical coordinates. 

As an example, we consider the Newtonian Galactic potential composite  by the sum of  three  components 
\begin{equation}
\Phi = \Phi_B + \Phi_D + \Phi_H,
\end{equation}
where $\Phi_B$  corresponds to  the central spherical bulge potential, 
 $\Phi_D$ describes the  thick disk and  $\Phi_H$ the 
spherical dark halo.
The bulge and disk potentials
 are given  by Plummer  and Miyamoto-Nagai models \cite{Plummer,Miyamoto,Nagai}
\begin{subequations}\begin{eqnarray}
\Phi_B &=& -\frac{GM_B}{\sqrt{r^2 + b_B^2}} ,  \\
&  & \nonumber   \\
\Phi_D &=& -\frac{GM_D}{\sqrt{R^2 + \left(a_D +\sqrt{ z^2 + b_D^2}\right)^2}} ,  \\
&  & \nonumber   
\end{eqnarray} \end{subequations}
where $r = \sqrt {R^2 + z^2}$, $M_B$ and $M_D$ are the masses of the components while
$b_B$, $a_D$, and $b_D$ are non-zero constants with dimensions of length. In addition,   $a_D$ and $b_D$   represent length and  height scales  of the disk-like distribution.  The dark matter  halo is represented by  NFW model  which can be cast as 
\begin{equation}
\Phi_H =  -  \frac{GM_H}{r}  \ln \left( 1 + \frac{r}{a_H}  \right),  
\end{equation}
where $M_H$ is the  dark halo mass and $a_H$ a scale radius.

\begin{figure}
$$
\begin{array}{cc}
\includegraphics[width=0.4\textwidth]{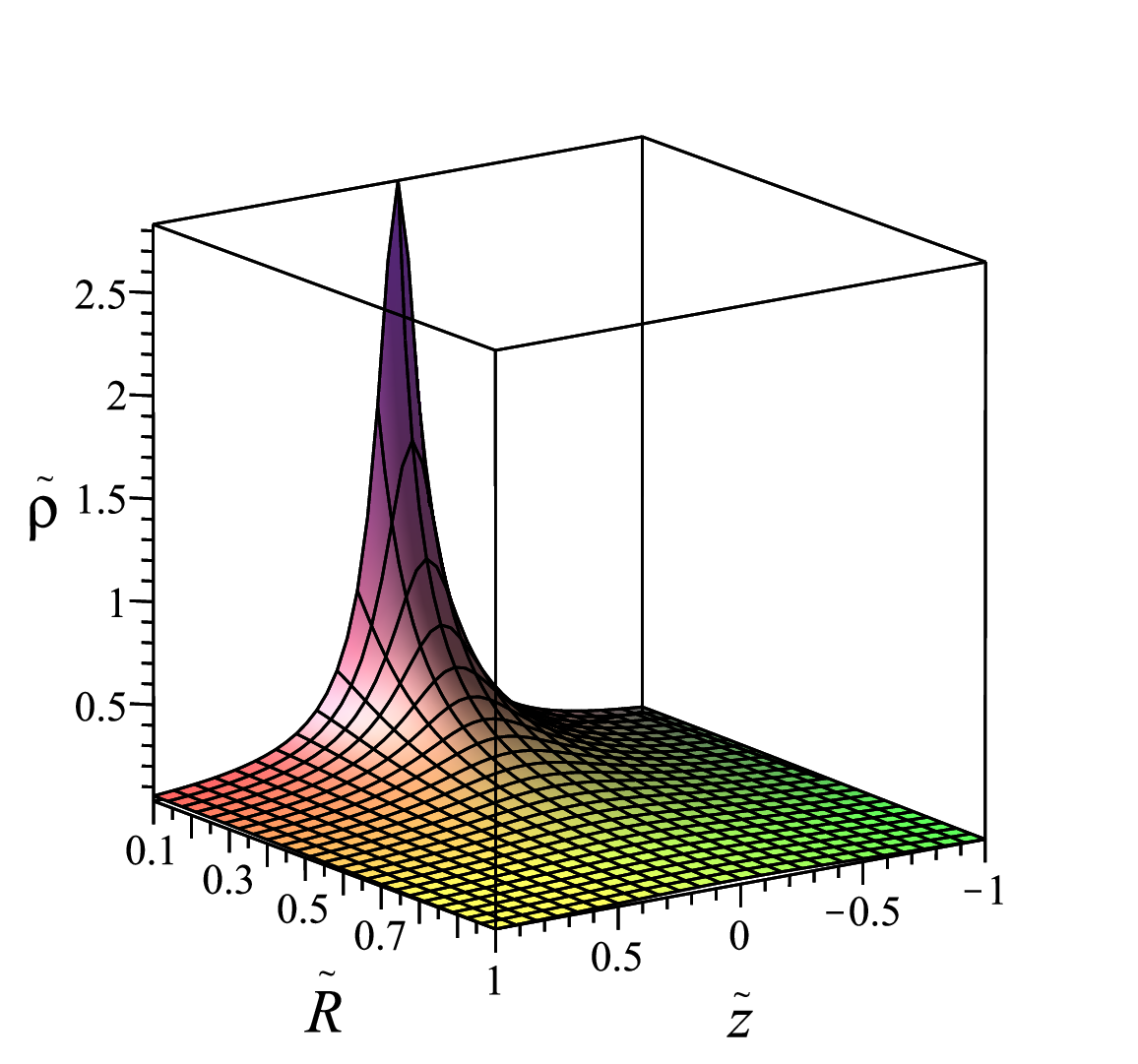} &
\includegraphics[width=0.32\textwidth]{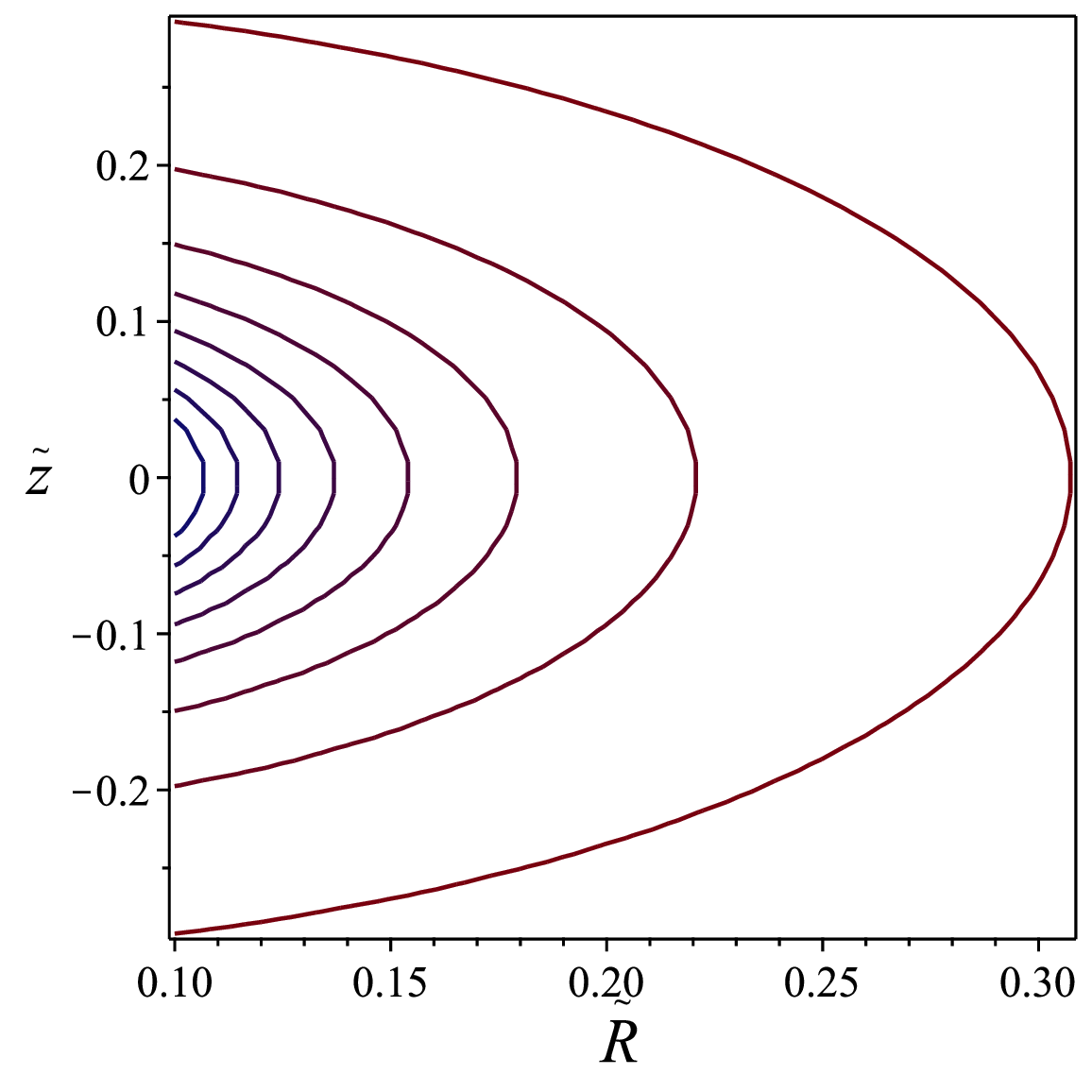}  \\
 (a) &  (b)  
\end{array}
$$	
\caption{  
$(a)$ The relativistic energy density profile $\tilde \rho$ and $(b)$  the isodensity curves for a   $\alpha = 0$ type relativistic galaxy model  composite by   three components  bulge,  thick disk and  dark matter halo, 
as functions of $\tilde R$ and $\tilde z$.
  }
\label{fig:fig7}
\end{figure}

\begin{figure}
$$
\begin{array}{cc}
\includegraphics[width=0.4\textwidth]{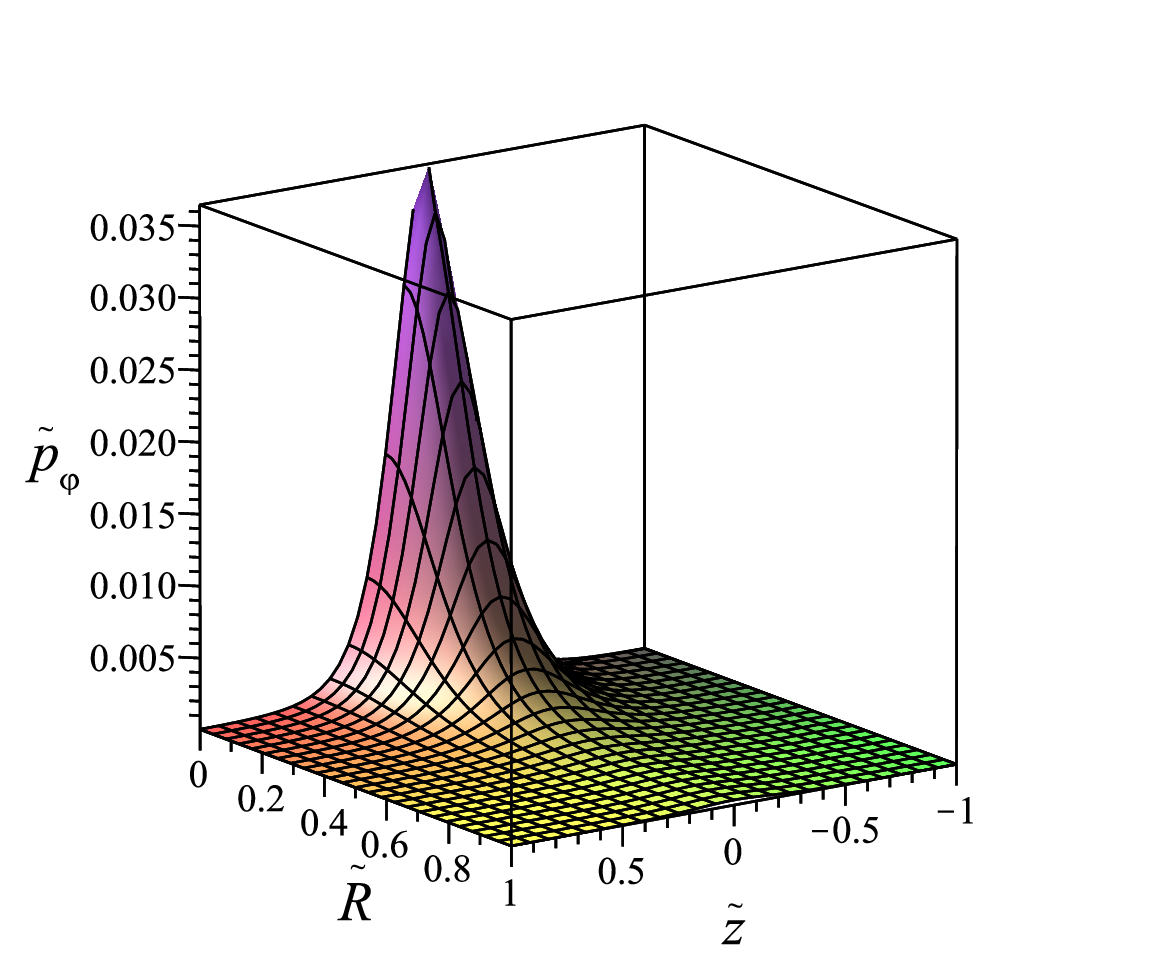} &
\includegraphics[width=0.32\textwidth]{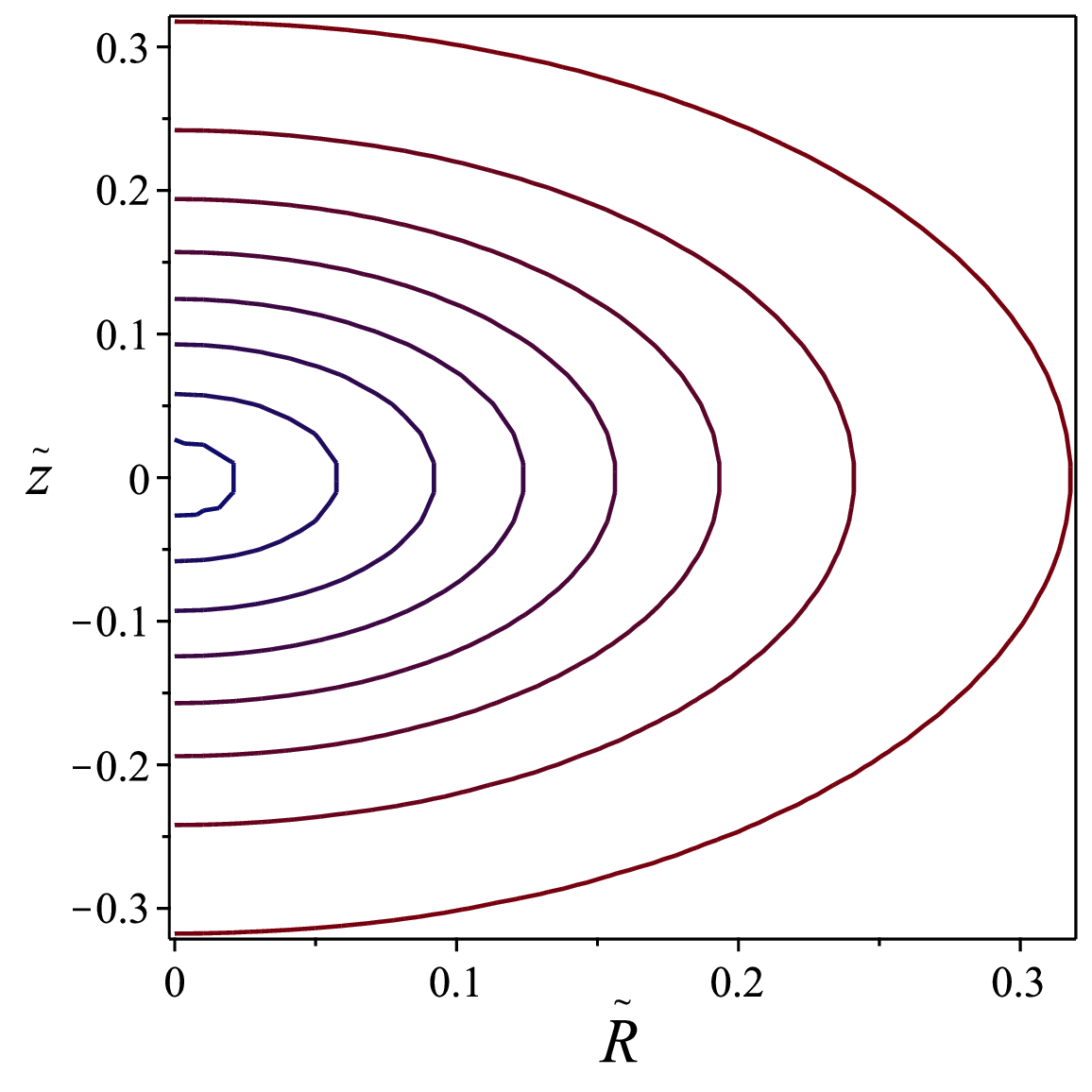}  
\\
 (a) &  (b)  \\
&  \\
\includegraphics[width=0.4\textwidth]{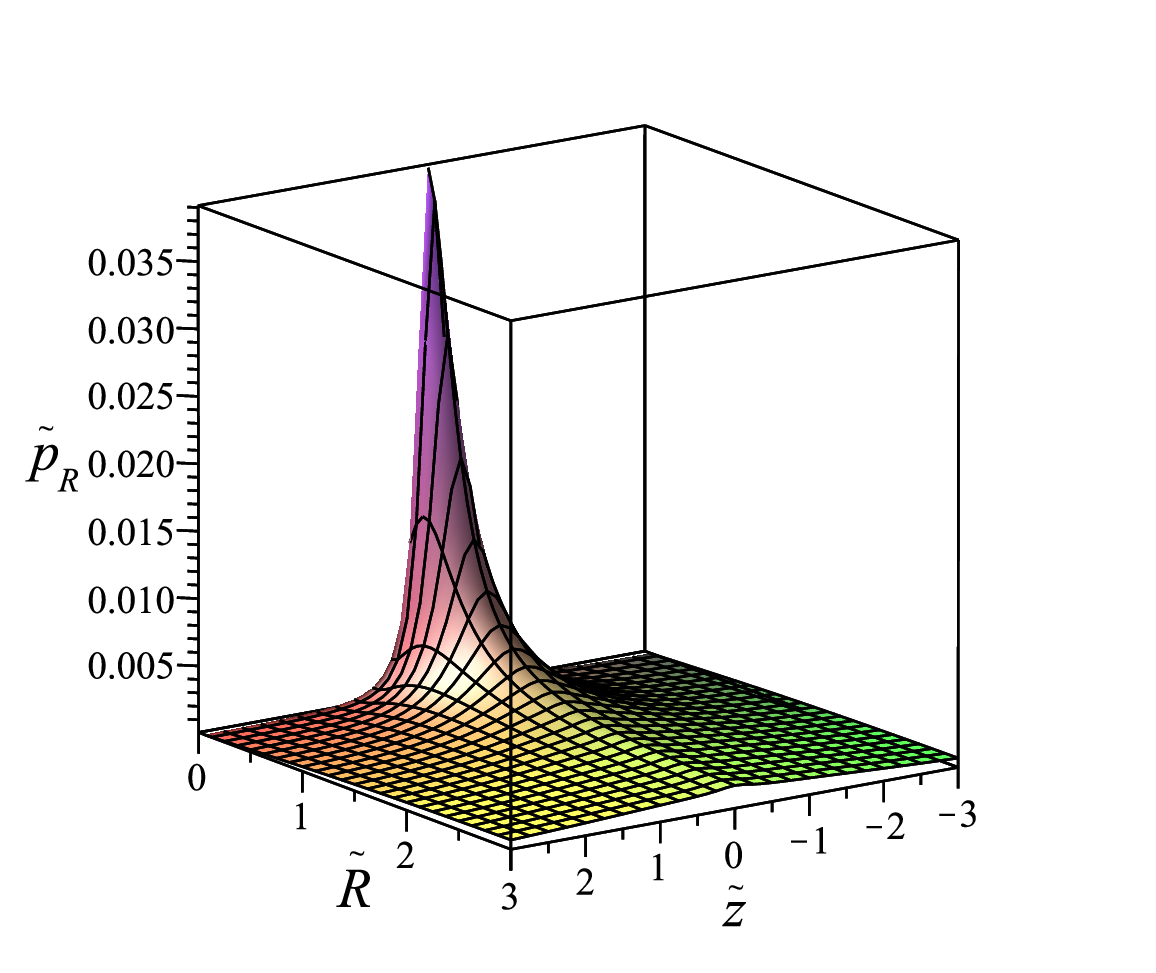} &
\includegraphics[width=0.32\textwidth]{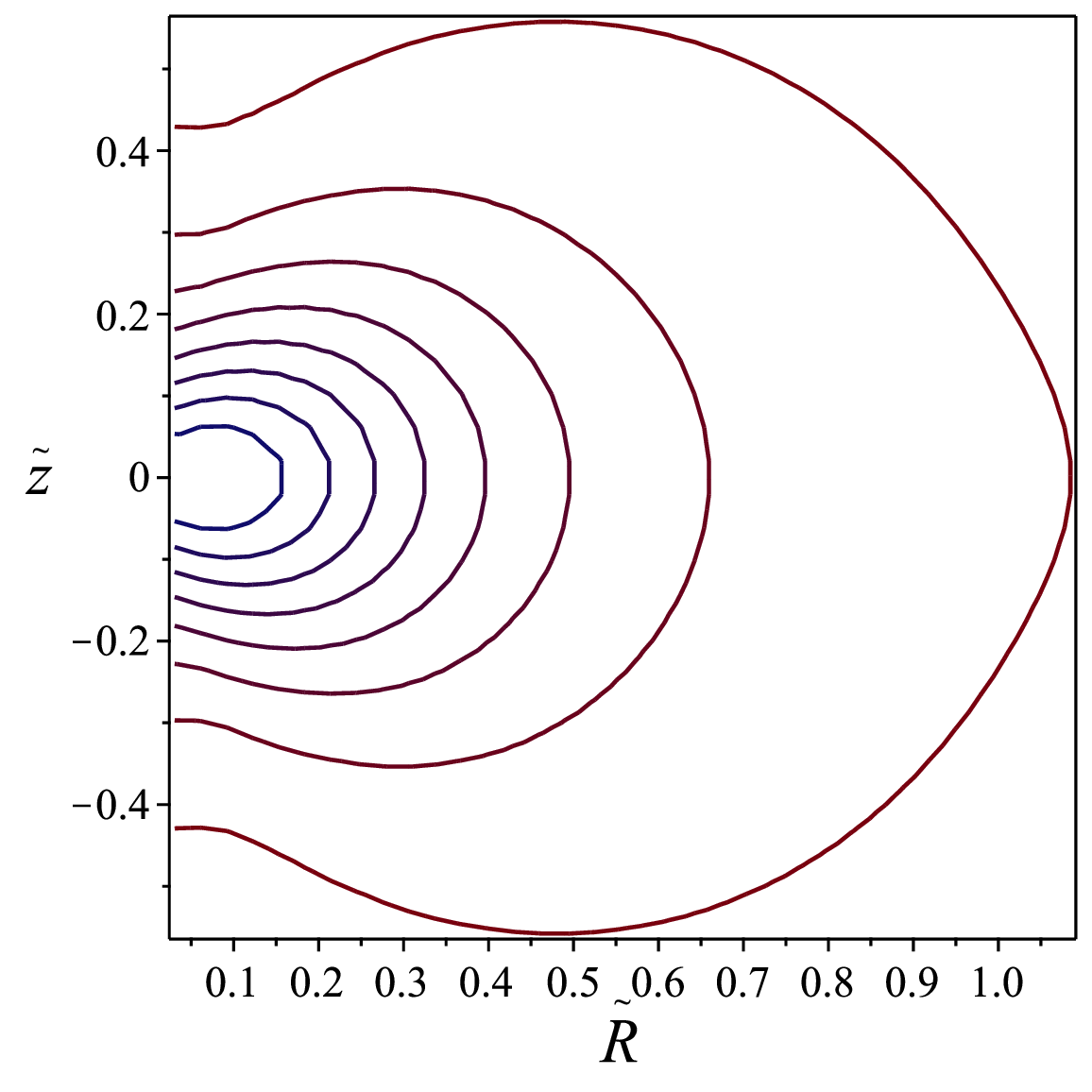} 
 \\
  (c) &  (d)  \\
&  \\
\includegraphics[width=0.4\textwidth]{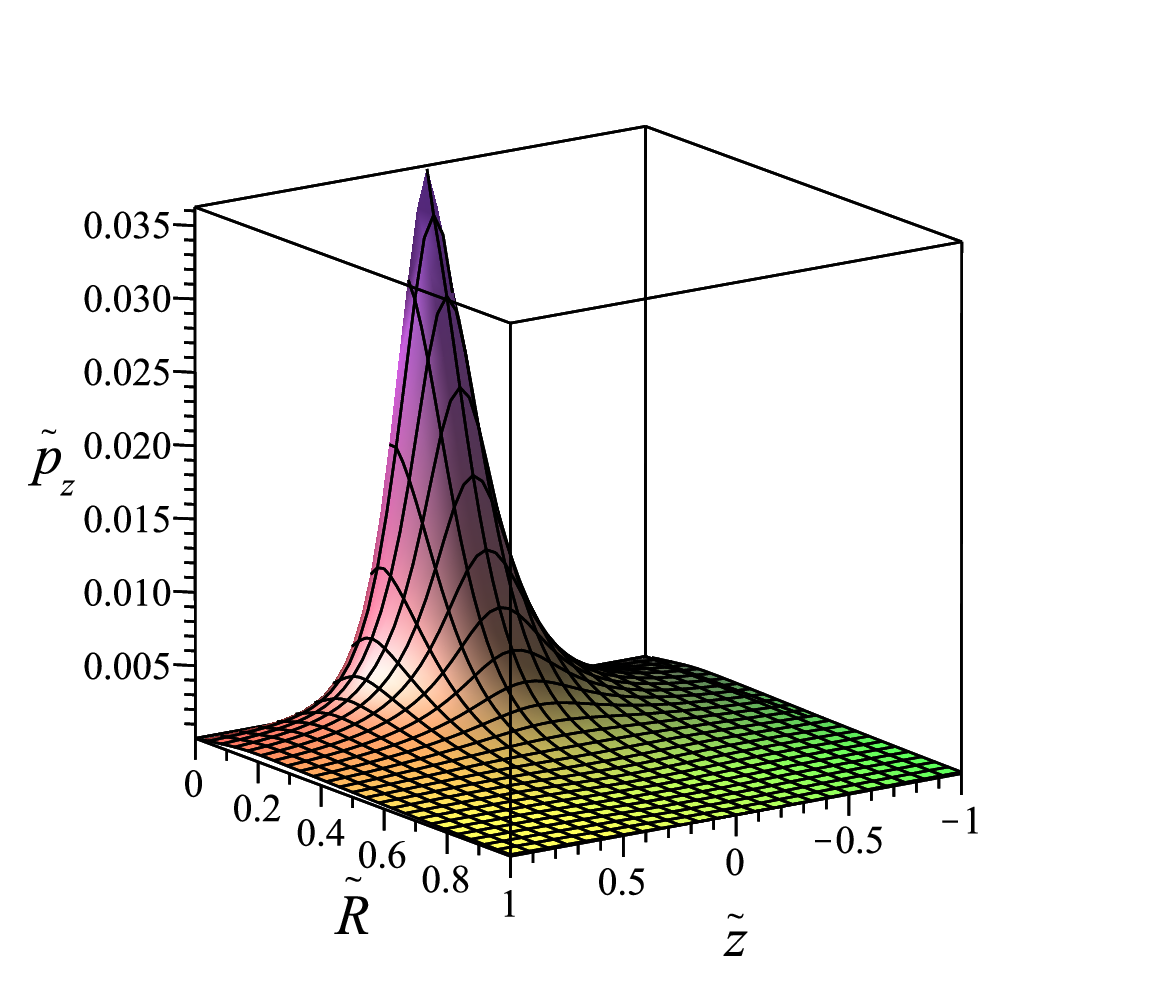} &
\includegraphics[width=0.32\textwidth]{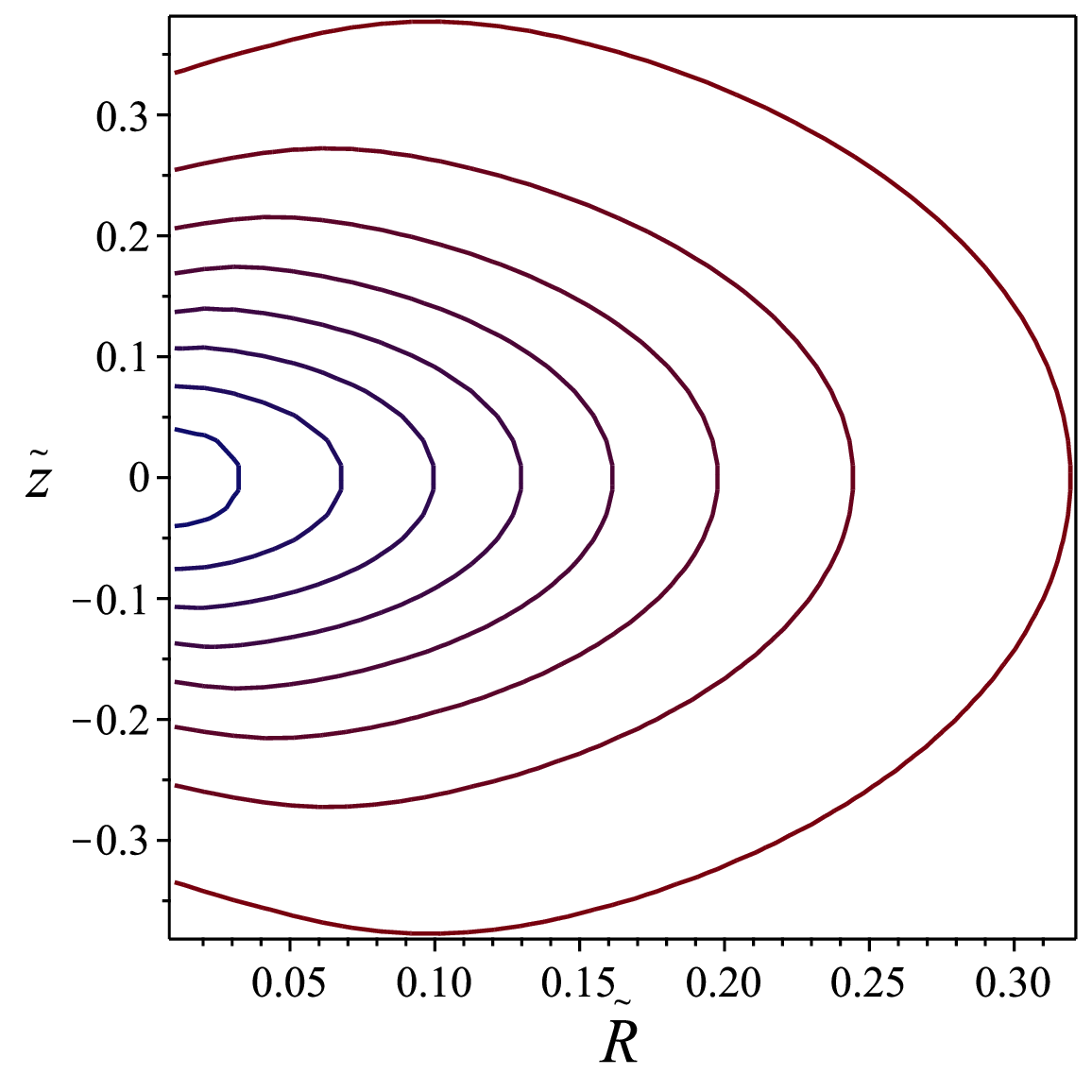} 
\\
 (e) &  (f)  
\end{array}
$$	
\caption{The surfaces and level curves of the pressures 
$\tilde p_\varphi$,  $\tilde p_R$ and  $\tilde p_z$, as functions of $\tilde R$ and $\tilde z$, for a    $\alpha = 0$ type relativistic galaxy model  composite by   three components  bulge,  thick disk and  dark matter halo.
}
\label{fig:fig8}
\end{figure}

In the relativistic  case, for simplicity let us consider again the metric potential 
$\nu$  (\ref{eq: nuani})  in cylindrical coordinates   for the first family of spacetimes when  
 $\alpha=0$ and $\alpha=1$. Since the expressions for the main physical quantities are huge, its analysis is better done graphically.  However, note that the  energy density  always  is a positive quantity in according to the  weak  energy condition.
The parameter values are choose using as reference 
the Newtonian potential parameters computed  in  \cite{Bajkova} $a_D=4.4$,   $a_H=7.7$, 
  $b_B=0.2672$,  $b_D=0.3084$, $M_b=443$,  $M_d=2798$
 and  $M_h=12474$.
 Thus,  in figures  \ref{fig:fig7}, \ref{fig:fig8},  \ref{fig:fig10} and  \ref{fig:fig11} we graph, as function of  $\tilde R = R / G M_D$ and  $\tilde z = z / G M_D$, the surfaces and
level curves of the energy density
$\tilde \rho = G^3 M_D^2 \rho $
and pressures
$\tilde p_\varphi = G^3 M_D^2 p_\varphi  $,
$\tilde p_R = G^3 M_D^2 p_R $
and 
$\tilde p_z = G^3 M_D^2 p_z  $
for  a   relativistic galaxy model  composite by   three components  bulge,  thick disk and  dark matter halo with  $\alpha = 0$ and parameters
$\tilde a_D= a_D / G M_D = 4.4$,   $\tilde a_H= a_H / G M_D = 7.7$, 
 $\tilde b_B= b_B / G M_D = 0.2672$,  $\tilde b_D= b_D / G M_D = 0.3084$, $\tilde M_B= M_B / M_D = 0.1583 $ 
 and  $\tilde M_H= M_H /  M_D =4.4582 $,
and $\alpha = 1$    
with parameters
$\tilde a_D=  10$,   $\tilde a_H=  15$, 
$\tilde b_B=0.5$,
$\tilde b_D= 0.3048$  and the same values of 
$\tilde M_B $ and $\tilde M_H $.
We observer that   the 
energy density, as in the Newtonian case, presents a central cusp and then it  decreases
rapidly with the radius.  The principal  stresses
are positive quantities  which means that we have pressure everywhere, but like the shell models, for the  other two families of solutions we find that 
far from  the central region of the structures  the tangential stresses   can take  very small negative values.

\begin{figure}
$$
\begin{array}{c}
\includegraphics[width=0.4\textwidth]{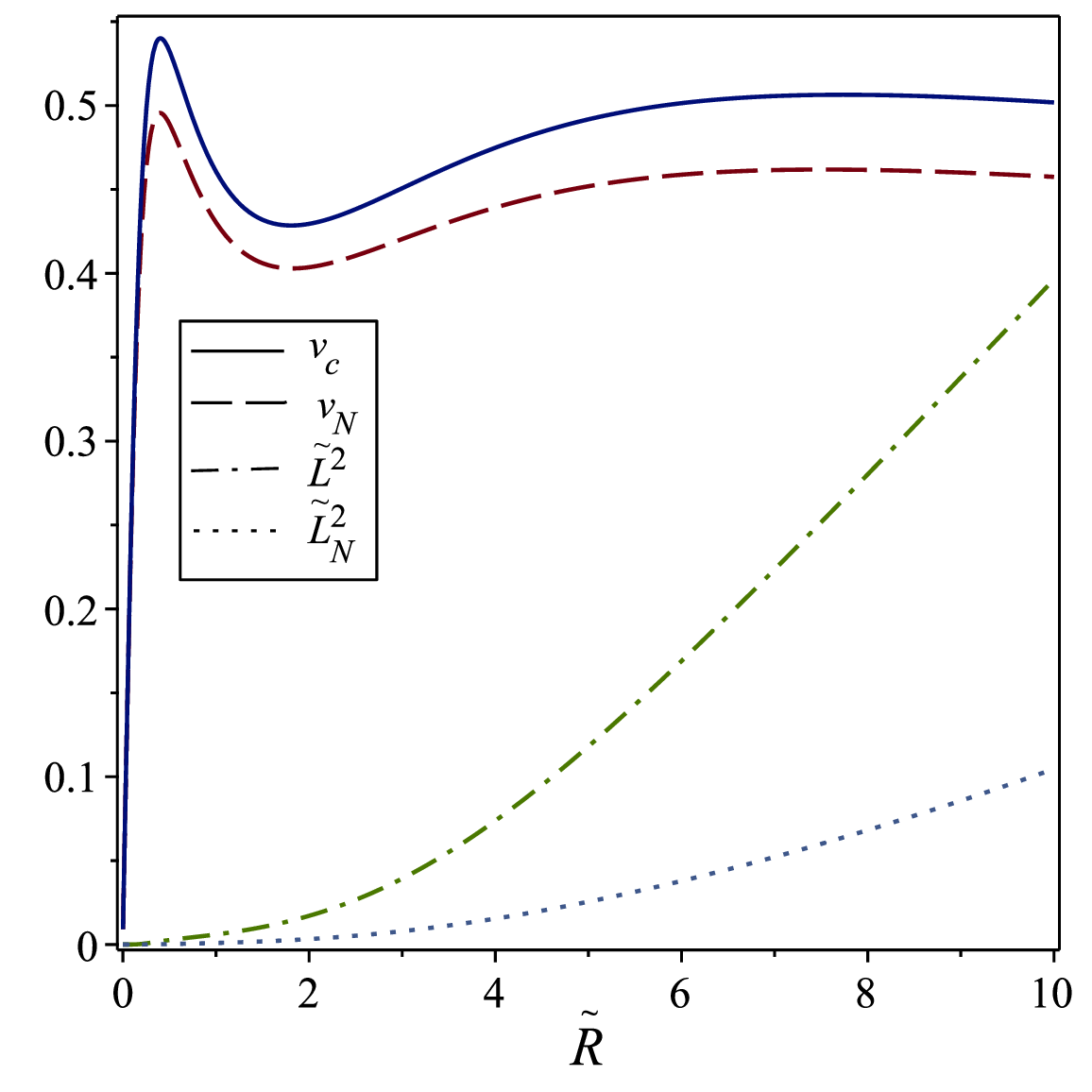} 
\end{array}
$$	
\caption{ The relativistic circular speed  $v_c$ (solid curve),
the Newtonian rotation curve  $v_{N}$ (dashed curve), the relativistic and Newtonian specific angular momenta $\tilde L^2$ (dashed-dotted curve)  and
 $\tilde L_N^2$ (dotted curve), scaled by 200,  for the system composite by   bulge,  thick disk and  dark matter halo, with  $\alpha = 0$, 
as functions of $\tilde R$.  }
\label{fig:fig9}
\end{figure}

\begin{figure}
$$
\begin{array}{cc}
\includegraphics[width=0.5\textwidth]{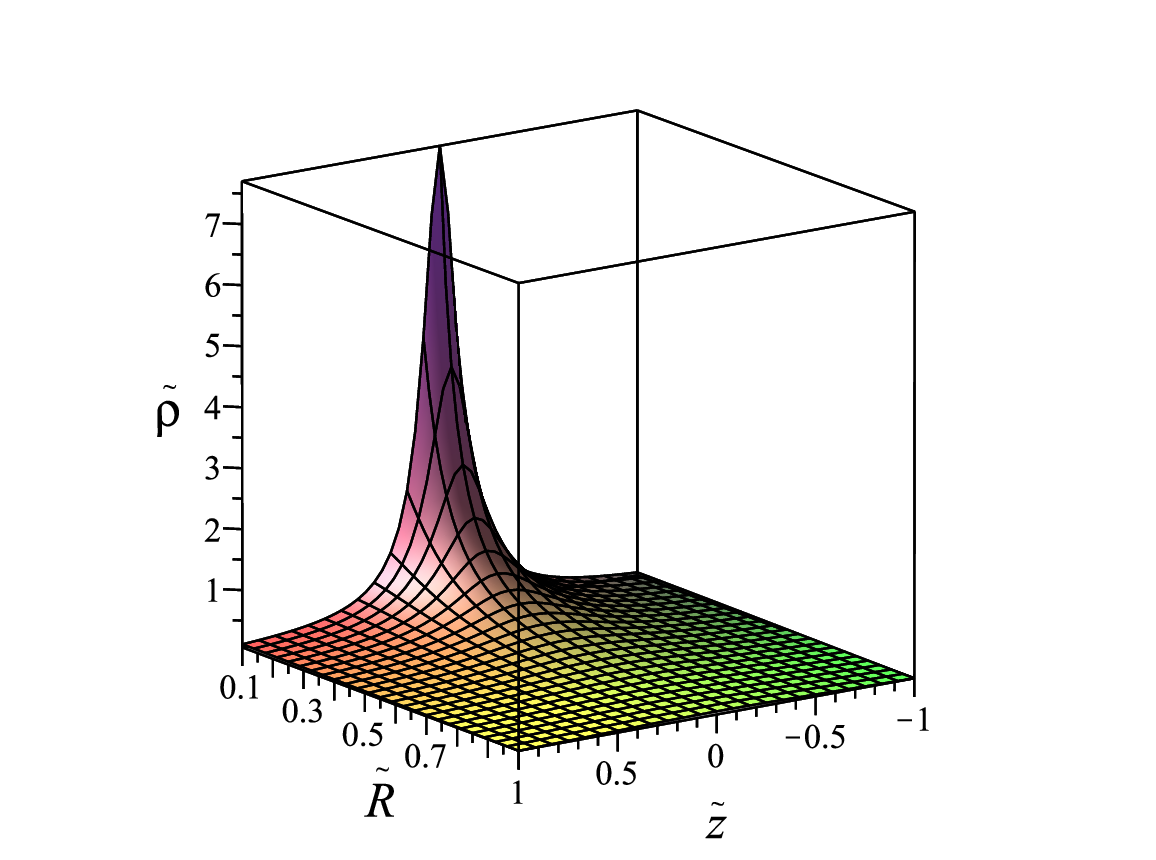} &
\includegraphics[width=0.32\textwidth]{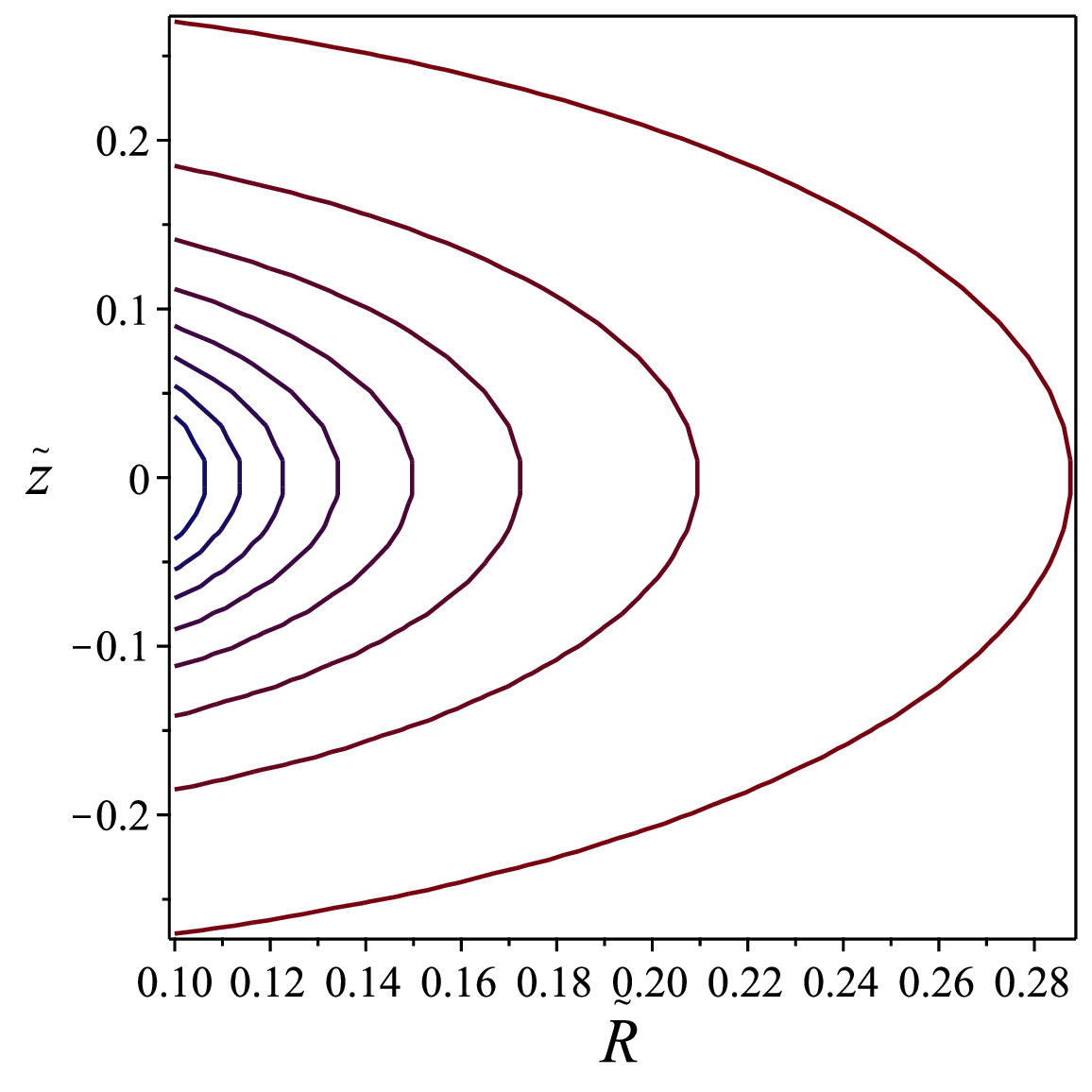}  \\
 (a) &  (b)  
\end{array}
$$	
\caption{  
$(a)$ The relativistic energy density profile $\tilde \rho$ and $(b)$  the isodensity curves for a   $\alpha = 1$ type relativistic galaxy model  composite by   three components  bulge,  thick disk and  dark matter halo, 
as functions of $\tilde R$ and $\tilde z$.
  }
\label{fig:fig10}
\end{figure}

\begin{figure}
$$
\begin{array}{cc}
\includegraphics[width=0.4\textwidth]{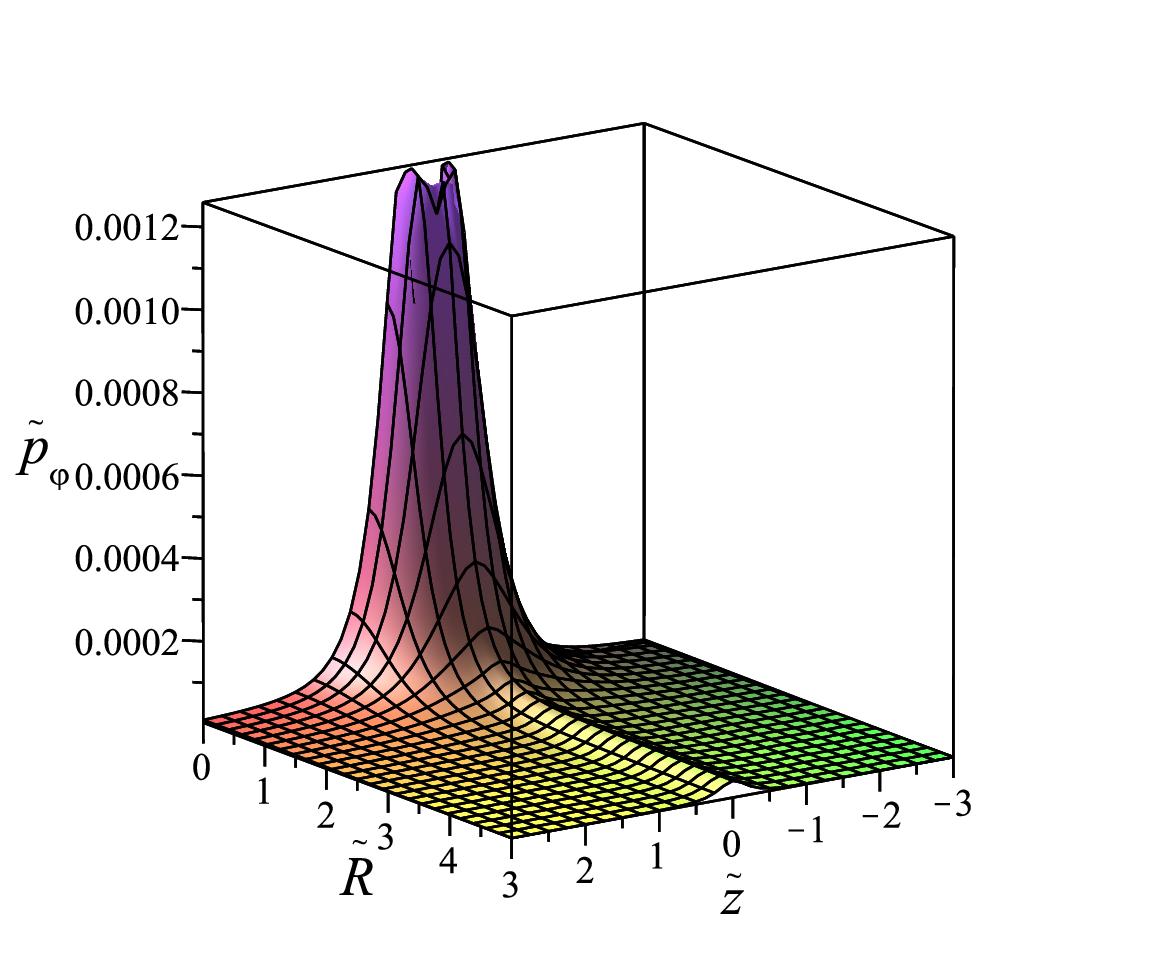} &
\includegraphics[width=0.32\textwidth]{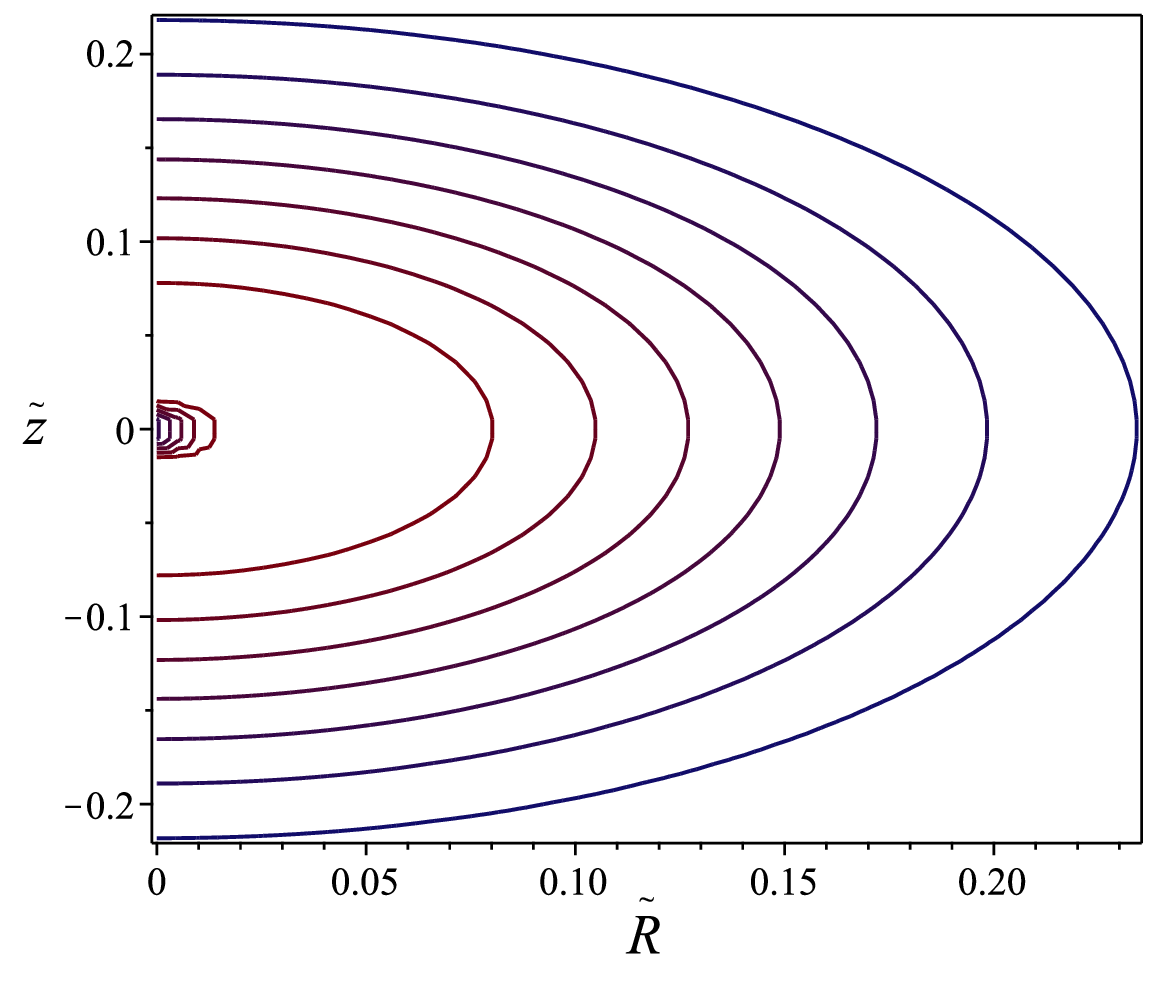}  
\\
 (a) &  (b)  \\
&  \\
\includegraphics[width=0.4\textwidth]{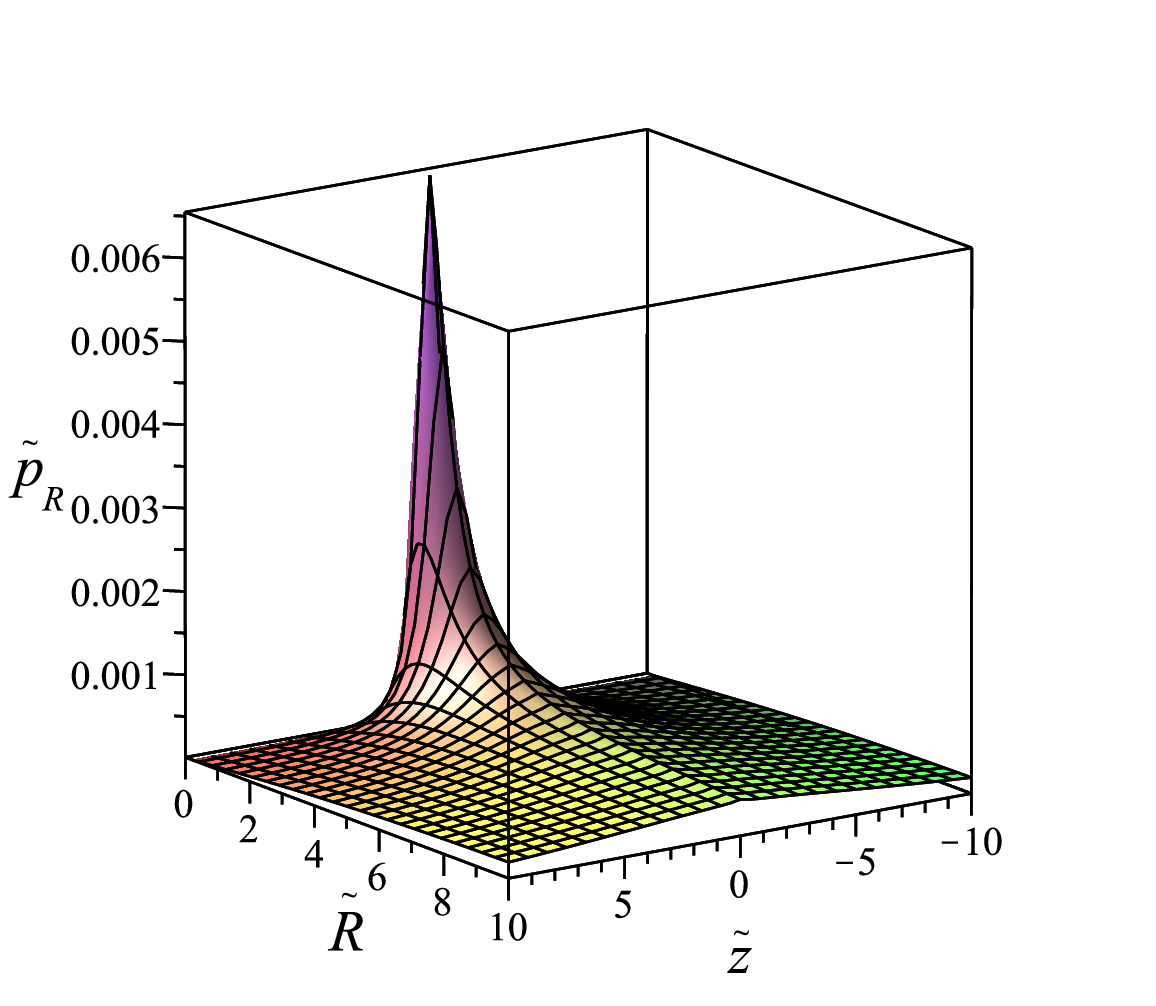} &
\includegraphics[width=0.32\textwidth]{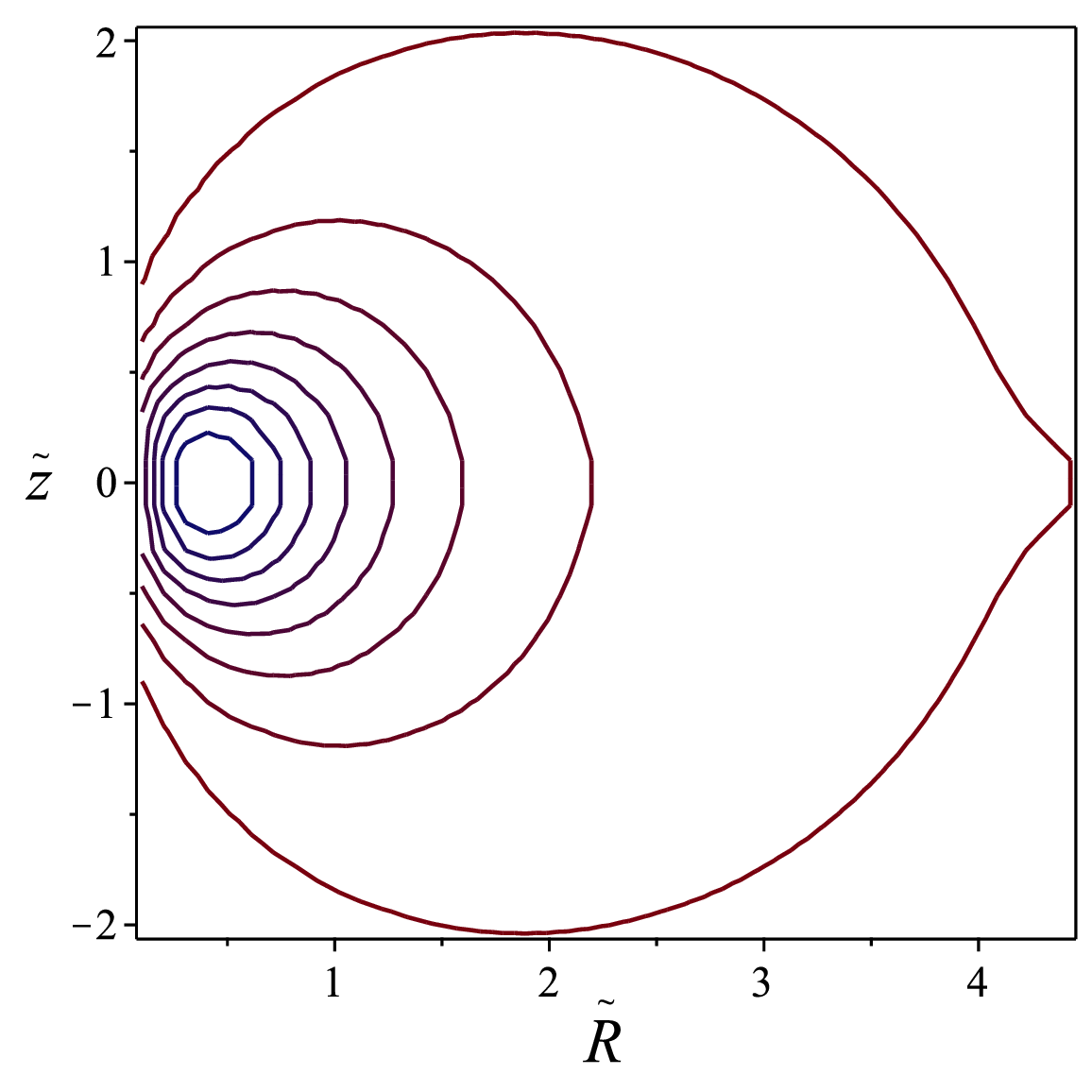} 
 \\
  (c) &  (d)  \\
&  \\
\includegraphics[width=0.4\textwidth]{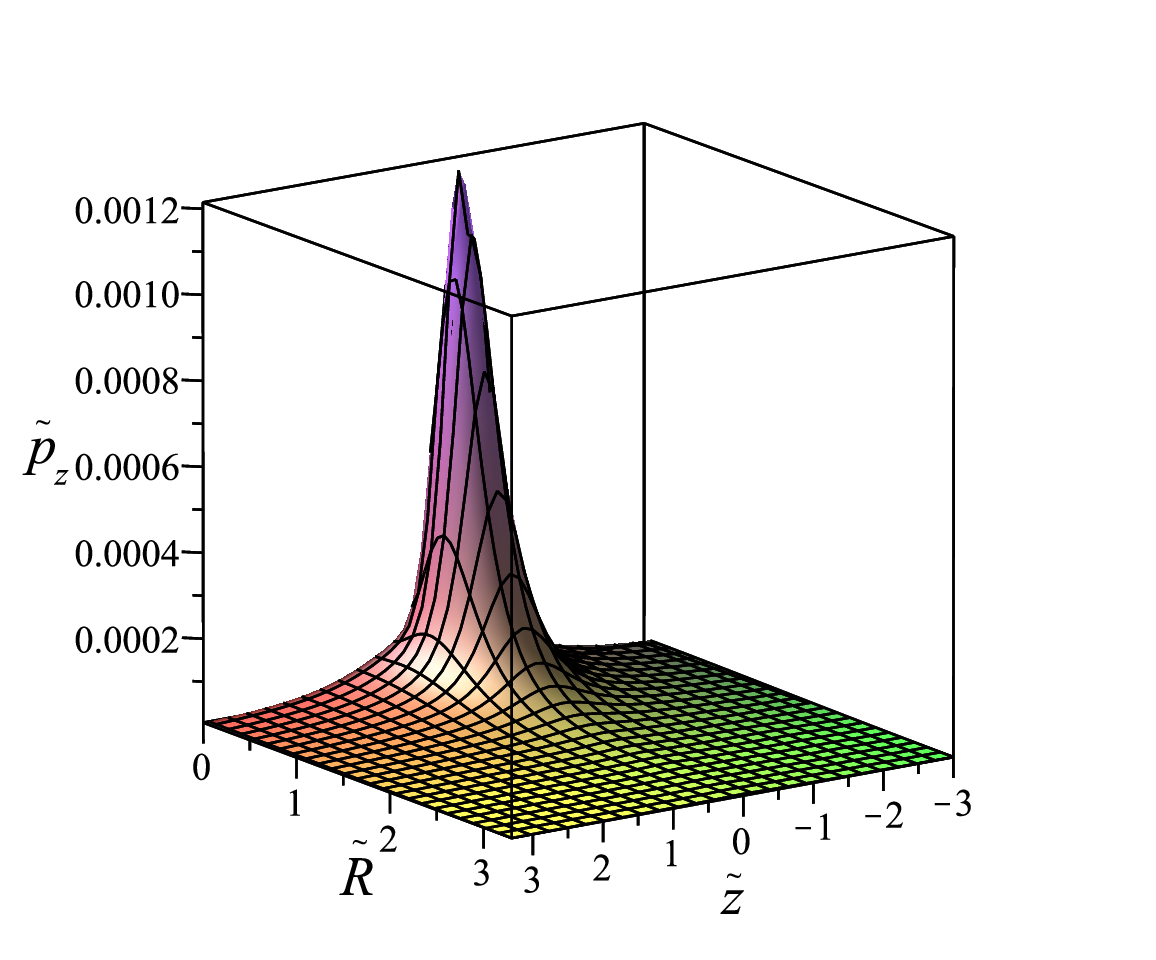} &
\includegraphics[width=0.32\textwidth]{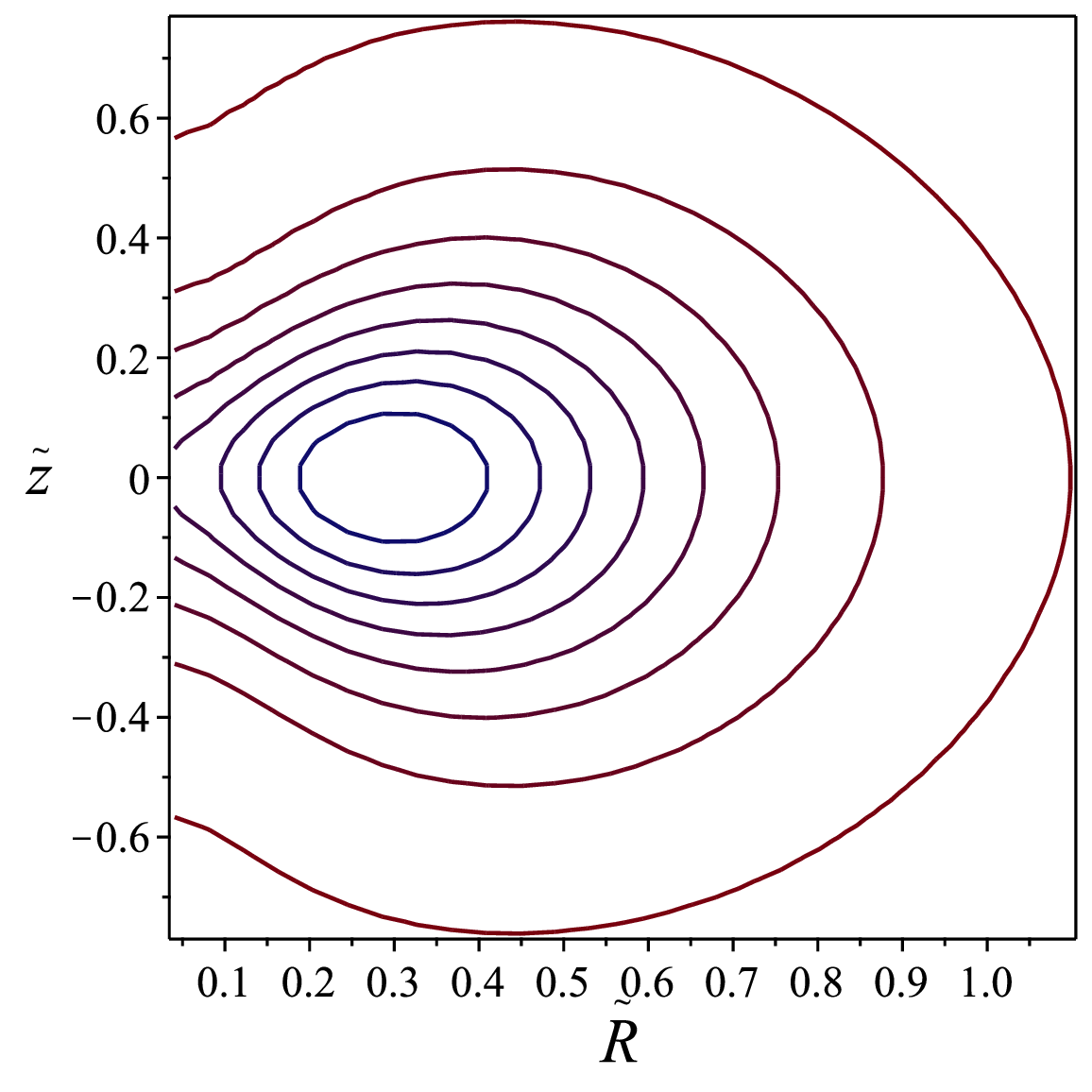} 
\\
 (e) &  (f)  
\end{array}
$$	
\caption{The surfaces and level curves of the pressures 
$\tilde p_\varphi$,  $\tilde p_R$ and  $\tilde p_z$, as functions of $\tilde R$ and $\tilde z$, for a    $\alpha = 1$ type relativistic galaxy model  composite by   three components  bulge,  thick disk and  dark matter halo.
}
\label{fig:fig11}
\end{figure}

\begin{figure}
$$
\begin{array}{c}
\includegraphics[width=0.4\textwidth]{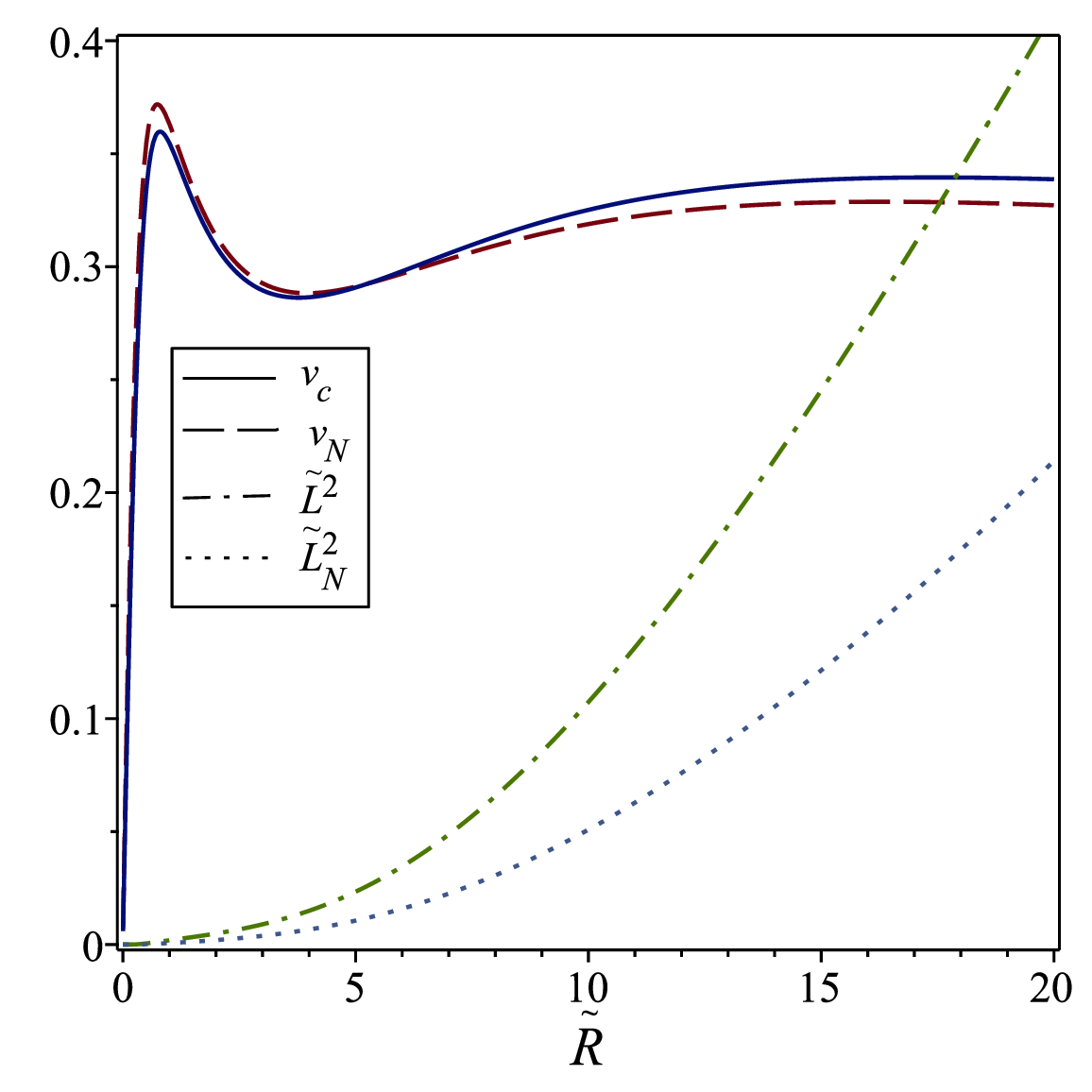} 
\end{array}
$$	
\caption{ The relativistic circular speed  $v_c$ (solid curve),
the Newtonian rotation curve  $v_{N}$ (dashed curve),
the relativistic and Newtonian specific angular momenta $\tilde L^2$ (dashed-dotted curve)  and
 $\tilde L_N^2$ (dotted curve), scaled by 200,  for the system composite by   bulge,  thick disk and  dark matter halo, with  $\alpha = 1$, 
as functions of $\tilde R$.  }
\label{fig:fig12}
\end{figure}

In figures  \ref{fig:fig9} and  \ref{fig:fig12}  we plot, as function of $\tilde R$,  the relativistic circular speed profile $v_c$, the Newtonian rotation curves $v_{N}$, the specific angular momentum  $\tilde L^2 = L^2 / G^2 M_D^2$ and
the Newtonian angular momentum  $\tilde L_N^2 = L_N^2 / G^2 M_D^2$,
for the same value of parameters.  
We observer that  when  $\alpha = 0$   the relativistic effects increase the circular speed of the particles  everywhere  whereas that when $\alpha = 1$ such effects initially  decrease the tangential speed  but after  a  certain value of $\tilde R$  they increase it.
Now for stars around  a typical galaxy 
the tangential velocity    is about   200 - 300 km/s
and accordingly, the geometric effects are expected to be small.  
For example, in the case of our Galaxy for a radius of   $R=0.4$ kpc  which corresponds to a  tangential velocity about $260.8$ 
km/s we obtain that   the relativistic corrections  
are in both cases of the order of $9.86  \times 10^{-5}$ km/s.
We believe that  for a sufficiently large distance traveled by a star such effects  could become significant especially in the central region  where the gravitational fields are strong.
The graphs also show  that the rotation curves  are  flattened after a certain value of the radial distance as observational data indicate.
We also observer that  equatorial, circular orbits of test particles moving around the structures are stables against radial
perturbations. 
In addition, since $d L ^2 / dr > d L_N^2/ dr$ everywhere, we see 
the relativistic  circular orbits    are  more stable    than  Newtonian ones.  
It should be noted that we focus our stability study only on radial perturbations because we are considering only circular orbits on the galactic plane $z=0$ and not the complete structure.
A more realistic stability analysis  of the  matter distribution should be based on perturbations  of the energy momentum tensor of the fluid \cite{Ujevic}  which may be not a trivial task  and will be left as a future research subject. 

On the other hand, when  $\alpha = 0$  we find that 
as the mass of the dark matter halo  or the central bulge mass are increased, keeping the other parameters constant, the models become more relativistic and after certain value of such parameters 
the dominant energy condition is not satisfied and consequently, the orbits are no longer stable.  
The opposite occurs when these parameters are decreased. 
In both scenario,   stresses are always  positive quantities (pressure).  

Meanwhile, when   $\alpha = 1$  we find that as the mass of the dark matter halo  is increased    the circular speed of the orbits initially decreases but after  a  certain value of $\tilde R$ the models become more relativistic. 
On the contrary,  when such parameter is decreased   the models initially become more relativistic  but then  
the tangential  speed decreases.
In turn, as  the central bulge mass  is increased one finds that  the circular speed also increases in all regions of the structures
and conversely  as it is  decreased  the rotation curves also decrease everywhere. 
In both situations, as those parameters are increased the stresses fast become negative (tension) in the central region of the structures.  For example, the models with  $m_h = 5.2$ or
$m_b=0.18$ have a central region with  tension but for 
$m_h \leq 5.1$ or $m_b \leq 0.17$ we still have pressure.
In turn, as such parameters are decreased, we always have pressure.
 
\section{ Conclusions}
Anisotropic fluid sources for conformastat spherically  symmetric  spacetimes 
from Newtonian potential-density pairs and its corresponding  tangential speed profile were constructed. 
As simple applications of the method, three analytical families of  spherically symmetric spacetimes
representing  anisotropic matter configurations
were presented  and  applied to the construction of  new  models  of relativistic   spherical  shells of finite thickness, and
of two relativistic galaxy models composite by a central spherical  bulge, the thick disk and the dark matter halo, expressing in this case  the metric in   cylindrical coordinates. 
The bulge and disk seed  potentials
 were  the  Plummer  and Miyamoto-Nagai models  and in the case of the  dark matter halo,   the NFW model. 
 
In all models considered,  the energy conditions are satisfied and were found stable circular orbits. However,  we observed  that for the first family of spacetimes it is always possible  to find in all the cases parameters for which the  principal stresses  are positive quantities  (pressure)  everywhere, but for the other two families,
 except for the cases  $\beta = 2$ and $\gamma = 1$, 
it was  found that far from  the central region of the structures  the azimuthal stresses  take  very small negative values.




\end{document}